\documentclass[a4paper,11pt]{article}
\pdfoutput=1 

\usepackage{amssymb,mathrsfs}
\usepackage{bbm}
\usepackage{siunitx}
\usepackage{amsmath}
\usepackage{braket}
\usepackage{mathtools}
\usepackage[usenames,dvipsnames,svgnames,table]{xcolor}
\usepackage [utf8] {inputenc}
\usepackage{color}
\usepackage{lmodern}
\usepackage{footnote}
\usepackage{slashed}
 \usepackage{mathrsfs} \usepackage[pdftex]{graphicx}
\usepackage{multirow}
\usepackage{jheppub} 
\usepackage{here}
\usepackage{hyperref}
\usepackage{float}
\usepackage{subfigure}
\newcommand{\eq}{\begin{eqnarray}}
\newcommand{\en}{\end{eqnarray}}

\title{Lellouch-L\"uscher factor for the $K\to 3\pi$ decays}

\author[1]{Jin-Yi Pang,}
\affiliation[1]{College of Science, University of Shanghai for Science and Technology, Shanghai 200093, China}
\emailAdd{jypang@usst.edu.cn}

\author[2]{Rishabh Bubna,}
\affiliation[2]{Helmholtz-Institut f\"ur Strahlen- und Kernphysik (Theorie)\\ and Bethe Center for Theoretical Physics, Universit\"at Bonn, 53115 Bonn, Germany}
\emailAdd{bubna@hiskp.uni-bonn.de}

\author[2]{Fabian M\"uller,}
\emailAdd{f.mueller@hiskp.uni-bonn.de}

\author[2,3]{Akaki Rusetsky,}
\affiliation[3]{Tbilisi State  University,  0186 Tbilisi, Georgia}
\emailAdd{rusetsky@hiskp.uni-bonn.de}

\author[4]{and Jia-Jun Wu}
\affiliation[4]{School of Physical Sciences, University of Chinese Academy of Sciences, Beijing 100049, China}
\emailAdd{wujiajun@ucas.ac.cn}

\abstract{

\noindent
We derive an explicit expression for the Lellouch-L\"uscher (LL)  factor in the $K\to 3\pi$
decays at leading order (without derivative couplings). Several important technical details 
are addressed, like a proper decomposition into the isospin amplitudes, the choice of a
minimal set of effective couplings and the renormalization, as well as the
algorithm for the solution of the pertinent Faddeev equations in the infinite volume
which is based on the contour deformation method. Most importantly, our numerical
results demonstrate that the three-body force contributes very little to the LL factor.
This result paves the way for the study of the $K\to 3\pi$ decays on the lattice.
  
}

\allowdisplaybreaks
\begin{document}
\maketitle

\section{Introduction}

In the recent years, first results of the lattice studies of the spectrum in the three-particle
systems have started to appear. These results have been analyzed by using three
different but conceptually equivalent formalisms known
as Relativistic Field Theory (RFT)~\cite{Hansen:2014eka, Hansen:2015zga},
Non-Relativistic Effective Field Theory (NREFT)~\cite{Hammer:2017uqm, Hammer:2017kms} and
Finite-Volume Unitarity (FVU)~\cite{Mai:2017bge,Mai:2018djl} approaches.
The references to the
activities in the field that include both formal developments as well as
actual simulations on the lattice are collected here~\cite{Kreuzer:2008bi,Kreuzer:2009jp,Kreuzer:2010ti,Kreuzer:2012sr,Briceno:2012rv,Polejaeva:2012ut,Jansen:2015lha,Hansen:2014eka,Hansen:2015zta,Hansen:2015zga,Hansen:2016fzj,Guo:2016fgl,Sharpe:2017jej,Guo:2017crd,Guo:2017ism,Meng:2017jgx,Briceno:2017tce,Hammer:2017uqm,Hammer:2017kms,Mai:2017bge,Guo:2018ibd,Guo:2018xbv,Klos:2018sen,Briceno:2018mlh,Briceno:2018aml,Mai:2019fba,Guo:2019ogp,Guo:2020spn,Blanton:2019igq,Pang:2019dfe,Jackura:2019bmu,Briceno:2019muc,Romero-Lopez:2019qrt,Konig:2020lzo,Brett:2021wyd,Hansen:2020zhy,Blanton:2020gha,Blanton:2020jnm,Pang:2020pkl,Hansen:2020otl,Romero-Lopez:2020rdq,Blanton:2020gmf,Muller:2020vtt,Blanton:2021mih,Muller:2021uur,Beane:2007es,Detmold:2008fn,Detmold:2008yn,Blanton:2019vdk,Horz:2019rrn,Culver:2019vvu,Fischer:2020jzp,Alexandru:2020xqf,Romero-Lopez:2018rcb,Blanton:2021llb,Mai:2021nul,Mai:2018djl,Muller:2020wjo,Muller:2022oyw,Hansen:2021ofl,Blanton:2021eyf,Severt:2022jtg,Baeza-Ballesteros:2023ljl,Draper:2023xvu,Bubna:2023oxo}.
For more information
on the subject, we refer the reader to the two recent reviews on the
subject~\cite{Hansen:2019nir,Mai:2021lwb}.

One of the most intriguing and challenging tasks in the three-particle sector is the study
of the three-particle decays. In this paper, we exclusively focus on the decays which
proceed via the weak or electromagnetic forces -- in other words, these particles would
be absolutely stable in pure QCD.
A classic example for this kind of decays is given by $K\to 3\pi$.
Here, one could also count the decays that proceed via the isospin
breaking (the most prominent example of this sort is given by the $\eta\to 3\pi$).
Putting differently, the $\eta$ is stable in pure {\em isospin-symmetric} QCD, with
$m_u=m_d$, and the decay amplitude is proportional to the quark mass difference
$(m_d-m_u)$ (the higher-order terms in this small parameter will be neglected). A
completely different picture emerges where both the formation and the decay of a
particle (resonance) is caused by strong interactions which are described by the QCD Lagrangian
alone. In this case, analytic continuation into the complex plane becomes inevitable.
We shall not consider such processes in the present paper.

The main conceptual problem in the determination of the decay amplitudes on the lattice
are caused by the presence of the final-state interactions. Since the mass of the decaying particle lies above the sum
of the masses of the decay products, the propagators in the Feynman diagrams that
describe final-state interaction may become singular in the integration region. This,
at a fixed energy, leads to an irregular dependence
of the measured matrix element on the lattice volume,
rather than to the exponentially suppressed finite-volume corrections which emerge in the
observables of stable particles. Lellouch and L\"uscher~\cite{Lellouch:2000pv} have
shown that, in case of two-particle decays, this singular behavior is contained in a single
function (the so-called LL factor), which relates the decay amplitudes in the infinite and in
a finite volume. A crucial property of this function is that it depends on the dynamics in the
final state (the two-body phase shift\textcolor{black}{s}, in this case\footnote{
\textcolor{black}{In general,
the LL factor receives contributions from an infinite tower of the partial waves, see,
e.g.,~\cite{Peterken:2023zwu}.}
}), but not on the interactions that lead
to the decay. Further development of these ideas can be found in
Refs.~\cite{Meyer:2011um,Hansen:2012tf,Bernard:2012bi} which include, in particular,
the generalization to the moving frames and the multi-channel decays.

An analog of the LL formula for the three-particle decays
has been derived only very
recently, independently by two
groups~\cite{Muller:2020wjo,Hansen:2021ofl,Muller:2022oyw}.
Albeit there is no substantial conceptual difference between the two- and three-particle
cases (for example, the functions that describe an irregular volume-dependence also
in the three-particle case depend
solely on the parameters of the final-state interactions), an algebraic structure of the
final expressions is much more cumbersome and obscure. In particular,
there exists only one
LL factor in the two-body decays, owing to the kinematic constraints (the magnitude of the
relative momentum in the two-body decays is fixed by energy-momentum conservation).
On the contrary, the three-body decays are characterized by an (infinite)
tower of effective couplings
that describe the dependence of the decay amplitudes on different kinematical
invariants. Consequently, the LL factor is not a single function but a matrix, which
should be truncated in actual calculations.

The aim of the present paper is to consider a single physical process, the kaon decay
into three pions, and to work out the LL factor
explicitly for this process in different isospin channels. \textcolor{black}
{In order to achieve this goal, the NREFT approach will be used. The choice
of the process was not completely arbitrary -- 
we believe that the three-pion decay of the kaon will likely be one of the first three-particle decays studied using lattice QCD,
not least since 
it is considered as one of the sources of information
about $CP$ violation in the light quark sector (see, e.g., \cite{Cirigliano:2011ny}).
Moreover, we believe that, taking into account the expected accuracy of lattice
calculations at the present stage, in the beginning
it will be reasonable to truncate all interactions in the two- and three-pion
sectors at the lowest order, i.e., to consider only non-derivative couplings in the
S-wave. 
These approximations will allow us to put the final result in a much more
compact and transparent form, suitable for a direct use by lattice practitioners, even
if the calculation of the most interesting $CP$-odd observables (for instance,
the asymmetry of slopes in the decays of $K^+$ and $K^-$~\cite{NA482:2007ucr})
will, at the end, require the inclusion of the higher-order derivative couplings at
 next-to-leading order.\footnote{For more details on the counting scheme
  in the effective theory, we refer to~\cite{Colangelo:2006va,Gasser:2011ju}.}
Such a generalization can be however performed in a straightforward manner, using the methods described in the present paper.\footnote{The only (small) complication that may arise here is related to the emergent spurious poles in the two-body pion-pion scattering amplitude that could be however removed, using the method described in Refs.~\cite{Ebert:2021epn,Pang:2022nim,Ebert:2023aio}.}
  In order not to overload the presentation with the technical details,
   we omit these higher-order terms in the following.}

Furthermore, there are several technical issues that were addressed only very briefly, or not addressed
at all in the previous work on the problem that was carried out within the NREFT framework.
Since the kaons can decay into different
isospin channels, one has to explicitly write down the Faddeev equations in these channels.\footnote{\textcolor{black}{Note that, in the RFT framework, the inclusion of the different isospin channels has been considered in Ref.~\cite{Hansen:2021ofl}.}}
Moreover, as it is well known~\cite{Bedaque:1998kg,Bedaque:1998km}, these Faddeev 
equations need to be renormalized. A choice of a minimal set of three-body couplings that
suffices to render amplitudes cutoff-independent is rather non-trivial and will be discussed
below. Moreover, we discuss an algorithm which will be used for a numerical solution
of these Faddeev equations in the infinite volume. This algorithm is based on the
deformation of the integration contour into the complex plane and has been known
for a long time in the non-relativistic scattering theory. The discussion in case of
relativistic kinematics in the literature is much more fragmented, and we shall attempt
to fill this gap in the present paper.

The main objective of this paper is however not of a technical nature. Namely, we shall
try to find an answer to the question, whether a prior knowledge of the exact values
of the three-body couplings, which describe short-range three-body interactions, is
essential for the determination of the LL factor.
If the answer to this question were
positive, it would substantially complicate the extraction of the decay amplitudes on
the lattice. Indeed, for this, one would have to first accurately
extract the three-pion couplings from the
measured spectrum of three pions, which is quite a challenging task. Fortunately,
it turns out that the LL factor shows very little dependence on the three-body force.
For this reason, even a rough estimate of the three-body
amplitude, based on Chiral Perturbation Theory (ChPT), will be sufficient for an accurate
calculation of the LL factor which is essentially determined through the S-wave $\pi\pi$
scattering lengths alone. This, in turn, paves the way for a direct extraction
of the $K\to 3\pi$ amplitudes in lattice QCD, circumventing, at the initial stage,
the determination of the three-pion couplings from the measured lattice spectrum.

The layout of the paper is the following. In Sect.~\ref{sec:isospin} we write down the most
general effective Lagrangian for the problem at hand (both in three-particle and
particle-dimer picture). The matching between these alternative
descriptions has been carried out, and an explicit expression  for the LL
factor in $K\to 3\pi$ decays is derived.
In Sect.~\ref{sec:Faddeev_LL},
the Faddeev equations in different isospin channels are explicitly written down
and the renormalization issues are addressed.
The matching to the relativistic amplitudes that will ultimately enable one to express
the three-body couplings in the Lagrangian through the three-body amplitudes calculated
in ChPT, is discussed.
In Sect.~\ref{sec:numerics}, these
Faddeev equations in the infinite volume are solved by using the contour rotation technique.
Furthermore, the finite-volume energy spectrum of the three-pion system is obtained
by solving the quantization condition and the finite-volume wave function are determined.
All these are necessary ingredients for the calculation of the LL factor. We finally check
the sensitivity of the calculated LL factor to the input values of the short-range part of the
three-body threshold amplitude and find that in a wide interval, the LL factor
practically does not depend on this input.
Sect.~\ref{sec:concl} contains our conclusions.

\section{Derivation of the $K\to3\pi$ LL formula}\label{sec:isospin}

\subsection{The Lagrangian in the three-particle picture}
\label{sec:TheLagrangian}
In the following, we will consider the decay of a positively charged kaon $K^+$ into three pions, which is induced via weak interactions. There are two decay channels:
$K^+\to \pi^0 \pi^0 \pi^+$ and $K^+ \to \pi^+\pi^+\pi^-$. In order to describe the decay $K \to 3\pi$ within the NREFT approach, we adapt the Lagrangian given in \cite{Gasser:2011ju}, rewriting it in an arbitrary frame defined by the four-velocity $v^\mu$:
\begin{align}
	\mathcal{L} = &\sum_{i_3}\, \pi^\dagger_{i_3} \, 2 w_v\left( i(v\partial) - w_v \right) \pi_{i_3} + \mathcal{L}_2 + \mathcal{L}_3  \\\nonumber
	&+ K^\dagger_{+} \, 2 W_v\left( i(v\partial) - W_v \right) K_{+} + \mathcal{L}_K \,,
\end{align}
where $w_v = \sqrt{M_\pi^2 + \partial^2 - (v\partial)^2}$ and $W_v = \sqrt{M_K^2 + \partial^2 - (v\partial)^2}$ and $M_\pi$ and $M_K$ denote the masses of pions and kaons respectively.
Note that we work in the basis with physical particles, so that the triplet of pion fields
is given by  $\pi_{\textcolor{black}{i_3}}=(\pi_+,\pi_0,\pi_-)$. 
At the leading order in the power counting the two-body Lagrangian reads as
\begin{align}\label{eq:2body_Lagrangian_particle_picture}
	\mathcal{L}_2 &= \frac{1}{2} \, C_1 \, \pi_0^\dagger \pi_0^\dagger \pi_0 \pi_0 + 2\, C_2 \left( \pi_+^\dagger \pi_0^\dagger \pi_+ \pi_0 + \pi_-^\dagger \pi_0^\dagger \pi_- \pi_0 \right) + C_3 \left(\pi_+^\dagger \pi_-^\dagger \pi_0 \pi_0 + \text{h.c.} \right) \nonumber \\
	&+ 2 \,C_4\, \pi_+^\dagger \pi_-^\dagger \pi_+ \pi_- + \frac{1}{2} \, C_5 \left( \pi_+^\dagger \pi_+^\dagger \pi_+ \pi_+ + \pi_-^\dagger \pi_-^\dagger \pi_- \pi_- \right) \,,
\end{align}
while the three-body Lagrangian is given by
\begin{align}\label{eq:L3}
	\mathcal{L}_3 &= D_1 \left( \pi^\dagger_+ \pi_+ + \pi^\dagger_0 \pi_0 +  \pi^\dagger_- \pi_- \right)^3 \nonumber \\
	&+D_2 \left( 2\pi^\dagger_+ \pi^\dagger_- - \pi^\dagger_0 \pi^\dagger_0 \right) \left( \pi^\dagger_+ \pi_+ + \pi^\dagger_0 \pi_0 +  \pi^\dagger_- \pi_- \right) \Big( 2\pi_+ \pi_- - \pi_0 \pi_0 \Big) \,.
\end{align}
The weak kaon decays are described by the Lagrangian 
\begin{align}\label{eq:LK}
	\mathcal{L}_K = G_1 \left( K_+^\dagger \pi_0 \pi_0 \pi_+ + \text{h.c.}\right) + G_2 \left( K_+^\dagger \pi_+ \pi_+ \pi_- + \text{h.c.}\right)\,.
\end{align}
At higher orders, all Lagrangians are amended by the terms that contain space derivatives
of all fields. A consistent power counting emerges, if one counts
three-momenta as
${\bf p}=\mathcal{O}(\delta)$, where $\delta$ stands for a generic small parameter.\footnote{In the manifestly covariant framework we are using here,
one has $p_\perp^\mu\doteq p^\mu-v^\mu(v\cdot p)=O(\delta)$ instead.\label{foot:5}}
Again, for consistency, one should count the difference $M_K-3M_\pi$ as a quantity of
order $\delta^2$~\cite{Gasser:2011ju}. As mentioned above, in this paper the
higher-order terms in $\delta$, corresponding to the derivative couplings,
are not considered.

\textcolor{black}{
To summarize, the following couplings emerge at leading order in the three-particle
picture:
\begin{itemize}
\item
  The couplings
  $C_i\,,~i=1,\ldots,5$, describing non-derivative pion-pion interactions in different
  isospin channels. These can be expressed through the S-wave $\pi\pi$ scattering
  lengths $a_0,a_2$ in a standard manner, through the matching condition.
\item
  The three-body couplings in the pion system $D_1,D_2$ (the three-body force). On the
  lattice, they can be determined from the fit to the three-body spectrum.
\item
  The couplings $G_1,G_2$ that describe the weak decays  of charged kaons into
  three pions.
\end{itemize}
}

\subsection{The Lagrangian in the particle-dimer picture}
In the particle-dimer picture, the most general Lagrangian at the leading order reads
as\footnote{\textcolor{black}{More details about particle-dimer formalism can be
found, e.g., in Refs.~\cite{Bedaque:1998mb,Bedaque:1998kg,Bedaque:1998km,Hammer:2017kms,Muller:2021uur,Muller:2022oyw}.}}
\begin{align}
   \tilde{\mathcal{L}}&  
                      =\sum_{i_3}\, \pi^\dagger_{i_3} \, 2 w_v\left( i(v\partial) - w_v \right) \pi_{i_3}
                      + \sum_{I, I_3}\, \sigma_I T^\dagger_{I I_3} T_{I I_3}
                      + \tilde{\mathcal{L}}_2 +
                      \tilde{\mathcal{L}}_3
  \\\nonumber
                    &+ K^\dagger_{+} \, 2 W_v\left( i(v\partial) - W_v \right) K_{+}
                      + \tilde{\mathcal{L}}_K  \,,
\end{align}
Here $\sigma_{I} = \pm1$, depending on the sign of the two-body scattering length. Furthermore, $i_3 = -1,0,1$ and $I_3 = -I,\dots,I$ denote the isospin projection of the pions and isospin-$I$ \textcolor{black}{S-wave} dimer field $T_{II_3}$, respectively. As two pions can couple to $I = 0,1,2$, all dimers with different isospins decouple and can be neglected. Furthermore, due to Bose-symmetry two pions can not be in an $I = 1$ state in the S-wave.  Therefore, at the leading order, the $I = 1$ dimer does not contribute,
and the two-body interaction is described by the Lagrangian
\begin{align}
	\tilde{\mathcal{L}}_2 = \sum_{I, I_3} \left(  T^\dagger_{I I_3} O_{I I_3} + \text{h.c.} \right) \,, \quad I = 0,2\,,
\end{align}
where the dimer operator is given by a sum over two pion field operators with pertinent Clebsh-Gordan coefficients: 
\begin{align}
	O_{I I_3} = \sum_{i_3, i'_3}\, \frac{1}{2}f_I \,\langle 1, i_3; 1, i'_3 | I, I_3 \rangle \, \pi_{i_3} \pi_{i'_3} \,.
\end{align}
Note also that $O_{1I_3} = 0$ for all $I_3 = -1,0,1$ follows due to Bose-symmetry.
\textcolor{black}{Furthermore, the couplings $f_I$ describe the decay of a dimer into two pions and can be expressed
through the S-wave $\pi\pi$ scattering lengths through the matching condition.}

The construction of the three-body Lagrangian proceeds by defining particle-dimer operators in the channels with a different total isospin $J$:
\begin{align}\label{eq:lincomb}
  \mathcal{O}^{(J,I)}_{J_3} =
  \sum_{I_3, i_3}\, \langle I, I_3; 1, i_3 |   J, J_3  \rangle \, \pi_{i_3} T_{I I_3} \,.
\end{align}
In the channels with the total isospin $J = 2$ and $J = 3$, there is only a single
set of operators with the dimers having $I = 2$. For isospin $J = 1$, there are
two independent operators, where the dimer has $I = 0$ and $I = 2$ respectively.
The most general Lagrangian is thus given by:
\begin{align}\label{eq:PD_Lagrangian_3body}
  \tilde{\mathcal{L}}_3 &= \sum_{J, J_3} \sum_{I, I'}\,  h_J^{(I,I')} \,
                        \left(\mathcal{O}^{(J,I)}_{J_3}\right)^\dagger
                        \mathcal{O}^{(J,I')}_{J_3} \,,
\end{align}
where, due to hermiticity,
$h_J^{(I,I')} = h_J^{(I',I)}$ (Note that $h_J^{(I,I')}$ are real due to the $T$-invariance.).
In the $J = 2$ and $J = 3$ channels, there is a single coupling, $h_2^{(2,2)}$ and $h_3^{(2,2)}$, respectively. For the channel with $J = 1$, there are three couplings $h_1^{(0,0)}$, $h_1^{(2,2)}$ and $h_1^{(2,0)} = h_1^{(0,2)}$.  

Finally, $\tilde{\mathcal{L}}_K$ describes the weak kaon decay. Due to charge conservation,
the positively charged kaon can only couple to the operators with $J_3=1$:
\begin{align}\label{eq:tildeLK}
  \tilde{\mathcal{L}}_K=\sum_{J,I}g^{(J,I)}\left(K_+^\dagger \mathcal{O}^{(J,I)}_1+\mbox{h.c.}\right)\, .
\end{align}
\textcolor{black}{
  Hence, in the particle-dimer picture, we have the following parameters in the lowest-order
  Lagrangian
  \begin{itemize}
  \item
    Two-particle-dimer couplings $f_0,f_2$. These correspond to the couplings $C_i$
    in the three-particle picture.
  \item
    Three-body force in the particle-dimer picture, described by the couplings
    $h_1^{(0,0)}$, $h_1^{(2,2)}$, $h_1^{(2,0)}$, $h_2^{(2,2)}$ and $h_3^{(2,2)}$.
    These correspond to the couplings $D_1,D_2$ in the three-particle picture.
  \item
    The weak couplings $g^{(1,0)},g^{(1,2)},g^{(3,2)}$, corresponding to the parameters
    $G_1,G_2$ in the three-particle picture.
  \end{itemize}
  Despite the fact that the number of the couplings in different formalisms differ,
  these formalisms are equivalent. This equivalence is, however,
  a rather subtle issue, and is discussed in the remaining part of this section.
}
\subsection{Matching in the two-body sector}
\begin{figure}[t]
	\centering
	\includegraphics[width=0.7\textwidth]{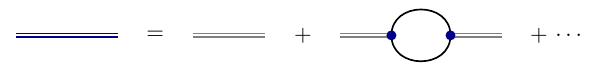}
	\caption{Full dimer propagator, obtained by summing up self-energy insertions to all orders. The blue double, gray double and black single lines denote the full dimer propagator, the free dimer propagator given by $-\sigma_{I}^{-1}\delta_{I I'}\delta_{I_3 I'_3}$, and the particle propagator. The blue dots represent the insertion of the vertex converting a dimer into particles.}
	\label{fig:dimer_propagator}
\end{figure}

The couplings $f_I$ can be matched to the two-body S-wave scattering length in the $I = 0$ and $I = 2$ isospin channels. Due to isospin symmetry, the dimer propagator is diagonal in the isospin space:
\begin{align}
i \langle 0 | T[ T_{II_3}(x) T^\dagger_{I'I'_3}(y) ] | 0 \rangle = \delta_{II'}\delta_{I_3 I'_3} \int\frac{d^4 P}{(2\pi)^4}\, e^{-iP(x-y)}\, S_I(P^2)\,.
\end{align}
Summing up the self-energy insertions (see Fig.~\ref{fig:dimer_propagator}),
for $I = 0,2$ we find
\begin{align}\label{eq:SI}
S_{I}(P^2) = \frac{f_I^{-2}}{-\sigma_I f_I^{-2} - \frac{1}{2} I(s)} \,, \quad s = P^2\, ,
\end{align}
where, for $s\geq 4M_\pi^2$,
\begin{align}\label{eq:Chew-Mandelstam}
I(s) = \frac{\sigma(s)}{16\pi^2}\ln\frac{\sigma(s)-1}{\sigma(s)+1} = J(s) + \frac{i\sigma(s)}{16\pi}\,, \quad \sigma(s) = \left( 1- \frac{4M_\pi^2}{s+i\varepsilon} \right)^{1/2} \,,
\end{align}
while for $I = 1$ trivially $S_1(P^2) = -\sigma_1^{-1}$. 

The two-particle scattering amplitude $\pi_{i'_3}(p_1) + \pi_{j'_3}(p_2) \to \pi_{i_3}(p_3) + \pi_{j_3}(p_4)$ is obtained from the dimer propagator by attaching the vertices that convert a dimer into a particle-pair. The isospin $I = 0,2$ the amplitudes are given by:
\begin{align}\label{eq:2body_amplitude}
T_I(p_1, p_2; p_3, p_4) = f_I^2 \,S_I(P^2) = \frac{16\pi\sqrt{s}}{16\pi\sqrt{s}[-\sigma_{I} f_I^{-2} - \frac{1}{2} J(s)] - i p(s)} \,,
\end{align}
where $P^2 = (p_1 + p_2)^2 = (p_3+p_4)^2$ and $s = P^2 = 4(M_\pi^2+p^2(s))$. Due to unitarity,
\begin{align}
16\pi\sqrt{s}[-\sigma_{I} f_I^{-2} - \frac{1}{2} J(s)] = p(s) \cot\delta_I(s) \,.
\end{align} 
Therefore, for $I=0,2$, the matching to the scattering length $a_I$ reads as:
\begin{align}\label{eq:matching_fI_aI}
\sigma_I^{-1} f_I^2 = 32\pi M_\pi a_I \,.
\end{align}
As discussed in appendix \ref{app:integrating_out_dimers}, integrating out the dimer fields at tree level merely amounts to the replacement
\begin{align}\label{eq:replacement}
T_{I I_3} \to -\sigma_{I}^{-1} O_{I I_3} \,.
\end{align}
The matching condition to the two-particle Lagrangian in the particle picture yields:
\begin{align}\label{eq:matching_Ci_fI}
&C_1 = -\frac{1}{6}\left( \sigma_0^{-1} f_0^{2} +  2 \sigma_2^{-1} f_2^{2} \right) = -16\pi M_\pi \,\frac{1}{3}(a_0 + 2 a_2) \,,  \nonumber\\ 
&C_2 = -\frac{1}{4} \sigma_2^{-1} f_2^{2} = -16\pi M_\pi \, \frac{1}{2}\, a_2 \,, \nonumber\\ 
&C_3 = -\frac{1}{6}\left( \sigma_2^{-1} f_2^2 - \sigma_0^{-1} f_0^2 \right) = -16\pi M_\pi \, \frac{1}{3}\,(a_2-a_0) \nonumber \\
&C_4 = -\frac{1}{12}\left(2\sigma_0^{-1} f_0^2 + \sigma_2^{-1} f_2^2 \right) =  -16\pi M_\pi \, \frac{1}{6}\,(2a_0+a_2) \,,  \nonumber \\
&C_5 = - \frac{1}{2}\,\sigma_2^{-1} f_2^2 =  -16\pi M_\pi \, a_2 \,.
\end{align}
This result agrees with \cite{Gasser:2011ju}.\footnote{Note a different sign convention is used in Ref.~\cite{Gasser:2011ju}, namely,
  $p\cot\delta = 1/a_I+\ldots$ instead of $p\cot\delta = -1/a_I+\ldots$.}

\subsection{Matching of the three-body and decay Lagrangians}

At tree level, carrying out the the replacement~(\ref{eq:replacement}) in
the particle-dimer operators, one straightforwardly gets:
\begin{align}\label{eq:limits}
  \mathcal{O}^{(1,0)}_{J_3}&\to -\frac{\sigma_0^{-1}f_0}{2\sqrt{3}}\,
  (2\pi_+\pi_--\pi_0\pi_0)\pi_{J_3}\, ,
                             \nonumber\\
  \mathcal{O}^{(1,2)}_{J_3}&\to -\frac{\sigma_2^{-1}f_2}{\sqrt{15}}\,
  (2\pi_+\pi_--\pi_0\pi_0)\pi_{J_3}\, ,
                            \nonumber\\
  \mathcal{O}^{(2,2)}_{J_3}&\to 0\, ,
                             \nonumber\\
   \mathcal{O}^{(3,2)}_{\pm 3}&\to -\frac{\sigma_2^{-1}f_2}{2}\,\pi_\pm\pi_\pm\pi_\pm\, ,
                            \nonumber\\
   \mathcal{O}^{(3,2)}_{\pm 2}&\to -\frac{\sigma_2^{-1}f_2\sqrt{3}}{2}\,\pi_\pm\pi_\pm\pi_0\, ,
                           \nonumber\\
  \mathcal{O}^{(3,2)}_{\pm 1}&\to -\frac{\sigma_2^{-1}f_2\sqrt{3}}{2\sqrt{5}}\,
 \left(\pi_\pm\pi_\pm\pi_\mp+2\pi_\pm\pi_0\pi_0\right)\, ,
                          \nonumber\\
  \mathcal{O}^{(3,2)}_0&\to -\frac{\sigma_2^{-1}f_2}{\sqrt{10}}\,
                               \left(3\pi_+\pi_-\pi_0+\pi_0\pi_0\pi_0\right)\, .
                               \end{align}
                               Furthermore, performing this replacement in the particle-dimer
                               Lagrangian, one can identify
                               the couplings $D_1,D_2$ from Eq.~(\ref{eq:L3}):
\begin{align}                               
  &  \sum_{J, J_3} \sum_{I, I'}\,  h_J^{(I,I')} \,
                        \left(\mathcal{O}^{(J,I)}_{J_3}\right)^\dagger
                        \mathcal{O}^{(J,I')}_{J_3}  \to                         
 D_1\left( \pi^\dagger_+ \pi_+ + \pi^\dagger_0 \pi_0 +  \pi^\dagger_- \pi_- \right)^3 \nonumber\\
  &\hspace*{2.cm}+\,
  D_2
  \left( 2\pi^\dagger_+ \pi^\dagger_- - \pi^\dagger_0 \pi^\dagger_0 \right) \left( \pi^\dagger_+ \pi_+ + \pi^\dagger_0 \pi_0 +  \pi^\dagger_- \pi_- \right) \Big( 2\pi_+ \pi_- - \pi_0 \pi_0 \Big) \,,
\end{align}
with
\begin{align}\label{eq:D1D2}
  D_1&= \frac{f_2^2}{4}\,h_3^{(2,2)}\, ,\nonumber\\
  D_2&= \frac{f_0^2}{12}\,h_1^{(0,0)}+\frac{\sigma_0^{-1}f_0\sigma_2^{-1}f_2}{3\sqrt{5}}\,
  h_1^{(2,0)}+\frac{f_2^2}{15}\,h_1^{(2,2)}-\frac{3f_2^2}{20}\,h_3^{(2,2)}\, .
\end{align}
Using the same replacement in the Lagrangian that describes the weak decays
of kaons~(\ref{eq:LK}), one could read off the couplings $G_1,G_2$:
\begin{align}
  \sum_{J,I}g^{(J,I)}\left(K_+^\dagger \mathcal{O}^{(J,I)}_1+\mbox{h.c.}\right)
  \to
  G_1
    \left(K_+^\dagger\pi_+\pi_+\pi_-+\mbox{h.c.}\right)
  +G_2
  \left(K_+^\dagger\pi_+\pi_0\pi_0+\mbox{h.c.}\right)\, ,
\end{align}
with
\begin{align}\label{eq:G1G2}
G_1&=-\frac{\sqrt{3}}{\sqrt{5}}\,\sigma_2^{-1}f_2g^{(3,2)}
  +\frac{1}{\sqrt{15}}\,\sigma_2^{-1}f_2g^{(1,2)}
     +\frac{1}{2\sqrt{3}}\,\sigma_0^{-1}f_0g^{(1,0)}\, .
       \nonumber\\[2mm]
  G_2&= -\frac{\sqrt{3}}{2\sqrt{5}}\,\sigma_2^{-1}f_2g^{(3,2)}
  -\frac{2}{\sqrt{15}}\,\sigma_2^{-1}f_2g^{(1,2)}
  -\frac{1}{\sqrt{3}}\,\sigma_0^{-1}f_0g^{(1,0)}\, ,
     \end{align}
     As seen from the above equations~(\ref{eq:D1D2}) and (\ref{eq:G1G2}), the number
     of the couplings in the particle-dimer picture is larger than in the three-particle
     picture. Namely, one could argue that only two couplings in each set
     $h_J^{(I,I')}$ and $g^{(J,I)}$ are independent, and others can be chosen freely.
     Note that establishing the number of the independent couplings is a subtle dynamical
     issue and
     is discussed in detail in Ref.~\cite{Muller:2022oyw}. Here, we merely state that in the
     S-wave $\pi\pi$ scattering no shallow dimers exist that justifies a naive counting
     presented above. To fix the freedom, we choose $h_1^{(2,0)}=h_1^{(2,2)}=h_2^{(2,2)}=0$ and $g^{(1,2)}=3g^{(3,2)}$, $g^{(2,2)}=\sqrt{5}g^{(3,2)}$ (The latter two conditions
     ensures that, at tree level, the operators $T_{20}\pi_+$ and $T_{21}\pi_0$ are absent
     in the Lagrangian $\tilde{\mathcal{L}}_K$, which describes the weak decay of a kaon
     into the particle-dimer pair.).

     An important question is, however, whether these redundant couplings, which are
     absent in the tree-level matching, re-emerge in the loops. We shall address this
     question in the following section.

\subsection{Reduction of the redundant couplings}

\begin{figure}[t]
	\centering
    \includegraphics[width=0.2\textwidth]{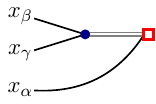}
    \caption{Tree level contribution to the decay matrix element. Black solid lines and gray double lines denote a particle propagator and the tree-level dimer propagator, res\-pec\-ti\-ve\-ly. The blue dot and the empty red square correspond to the particle-dimer conversion vertex and the kaon initial decay coupling, respectively. Furthermore, $(\alpha\beta\gamma)$
stands for some permutation of $(123)$.}
	\label{fig:vertex_function_G}
\end{figure}

Following Eq.~(\ref{eq:limits}), one can trivially define a linear combination of
the operators $\mathcal{O}^{(1,0)}_{J_3}$ and $\mathcal{O}^{(1,2)}_{J_3}$,
which vanishes under the replacement~(\ref{eq:replacement}). One need not
display an explicit form of this linear combination here which, together with
$\mathcal{O}^{(2,2)}_{J_3}$, forms a set of the {\em irrelevant} operators
$\hat{\mathcal{O}}^{(a)}_{J_3},~a=1,2$. The contribution of the irrelevant operators
to the physical matrix elements at tree-level vanishes. An orthogonal linear combination
of  $\mathcal{O}^{(1,0)}_{J_3}$ and $\mathcal{O}^{(1,2)}_{J_3}$ and 
$\mathcal{O}^{(3,2)}_{J_3}$ form a set of {\em relevant} operators
$\mathcal{O}^{(a)}_{J_3},~a=1,2$. The question which will be addressed below,
can be stated as follows: do the irrelevant operators contribute to the physical
observables beyond the tree level? We shall demonstrate that this is not the case,
and the irrelevant operators can be safely dropped from the beginning.

Let us start from the decay of a kaon into three pions, and consider the following
vertex function
\begin{align}\label{eq:vertex_function_G}
  \hat{G}_{i_1 i_2 i_3}^{(a)}(x_1, x_2, x_3)
  = \langle 0 | T\left[ O^{i_1 i_2 i_3}_{3\pi}(x_1, x_2, x_3)  \,
  \left(\hat{\mathcal{O}}^{(a)}_1\right)^\dagger(0)  \right] | 0 \rangle\,,
\end{align}
where $O_{3\pi}$ denotes a three-pion operator:
\begin{align}\label{eq:three_pion_operator}
	O^{i_1 i_2 i_3}_{3\pi}(x_1, x_2, x_3) = \pi_{i_1}(x_1) \pi_{i_2}(x_2) \pi_{i_3}(x_3)\, .
\end{align}
At tree level, this vertex, shown in Fig.~\ref{fig:vertex_function_G}, vanishes, because of
the Bose-symmetry of three pions in the final state. Now, note that the same vertex
appears in any loop diagram that describes the kaon decay. It is straightforward to
check that these loop diagrams vanish as well, since the tree-level vertex does not
depend on the momenta of the final pions -- in other words, it does not distinguish
between the real an virtual pions. As a result, the contribution of the irrelevant operators
to the pion decay amplitude vanishes to all orders in perturbation theory.

\begin{figure}[t]
	\centering
	\includegraphics[width=0.4\textwidth]{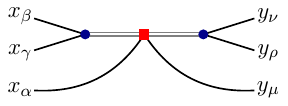}
	\caption{The tree-level contribution to the matrix element of the three-pion scattering. Black solid lines
          and gray double lines denote the particle propagator and the tree-level dimer
          propagator, respectively. The blue dots and the red rectangle correspond to
          the particle-dimer conversion vertex and the particle-dimer interaction vertex,
          respectively. The labels $(\alpha\beta\gamma)$ and $(\mu\nu\rho)$ denote the
         permutations of $(123)$.}
	\label{fig:vertex_function_V}
\end{figure}

\begin{figure}[t]
	\centering
	\includegraphics[width=0.5\textwidth]{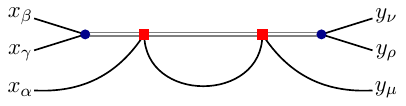}
	\caption{The particle-dimer loop diagram that could potentially contribute to
          the six-pion amplitude.  Black solid lines and gray double
          lines denote a particle propagator and the tree-level dimer propagator,
          respectively. The blue dots and the red rectangle correspond to the
          particle-dimer conversion vertex and the particle-dimer interaction vertex,
          respectively.}
	\label{fig:possible_contribution_particle_dimer}
\end{figure}

A similar argument applies for the three-pion scattering.
The tree-level contribution to the quantity
\begin{align}
  &\hspace*{1.5cm}V_{i_1 i_2 i_3; j_1 j_2 j_3}^{(a,b)}(x_1,x_2, x_3; y_1, y_2, y_3)
  \nonumber\\
  =& \langle 0 | T \bigg[ O^{i_1 i_2 i_3}_{3\pi}(x_1, x_2, x_3) O^{j_1 j_2 j_3}_{3\pi}(y_1, y_2, y_3)^\dagger \sum_{J_3}\left(\mathcal{O}^{(a)}_{J_3}(0)\right)^\dagger
    \left(\hat{\mathcal{O}}^{(b)}_{J_3}(0)\right) \bigg] | 0 \rangle
\end{align}
is shown in Fig.~\ref{fig:vertex_function_V}. Here, for simplicity, we assume that only
one irrelevant operator appears, but the discussion in case of two irrelevant operators
follows exactly the same path. The tree-level contribution, where each dimer line
is equipped by two particle lines prior to escaping, obviously vanishes. The question,
whether the irrelevant operators contribute in the loops reduces to the question, whether
all internal dimer lines end up in the two-pion vertex. There is only one diagram,
shown in Fig.~\ref{fig:possible_contribution_particle_dimer}, where this is not the case.
However, the tree-level dimer propagator $S^{(0)}_I(x) = -\sigma_I^{-1} \delta^4(x)$
is local in position space. Hence, a closed loop over the
non-relativistic pion propagator emerges,
which vanishes due to the pole structure of the latter.
To summarize, the irrelevant operators contribute neither to the kaon decay amplitudes,
nor to the three-pion scattering amplitudes to all orders and, hence, can be safely
discarded from the beginning.

\section{Faddeev equations and derivation of the LL factor}\label{sec:Faddeev_LL}

\subsection{Faddeev equation for particle-dimer amplitude}
\label{sec:FaddeeveEquation}
We start with writing down the Faddeev equation for the particle-dimer amplitude.\footnote{\textcolor{black}{The Faddeev equations in the particle-dimer picture (both the non-relativistic and relativistic cases) has been considered in detail in the following papers~\cite{Bedaque:1998mb,Bedaque:1998kg,Bedaque:1998km,Hammer:2017kms,Muller:2021uur,Muller:2022oyw}.}}
There are three isospin channels with $J = 1, 2, 3$ (In the
decays of charged kaons, the channel with $J=0$ is excluded due to charge
conservation.). It is
important to note that, due to the symmetry properties of the three-pion wave function,
the isospin channel $J = 2$ does not contribute to the kaon decay
at the leading order.
The particle-dimer amplitude $\mathcal{M}_{J;II'}(p, q; P)$ in an arbitrary reference frame defined by a unit vector $v^\mu$ obeys the equation
\begin{align}\label{eq:Faddeev_IV}
    \mathcal{M}_{J;II'}({ p},{ q};{ P}) &= Z_{J;II'}({ p},{ q};{ P}) \nonumber\\&+\sum_{I''} \int^{\Lambda_{v}} \frac{d^3 k_{\perp}}{(2\pi)^{3}2w_{v}({ k})}Z_{J;II''}({ p},{ k};{ P})\tau_{I''}(({ P} - { k})^2)\mathcal{M}_{J;I''I'}({ k},{ q};{ P})\, .
\end{align}
Here, $I$ and $I'$ are the incoming and outgoing dimer isospin indices,  ${p}$ and ${q}$ represent the on-shell four-momenta of outgoing/incoming particles, respectively, ${P}$ is the total four-momentum of the three-pion system \textcolor{black}{and $k_\perp^\mu
  =k^\mu-v^\mu(vk)$ denotes the perpendicular component of any vector $k^\mu$
in a frame defined by the unit vector $v^\mu$, see also footnote~\ref{foot:5}}. Furthermore, $\Lambda_{v}$ denotes the ultraviolet cutoff which is defined by:
\begin{align}
	\int^{\Lambda_v} \frac{d^3 k_\perp}{(2\pi)^3} \, F(k) = \int \frac{d^4 k}{(2\pi)^3} \, \delta^4(k^2-m^2) \theta(\Lambda^2 + k^2 - (vk)^2) F(k) \,.
\end{align}
The Faddeev equation is diagrammatically illustrated in Fig.~\ref{fig:faddeev}.
Note that, in order to streamline the notations,
  we have changed the normalization of the amplitude, according to
  $\mathcal{M}_{J;II'}\to f_I^{-1}\mathcal{M}_{J;II'}f^{-1}_{I'}$. The quantity $Z_{J;II'}$
  represents the driving term of the Faddeev equation while $\tau_I$ stands for the
  dimer propagator:
\begin{figure}[t]
    \centering
    \includegraphics[width=0.7\textwidth]{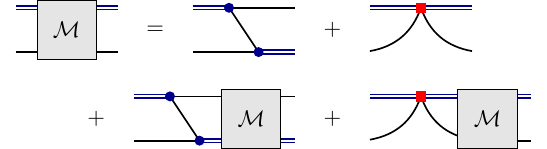}
    \caption{Faddeev equation for the particle-dimer scattering amplitude. The
      double blue line and the solid black line correspond to the 
      full propagator of the dimer field the  pion propagator, respectively.
      The blue circle denotes the vertex, converting the dimer to particles and the red
      box denotes the dimer-particle contact vertex. Isospin indices are implicit.}
    \label{fig:faddeev}
\end{figure}
\begin{align}\label{eq:fullZ}
    Z_{J;II'}(p,q;P) &= \frac{c_{J;II'}}{2w_v(P-p-q)(w_v(P-p-q) + w_v(p) + w_v(q) - vP - i\varepsilon)} + \frac{H_{J;II'}(\Lambda)}{\Lambda^2}\,, \nonumber\\[2mm]
    \tau_I((P-k)^2) &= f_I^2 S_I((P-k)^2)\, .
\end{align}
The explicit expression for the quantity $\tau_I$, \textcolor{black}{which is directly related
to the dimer propagator defined in Eq.~(\ref{eq:SI}),} can be read off from Eq.~(\ref{eq:2body_amplitude}).
Furthermore, $c_{J;II'}$ are expressed through the Clebsh-Gordan coefficients
and emerge after the projection onto the states with total isospin $J$:
\begin{align}
  c_{1;00} = \frac{1}{3}, \hspace{5pt}
  c_{1;02} = c_{1;20} = \frac{\sqrt{5}}{3},
  \hspace{5pt}
  c_{1;22} = \frac{1}{6}\, ,\hspace{5pt}
  c_{2;22} =  -\frac{1}{2},
  \hspace{5pt}
  c_{3;22} = 1\, .
\end{align}
	Finally 
	\begin{align}
		H_{J;II'}(\Lambda) = \Lambda^2 \, f_{I}^{-1} \, h^{(I,I')}_{J} \,  f_{I'}^{-1}\,.
	\end{align}
	According to the choice $h_1^{(2,0)}=h_1^{(2,2)}=h_2^{(2,2)}=0$,
        only the couplings $H_{1;00}$ and $H_{3;22}$ are non-zero. Furthermore, as shown in Refs.~\cite{Bedaque:1998kg,Bedaque:1998km}, if one restricts oneself only to the first
	term in $Z_{J;II'}$ and sets all  $H_{J;II'}(\Lambda)$ to zero,
	the solution of the Faddeev equation is cutoff-dependent and shows oscillatory
	dependence on $\Lambda$. This cutoff-dependence is eliminated by adding the
	contribution from $H_{J;II'}(\Lambda)$. The $\Lambda$-dependence of these couplings
	is such that it exactly cancels the oscillatory behavior coming from the first term.
	For our case, two remarks are in order. First, as shown in  Ref.~\cite{Bedaque:1998km}, since the
        coefficient in the isospin-two channel, $c_{2;22} = -\frac{1}{2}$,
        has a negative sign, the amplitude even without the inclusion of $H_{2;22}(\Lambda)$
        is cutoff-independent for all momenta $p\ll\Lambda$. Hence,
        one does not need to introduce the particle-dimer contact term in the $J = 2$
        channel altogether.
        Second there are no physical dimers and hence there is only one independent
        three-pion amplitude in the $J=1$ channel. Consequently, in this channel,
        it is sufficient to match the coupling $H_{1;00}$ only. To summarize, as expected
        from the beginning,
        all observable three-pion amplitudes can be made cutoff-independent by matching
        only two couplings  $H_{1;00}$ and  $H_{3;22}$, albeit the original particle-dimer 
Lagrangian contained four independent couplings. This statement does not hold, in general,
for the (unobservable) particle-dimer amplitudes (We remind the reader that there
are no shallow \textcolor{black}{bound states} in our case.).

For matching of $H_{J;II'}(\Lambda)$, we need the three-particle threshold amplitude and,
equivalently, the particle-dimer threshold amplitude, which is obtained by
setting ${\bf p},{\bf q} = 0$ in the CM frame. This amplitude is singular at $E=3M_\pi$.
To get the regular part of this amplitude, we start with evaluating the
amplitude slightly below threshold, assuming that $E=3M_\pi-\varepsilon$, and consider
the limit $\varepsilon \to 0$ at the end. Adding loops makes the singularity at
$\varepsilon=0$ weaker, and evaluating the diagrams up to two loops suffices
for finding all singularities. Next, the singularities in $\varepsilon$ are isolated
and subtracted from the threshold amplitude, in order to obtain the regular part.
This process is discussed in detail in Ref.~\cite{Pang:2019dfe} in a purely
non-relativistic setting. The singularities in our case are the same and can be easily read off from Ref.~\cite{Pang:2019dfe}:
\begin{align}\label{eq:singular-M}
    \mathcal{S}_{J;II'}(\varepsilon) = &\frac{1}{2M_\pi\varepsilon}c_{J;II'} - \frac{1}{2\sqrt{M_\pi\varepsilon}} \sum_{I''} (c_{J;II''}~a_{I''}~c_{J;I''I'})\nonumber\\&+ \frac{\sqrt{3}}{2\pi}\,\log{\frac{\varepsilon}{M_\pi}} \sum_{I''} (c_{J;II''}~a_{I''}^{2}~c_{J;I''I'})\nonumber\\
    &- \frac{2}{3}\,\log{\frac{\varepsilon}{M_\pi}} \sum_{I''; I'''}(c_{J;II''}~a_{I''}~c_{J;I''I'''}~a_{I'''}~c_{J;I'''I'})\, .
\end{align}
In an arbitrary frame, the quantity $\varepsilon$ should be defined as
$\varepsilon=\sqrt{P^2}-3M_\pi$, whereas ${\bf p},{\bf q}$ are replaced by $p_\perp^\mu,q_\perp^\mu$.  By subtracting $\mathcal{S}_{J;II'}(\varepsilon)$ from the threshold amplitude $\mathcal{M}_{J;II'}$ and taking the limit $\varepsilon \to 0 $, we get the regular particle-dimer threshold amplitude.

The infinite-volume three-particle amplitude\footnote{We would like to stress here
  that this is not a physical amplitude. The physical states with a given full isospin
  $J$ can be built up
  from a pion and a dimer with a isospin $I$ in different ways. in the quantity $T_{J;II'}$,
the full isospin $J$, as well as $I,I'$ are fixed.} is related to the particle-dimer amplitude in the following way:
\begin{align}
T_{J;II'}(\{p\},\{q\};P) &= \sum_{\alpha,\beta = 1}^{3} \Big[ (2\pi)^{3}\delta^{3}(p_{\alpha\perp} - q_{\beta\perp})\delta_{II'}2w_{v}(p_{\alpha})\tau_{I}((P - p_{\alpha})^{2}) \nonumber\\
&+ \tau_{I}((P - p_{\alpha})^{2})  \mathcal{M}_{J;II'}(p_{\alpha},q_{\beta};P)\tau_{I'}((P - q_{\beta})^{2}) \Big]\,.\label{eq:T_inf}
\end{align}
Here, $\{p\}$ represents the set of all four-particle momenta $p_{\alpha}$ with $\alpha = 1,2,3$. The regular part of the three-particle amplitude can be related to the regular
part of the particle-dimer amplitude in an obvious manner.

In a finite volume, the counterpart of the Faddeev equation can be written down as follows
\begin{align}
  \mathcal{M}^{L}_{J;II'}(p,q;P) =& Z_{J;II'}(p,q;P)
  \nonumber\\&+ \sum_{I''}\frac{1}{L^{3}}\sum^{\Lambda_v}_{\bf k} \frac{1}{2w(\mathbf{k})} Z_{J;II'}(p,k;P) \tau^{L}_{I''}(P-k) \mathcal{M}^{L}_{J;I''I'}(k,q;P)\, ,
\end{align}
where $L$ is the spatial extent of the box and finite-volume quantities are represented by the superscript $L$. The summation is over the discrete values of momentum
${\bf k} = 2\pi{\bf n}/L$, ${\bf n} \in \mathbb{Z}^{3}$. The finite volume propagator is given by
\begin{align}
	\tau^{L}_{I}(P) &= \frac{16\pi\sqrt{s}}{16\pi\sqrt{s}[-\sigma_I f_I^{-2} - \frac{1}{2}J(s)] - \dfrac{2}{\sqrt{\pi} L \gamma } Z^{\bf d}_{00}(1;q^{2}_{0})}\, ,
	\end{align}
        where $s = P^2$ and the real part of the Chew-Mandelstam function, $J(s)$,
is defined in Eq.~(\ref{eq:Chew-Mandelstam}).
Furthermore, the L\"uscher zeta-function
$Z^{\bf d}_{00}(1;q^{2}_{0})$ is defined as
\begin{align}
    Z^{\bf d}_{00}(1;q^{2}_{0}) = \frac{1}{\sqrt{4\pi}}\sum_{{\bf r^{2}}\in P_{d}}\frac{1}{{\bf r^{2}} - q^{2}_{0}}\, ,
\end{align}
with $\gamma = \dfrac{P_{0}}{\sqrt{s}}$, ${\bf d} = \dfrac{L}{2\pi}\,{\bf P}$, $q_{0}^{2} = \dfrac{L^{2}}{4\pi^{2}}\biggl(\dfrac{s}{4} - m^{2}\biggr)$ and
\begin{align}
    P_{d} = \big\{ {\bf r} = \mathbb{R}^{3}| r_{\parallel} = \gamma^{-1}\Big( n_{\parallel} - \frac{1}{2}|{\bf d}| \Big),{\bf r}_{\perp} = {\bf n}_{\perp},{\bf n}\in \mathbb{Z}^{3} \big\}\, .\nonumber
\end{align}
\begin{figure}[t]
    \centering
    \includegraphics[width=0.9\textwidth]{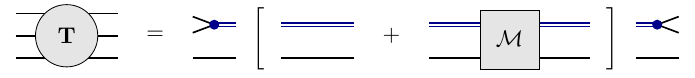}
    \caption{The three-particle amplitude in terms of the particle-dimer scattering amplitude. The sum over spectator momenta and the isospin indices are implicit.}
    \label{fig:three-particle}
\end{figure}
The quantization condition,
\textcolor{black}{which now explicitly includes different isospin channels,}  is given by
\begin{align}
    \text{det}(A) = 0\, , \hspace{20pt} A_{J; I I'}(p,q;P) = 2w({\bf k})L^{3}\delta_{{\bf p q}}\delta_{I I'}[\tau^{L}_{I}(P-k)]^{-1} - Z_{J; I I'}(p,q;P)\, .
\end{align}
The finite-volume spectrum $E_n$
of the three-particle system is determined by the discrete solutions of the
quantization condition. The particle-dimer amplitude is factorized
near the pole of the quantization condition as follows,
\begin{align}
    \mathcal{M}^{L}_{J;II'}(p,q;P)\Big|_{E \to E_{n}} = \frac{\phi^{(n)}_{J;I}(p)\phi^{(n)}_{J;I'}(q)}{P_{n,\parallel} - P_\parallel - i\varepsilon} + \, \text{regular}\, ,\nonumber
\end{align}
where $\phi^{(n)}_{J;I}(p)$ is the finite-volume particle-dimer wave function and it obeys the homogeneous equation:
\begin{align}
    \phi^{(n)}_{J;I}(p) = \sum_{I'}\frac{1}{L^{3}}\sum_{{\bf k}}^{\Lambda_v}\frac{1}{2w(\mathbf{k})}Z_{J;II'}(p,k;P)\tau^{L}_{I'}(P-k)\phi^{(n)}_{J;I'}(k)\, .
\end{align}
Note that $Z_{J;II'}(p,k;P)$ and $\tau^{L}_{I}(P-k)$ are energy-dependent quantities and hence the wave function is not normalized to unity but rather obeys the following normalization condition
\begin{align}
   \sum_{II'} \frac{1}{L^{6}}&\sum_{{\bf p},{\bf k}}^{\Lambda_v} \phi^{(n)}_{J;I}(p)\frac{\tau^{L}_{I}(P-p)}{2w({\bf p})}\frac{d Z_{J;II'}(p,k;P)}{d P_{n,\parallel}}\frac{\tau^{L}_{I'}(P-k)}{2w({\bf k})}\phi^{(n)}_{J;I'}(k)\nonumber\\
    &+ \sum_I\frac{1}{L^{3}}\sum_{{\bf p}}^{\Lambda_v} \phi^{(n)}_{J;I}(p)\frac{1}{2w({\bf p})}\frac{d \tau^{L}_{I}(P-p)}{d P_{n,\parallel}}\phi^{(n)}_{J;I}(p) = 1\, ,
\end{align}
where one substitutes $P_\parallel = P_{n,\parallel}$ after differentiation.

\subsection{Matching of the three-pion coupling}
\label{sec:three-bodyforce}
Below, we shall perform the matching of the particle-dimer couplings to the three-pion
threshold amplitude. These amplitudes will be evaluated at tree level in perturbation
theory.\footnote{Note that, within the RFT approach, this matching has been carried out at one loop recently~\cite{Baeza-Ballesteros:2023ljl}. Here, we
  restrict ourselves to the tree-level calculations. Anyway, it will be demonstrated
  below that, in the LL factor we are after,  there is barely any dependence on the
  exact value of the threshold tree-body amplitude.} We start from calculating
the physical three-particle amplitudes
in the particle basis.\footnote{There are only two independent six-pion couplings at lowest order. This means that one can use any two linearly independent physical amplitudes to perform the matching. We use $3\pi^+\to 3\pi^+$ and $3\pi^0\to 3\pi^0$ amplitudes for this purpose here.} In order to express these in terms of particle-dimer amplitudes,
one has to equip the latter by the full dimer propagators for the outgoing legs
and the vertex functions
that describe the transition of a dimer into a two-pion pair, see Eq.~(\ref{eq:T_inf}).
The (connected part of the)
threshold amplitude for the scattering of three charged pions is straightforward
to obtain, since it contains only the isospin $J=3$ contribution: 
\begin{align}\label{eq:conn-plus}
  T^{\sf conn}(3\pi^+\to 3\pi^+)
= 9\mathcal{M}_{3;22}(\varepsilon)\tau_{2}(\varepsilon)^{2}\, .
\end{align}
Here, the factor 9 comes after summation over all permutations of the external lines.
Similarly, the connected part of the amplitude with three neutral pions is given by
\begin{align}
  T^{\sf conn}(3\pi^0\to 3\pi^0) =\,\, &
  3\mathcal{M}_{1;00}(\varepsilon)\tau_{0}^{2}(\varepsilon)
  + \frac{12}{\sqrt{5}}\mathcal{M}_{1;02}(\varepsilon)\tau_{0}(\varepsilon)\tau_{2}(\varepsilon)\nonumber\\
  &  + \frac{12}{5}\mathcal{M}_{1;22}(\varepsilon)\tau_{2}(\varepsilon)^{2}
  + \frac{18}{5}\mathcal{M}_{3;22}(\varepsilon)\tau_{2}(\varepsilon)^{2}\, .\label{eq:t1pi}
\end{align}
The amplitudes entering the above equations are defined as
\begin{align}
  \mathcal{M}_{J;II'}(\varepsilon)&=\mathcal{M}_{J;II'}(\varepsilon)(\bar p,\bar q;\bar P)\, ,
\quad\quad
  \tau_{I}(\varepsilon)=\tau_I((\bar P-\bar p)^2)\, ,
\nonumber\\
  &\bar p=\bar q=(M_\pi,{\bf 0})\, ,\quad \bar P=(3M_\pi-\varepsilon,{\bf 0})\, .
\end{align}
Furthermore, according to Eqs.~(\ref{eq:singular-M}) and (\ref{eq:2body_amplitude}),
the expansion of  $\mathcal{M}_{J;II'}(\varepsilon)$ and $\tau_{I}(\varepsilon)$ in
$\varepsilon$ takes the following form:
\begin{align}
\mathcal{M}_{J;II'}(\varepsilon)&=\frac{c_{J;II'}}{2M_\pi\varepsilon}
+\frac{d_{J;II'}}{2\sqrt{M_\pi\varepsilon}}+e_{J;II'}\log\frac{\varepsilon}{M_\pi}
+\bar{\mathcal{M}}_{J;II'}+\mathcal{O}(\sqrt{\varepsilon})\, ,
\nonumber\\
\tau_{I}(\varepsilon)&=-32\pi M_\pi a_I\left(1+a_I\sqrt{M_\pi\varepsilon}+
a_I^2M_\pi\varepsilon-\frac{2a_I\varepsilon}{\pi}\right)\, ,
\end{align}
where $d_{J;II'}$ and $e_{J;II'}$ are coefficients that are proportional to the powers of the
two-body scattering length. Therefore, applying chiral power counting,
from Eq.~(\ref{eq:singular-M}) it directly follows that
\begin{align}
  c_{J;II'}=\mathcal{O}(1)\, ,\quad\quad
d_{J;II'}=\mathcal{O}(M_\pi^2)\, ,\quad\quad
e_{J;II'}=\mathcal{O}(M_\pi^4)\, .
\end{align}
Taking into account the fact that the tree-level amplitude in ChPT,
which will be matched to the six-point amplitudes given in Eqs.~(\ref{eq:conn-plus})
and (\ref{eq:t1pi}), is of order $M_\pi^2$ (see appendix~\ref{sec:xpt}),
it follows that, at the accuracy we are
working, the regular part of these amplitudes (i.e., the piece that is obtained from the
amplitudes after dropping all divergent pieces in $\varepsilon$ and performing the
limit $\varepsilon\to 0$) is given by
\begin{align}\label{eq:reg}
  T_{\sf reg}^{\sf conn}(3\pi^+ \to 3\pi^+)
&= 9\bar{\mathcal{M}}_{3;22}\bar\tau_{2}^{2}\, ,
\nonumber\\
T_{\sf reg}^{\sf conn}(3\pi^0 \to 3\pi^0) &=
  3\bar{\mathcal{M}}_{1;00}\bar\tau_{0}^{2}
  + \frac{12}{\sqrt{5}}\bar{\mathcal{M}}_{1;02}\bar\tau_{0}\bar\tau_{2}
  + \frac{12}{5}\bar{\mathcal{M}}_{1;22}\bar\tau_{2}^{2}
  + \frac{18}{5}\bar{\mathcal{M}}_{3;22}\bar\tau_{2}^{2}\, ,
\end{align}
where $\bar\tau_I=32\pi M_\pi a_I$, and $\bar{\mathcal{M}}_{J;II'}$ denotes the regular
part of the particle-dimer scattering amplitude.

These amplitudes can be matched to the ones obtained from ChPT,
see the appendix~\ref{sec:xpt}. The result there is given by
\begin{align}
    T_+^{\chi} = \frac{18 M_{\pi}^{2}}{F_{\pi}^{4}}\, ,\quad\quad
    T_0^{\chi} =  -\frac{9M_{\pi}^{2}}{8 F_{\pi}^{4}}\, \label{eq:txpt}.
\end{align}
The three-pion couplings $H_{1;00}(\Lambda)$ and $H_{3;22}(\Lambda)$ can
be determined by numerically solving the Faddeev equation, extracting the
threshold particle-dimer amplitudes $\bar{\mathcal{M}}_{J;II'}$ from these
solutions after subtracting the divergent pieces, and then equating
the result given in Eq.~(\ref{eq:reg}), to $T_{+,0}^{\chi}$.

\subsection{Derivation of the LL factor}

The LL factor connects the decay amplitudes between the
infinite volume and a finite volume. Therefore, in order to obtain the LL factor,
we need to calculate
the amplitude of $K\to3\pi$ twice, separately in the infinite volume and in a finite
volume.

In the infinite volume, we calculate the decay amplitude in two steps.
First, $K$ decays into a dimer and a spectator pion. Then, the
dimer further decays into two pions. The final state at threshold can have
the total isospin $J=1,3$. Assuming, for convenience that $J_3=1$, we get
\begin{align}
  \langle\pi^{i_{1}}(p_{1})\pi^{i_{2}}(p_{2})\pi^{i_{3}}(p_{3})|K^{+}\rangle & =
 \sum_{\alpha}\sum_{I}\mathcal{A}(K^{+}\to T_{I,(1-i_{\alpha})}\pi^{i_{\alpha}}(p_\alpha))
  \langle1,i_\beta;1,i_\gamma|I,(1-i_\alpha)\rangle
  \, ,\label{eq:infinite-amp-1}
\end{align}
with $\alpha\beta\gamma=(123),(231),(312)$. The amplitude for $K^+$ decaying into a dimer and a pion can be expressed as 
\begin{align}
  \mathcal{A}(K^{+}\to T_{I,(1-i_\alpha)}\pi^{i_\alpha})
 = \sum_{J}\langle I,(1-i_\alpha);1,i_\alpha|J,1\rangle
 \Big(g^{(J,I)}\tau_I+ \sum_{I'}\tau_I\Phi_{J;II'}g^{(J,I')}\Big)\, .\label{eq:infinite-amp-3}
\end{align}
where the amplitude $\Phi$ in the CM frame is defined as 
\begin{align}
  \Phi_{J;II'}({\bf p}) & =
  \int^{\Lambda}\frac{d^{3}{\bf q}}{(2\pi)^{3}2w({\bf q})}\mathcal{M}_{J;II'}({\bf p},{\bf q};E)\tau_{I'}({\bf q};E)\, ,
\end{align}
where $\tau_{I'}({\bf q};E)$ stands for $\tau_{I'}((P-q)^2)$ with $P^\mu=(E,{\bf 0})$
and $q^\mu=(\sqrt{M_\pi^2+{\bf q}^2},{\bf q})$.
From the Faddeev equation for the particle-dimer scattering amplitude~(\ref{eq:Faddeev_IV}), the equation for the amplitude $\Phi$ can be derived:
\begin{align}
\Phi_{J;II'}({\bf p}) & =\sum_{I''}\int^{\Lambda}\frac{d^{3}{\bf q}}{(2\pi)^{3}2w({\bf q})}Z_{J;II''}({\bf p},{\bf q};E)\tau_{I''}({\bf q};E)\Big(\delta_{I''I'}+\Phi_{J;I''I'}({\bf q})\Big),\label{eq:infinite-volume-equation}
\end{align}
Using Eqs. (\ref{eq:infinite-amp-1}), (\ref{eq:infinite-amp-3}) and performing the
projection onto the S-wave, for the charged (``c'') and the neutral (``n'') channels
we obtain 
\begin{align}
\begin{pmatrix}\langle\pi^{+}(p_{1})\pi^{+}(p_{2})\pi^{-}(p_{3})|K^{+}\rangle\\
\langle\pi^{0}(p_{1})\pi^{0}(p_{2})\pi^{+}(p_{3})|K^{+}\rangle
\end{pmatrix} & =\begin{pmatrix}X_{c0} & X_{c2}\\
X_{n0} & X_{n2}
\end{pmatrix}\begin{pmatrix}g^{(1,0)}\\
g^{(3,2)}
\end{pmatrix},\label{eq:infinite-volume-matrix}
\end{align}
where we took into account the constraints
$g^{(1,2)}=3g^{(3,2)}$ and $g^{(2,2)}=\sqrt{5}g^{(3,2)}$ already. The entries of the matrix
$X$ are:
\begin{align}\label{eq:X}
X_{c0}= & \frac{\tau_{0}(p_{1})}{\sqrt{3}}\biggl[1+\Phi_{1;00}(p_{1})\biggr]+\frac{\tau_{2}(p_{1})}{2\sqrt{15}}\Phi_{1;20}(p_{1})+(p_{1}\to p_{2})
 +\frac{3\tau_{2}(p_{3})}{\sqrt{15}}\Phi_{1;20}(p_{3}),\nonumber\\
 X_{c2}= & \sqrt{3}\tau_{0}(p_{1})\Phi_{1;02}(p_{1})
                        +\sqrt{\frac{3}{20}}\tau_{2}(p_{1})\biggl[\frac{5}{3}+\Phi_{1;22}(p_{1})
                        +\frac{2}{3}\Phi_{3;22}(p_{1})\biggr]
 +(p_{1}\to p_{2})\nonumber \\
 & +\sqrt{15}\tau_{2}(p_{3})\biggl[\frac{2}{3}+\frac{3}{5}\Phi_{1;22}(p_{3})+\frac{1}{15}\Phi_{3;22}(p_{3})\biggr],
\nonumber\\
  X_{n0}= & -\frac{3\tau_{2}(p_{1})}{2\sqrt{15}}\Phi_{1;20}(p_{1})+(p_{1}\to p_{2})
   -\frac{\tau_{0}(p_{3})}{\sqrt{3}}\biggl[1+\Phi_{1;00}(p_{3})\biggr]+\frac{\tau_{2}(p_{3})}{\sqrt{15}}\Phi_{1;20}(p_{3}),\nonumber\\
X_{n2}= & -\sqrt{\frac{27}{20}}\tau_{2}(p_{1})\biggl[\frac{5}{9}+\Phi_{1;22}(p_{1})-\frac{4}{9}\Phi_{3;22}(p_{1})\biggr]+(p_{1}\to p_{2})\nonumber \\
        & -\sqrt{3}\tau_{0}(p_{3})\Phi_{1;02}(p_{3})+\sqrt{\frac{3}{5}}\tau_{2}(p_{3})\biggl[
          \frac{5}{3}+\Phi_{1;22}(p_{3})+\frac{2}{3}\Phi_{3;22}(p_{3})\biggr].
\end{align}
Note also that we denote $p_i=|{\bf p}_i|$.

In a finite volume, our aim is to compute the matrix element $\langle\Gamma n,J|K^{+}\rangle$.
Here, $\langle\Gamma n,J|$ refers to the state on the lattice, carrying total isospin $J$, total
momentum $\mathbf{d}$ (in units of $2\pi/L$), and residing in the irrep $\Gamma$, corresponding
to the $n$-th energy level\footnote{Here, the size of the lattice is $L$. This parameter
  should be adjusted so that the energy $E_n$ exactly equals to $M_K$.
  The label $n$ is used for different eigenvalues. In order to avoid the clutter of indices,
  we have opted for lumping the irrep index $\Gamma$, as well as the total momentum
$\mathbf{d}$ together with the level index $n$ and hope that this will not lead to a confusion.}
This matrix element can be calculated using
the wave function in a finite volume \cite{Muller:2022oyw}:
\begin{align}
\langle\Gamma n,J|K^{+}\rangle & =\frac{1}{L^{3/2}}\left(\frac{M_{K}}{\sqrt{M_{K}^{2}+\left(\dfrac{2\pi{\bf d}}{L}\right)^{2}}}\right)^{1/2}
\sum_{I}\frac{1}{L^{3}}\sum_{\mathbf{k}}^{\Lambda_{v}}\frac{1}{2w({\bf k})}\,
\phi_{J;I}^{(n)}(\mathbf{k})\tau_{I}^{L}(\mathbf{k};E)
g^{(J,I)}.
\end{align}
Here, $\phi_{J;I}^{(n)}(\mathbf{k})$ represents
the Bethe-Salpeter wave function describing the state $|\Gamma n,J\rangle$,
Furthermore, $\mathbf{k}$ is the momentum of the spectator, and $I$ is
the isospin of the dimer, $\Lambda_{v}$ means that cutoff on $\mathbf{k}$
is imposed 
in a moving system with velocity $v$~\cite{Muller:2021uur}.
To calculate the wave function $\phi_{JI}^{(n)}(\mathbf{k})$,
we first project the Faddeev equation onto irreps $\Gamma$, i.e.,
\begin{align}
\mathcal{M}_{J;II'}^{(\Gamma)}(r,r') & =Z_{J;II'}^{(\Gamma)}(r,r')\nonumber \\
+\sum_{I''} & \frac{1}{GL^{3}}\sum_{s}^{\Lambda_{v}}\left(\frac{\vartheta_{s}}
  {2w_{s}}\right)Z_{J;II'}^{(\Gamma)}(r,s)\tau_{I''}^{L}(s)\mathcal{M}_{J;I''I'}^{(\Gamma)}(s,r^{'})\, .
\end{align}
In the projected equation, the momenta of the spectator particles
are replaced by shell indices $r$, $r'$ and $s$. We use $\vartheta_{s}$
to represent the multiplicity of the shell $s$, while $G$ denotes the number of elements in the discrete symmetry group that leaves the total three-momentum of the system invariant. The projection of
$\tau$ and $Z$ is performed, according to the method of Ref.~\cite{Doring:2018xxx}
\begin{align}
  \tau^{L}(s)=\tau^{L}(\mathbf{k}_{0}(s)),\quad &
  Z_{J;II'}^{(\Gamma)}(r,s)=
  \sum_{g\in\mathcal{G}}\left(T^{(\Gamma)}(g)\right)^{\dagger}
  Z_{J;II'}(g\mathbf{p}_{0}(r),\mathbf{k}_{0}(s)).
\end{align}
Here $T^{(\Gamma)}(g)$ is the representation matrix and the momenta $\mathbf{p}_{0}(r)$
and $\mathbf{k}_{0}(s)$ are the reference vectors of the $r$-shell and the $s$-shell, respectively.
Based on this, it is not difficult to deduce that the wave function
is a solution to the homogeneous equation,
\begin{align}\label{eq:homogenous}
  \phi_{J;I}^{(n)}(r) & =\sum_{I'}\frac{1}{GL^{3}}\sum_{s}\left(\frac{\vartheta_{s}}
      {2w_{s}}\right)Z_{J;II'}^{(\Gamma)}(r,s)\tau_{I'}^{L}(s)\phi_{J;I'}^{(n)}(s)\, .
\end{align}
The solutions of
the equation should be normalized, according to~\cite{Muller:2022oyw}
\begin{align}
 & \sum_{r,s}\sum_{I,I^{'}}\frac{\vartheta_{r}}{2w_{r}L^{3}}\,\frac{\vartheta_{s}}{2w_{s}L^{3}}\,\phi_{JI}^{(n)}(r)\tau_{I}^{L}(r)\mathbb{P}_{J;II'}(r,s)
  \tau_{I'}^{L}(s)\phi_{JI'}^{(n)}(s)
\nonumber\\
  &  -\sum_{r}\sum_{I}
\frac{\vartheta_{r}}{2w_{r}L^{3}}\,
  \phi_{JI}^{(n)}(r)\tau_{I}^{L}(r)
  \mathbb{Q}_{I}(r)
  \tau_{I}^{L}(r)\phi_{JI}^{(n)}(r)
  =1.
\end{align}
Here 
\begin{align}
  \mathbb{P}_{J;II'}(r,s) =\frac{\partial}{\partial P_{\parallel}}Z_{J;II^{'}}^{\Gamma}(r,s)\, ,
  \quad\quad
\mathbb{Q}_{I}(r) =\frac{\partial}{\partial P_{\parallel}}\left(\tau_{I}^{L}(r)\right)^{-1}.
\end{align}
The time-like component $P_{\parallel}$ is obtained from total four-momentum $P^\mu$:
\begin{align}
  P_{\parallel} & =v\cdot P=\sqrt{P^{2}}.
\end{align}
Similar to the infinite-volume case, we can define the finite-volume
amplitude $\Phi$:
\begin{align}\label{eq:PhinJI}
  \Phi^{(n)}_{J;I} & =\frac{1}{L^{3}}\sum_{\mathbf{k}}^{\Lambda_{v}}\frac{1}
      {2w({\bf k})}\phi_{J;I}^{(n)}(\mathbf{k})\tau_{I}^{L}(\mathbf{k};E)
      =\frac{1}{L^{3}}\sum_{r}^{\Lambda_{v}}\left(\frac{\vartheta_{r}}{2w_{r}}\right)
      \phi_{J;I}^{(n)}(r)\tau_{I}^{L}(r)\, .
\end{align}
Therefore, we have
\begin{align}
  \langle\Gamma n,J|K^{+}\rangle & =\frac{1}{L^{3/2}}\left(\frac{M_{K}}
                {\sqrt{M_{K}^{2}+\left(\dfrac{2\pi{\bf d}}{L}\right)^{2}}}\right)^{1/2}
 \sum_{I}\Phi^{(n)}_{J;I}g^{(J,I)}\, .
\end{align}
Finally, one obtains:
\begin{align}
\begin{pmatrix}L^{3/2}\langle\Gamma n,1|K^{+}\rangle\\
L^{3/2}\langle\Gamma n,3|K^{+}\rangle
\end{pmatrix} & =\begin{pmatrix}A_{10} & A_{12}\\
A_{30} & A_{32}
\end{pmatrix}\begin{pmatrix}g^{(1,0)}\\
g^{(3,2)}
\end{pmatrix}.\label{eq:finite_volume_dependenc}
\end{align}
Here, matrix element $A_{JI}$ is given by 
\begin{align}\label{eq:AJI}
A_{JI} & =\left(\frac{M_{K}}{\sqrt{M_{K}^{2}+\left(\dfrac{2\pi{\bf d}}{L}\right)^{2}}}\right)^{1/2}a_{JI}\Phi^{(n)}_{J;I}\, ,
\end{align}
and the coefficients $a_{JI}$ are
\begin{align}
a_{10}=1,\quad a_{12}=3,\quad a_{30}=0,\quad  a_{32}=1.
\end{align}
The decay amplitudes both in a finite and in the infinite volume are expressed through
the same couplings $g^{(1,0)}$ and $g^{(3,2)}$, see Eqs.~(\ref{eq:infinite-volume-matrix}) and (\ref{eq:finite_volume_dependenc}). Excluding these couplings, we
 obtain the
LL factor for $K\to3\pi$ which, in this case, is a $2\times 2$ matrix: 
\begin{align}
\begin{pmatrix}\langle\pi^{+}\pi^{+}\pi^{-}|K^{+}\rangle\\
\langle\pi^{0}\pi^{0}\pi^{+}|K^{+}\rangle
\end{pmatrix} & =\begin{pmatrix}\mathbbm{L}_{c1} & \mathbbm{L}_{c3}\\
\mathbbm{L}_{n1}& \mathbbm{L}_{n3}
\end{pmatrix}\begin{pmatrix}L^{3/2}\langle\Gamma n,1|K^{+}\rangle\\
L^{3/2}\langle\Gamma n,3|K^{+}\rangle
\end{pmatrix},
\end{align}
where 
\begin{align}
\begin{pmatrix}\mathbbm{L}_{c1} & \mathbbm{L}_{c3}\\
\mathbbm{L}_{n1}& \mathbbm{L}_{n3}
\end{pmatrix}
                & =\begin{pmatrix}X_{c0} & X_{c2}\\
X_{n0} & X_{n2}
\end{pmatrix}\begin{pmatrix}A_{10} & A_{12}\\
A_{30} & A_{32}
\end{pmatrix}^{-1}\, .
\end{align}
\textcolor{black}{The explicit expression for the 3-particle LL factor, given by the
above formulae, represents one of the main results of the present work.}

\section{Numerical calculation of LL factor}\label{sec:numerics}

\subsection{Solution of the Faddeev equation in the infinite volume}

In order to solve Eq.~(\ref{eq:infinite-volume-equation}) in the infinite volume,
we need to first study the analytic properties of the kernel. For simplicity, we
restrict ourselves to
the CM system and use the notation $p=|{\bf p}|$, $q=|{\bf q}|$ and $z=\cos\theta$,
where $\theta$ is the angle between ${\bf p}$ and ${\bf q}$. It is
straightforwardly seen that the singularity of $Z({\bf p},{\bf q};E)$ 
is determined by the three-body on-shell condition\footnote{In
principle, the dimer propagator $\tau_{L}$ also has singularities,
but they are irrelevant to our calculation.}, 
\begin{align}\label{eq:sing}
\sqrt{p^{2}+q^{2}+2pqz+M_\pi^{2}}+\sqrt{p^{2}+M_\pi^2}+\sqrt{q^{2}+M_\pi^{2}} & =E,
\end{align}
with $-1\leq z\leq 1$. As known~\cite{Glockle:1978zz,Cahill:1971ddy,Garofalo:2022pux,Dawid:2023jrj}, one can avoid these (logarithmic) singularities
by deforming the contour into the complex plane
(see Fig.~\ref{fig:Contour-and-singularity}).
In our case, the contour can be chosen as follows, 
\begin{align}
q & =\begin{cases}
t-i\mu(1-e^{-t/\sigma})(1-e^{(t-B_{\max})/\sigma}), & (0<t<B_{\max})\\
t, & (B_{\max}<t<\Lambda)
\end{cases}
\end{align}
where the parameters are 
\begin{align}
\delta=0.5M_\pi,\;\sigma=M_\pi,\;\mu=0.1M_\pi,\; & B_{\text{max}}=p_{t}+\delta.
\end{align}
Here $p_{t}$ is the magnitude of the momentum of the spectator, for which 
the two-body system is exactly at threshold. If $|{\bf p}|>p_t$, the two-body system
moves below threshold. This gives
\begin{align}
(E-w({\bf p}))^{2}-\mathbf{p}^{2} & \leq 4M_\pi^{2},
\end{align}
and
\begin{align}
p\geq\sqrt{\left(\frac{E^{2}-3M_\pi^{2}}{2E}\right)^{2}-M_\pi^{2}} & = p_{t}.
\end{align}
The solution of the equation~(\ref{eq:infinite-volume-equation}) proceeds step by step.
First, after projecting onto the S-wave,
one solves this equation with both the momenta $p$ and $q$ belonging
to the contour $C$: 
\begin{align}\label{eq:on-the-contour}
  \Phi_{J;II^{'}}(p) & =\sum_{I^{''}}\int_{C}\frac{q^2dq}{2\pi^{2}2w(q)}
    Z_{J;II^{''}}(p,q;E)\tau_{I^{''}}(q;E)\Big(\delta_{I^{''}I^{'}}+\Phi_{J;I^{''}I^{'}}(q)\Big)\, .
\end{align}
The integrand is never singular on the integration contour $C$. This can be visualized
in Fig.~\ref{fig:Contour-and-singularity}, where the shaded area is obtained as follows.
The total energy $E=M_K$ is fixed.
We first choose the momentum $p$ on the contour and plot the curve for $q$,
obtained by the solution
of Eq.~(\ref{eq:sing}) for $-1\leq z\leq 1$. Changing then $p$ with a very
small step along the contour, we arrive at a new curve. Repeating this procedure many
times, we arrive at a shaded area that does not intersect with our integration curve. Hence,
for $p$ and $q$ both on the curve, the kernel $Z$ never becomes singular, and the
integral equation
can be straightforwardly transformed into a set of linear equations by discretizing
the integration with the use of the
fixed mesh points and weights. We are ultimately interested, however,
in the amplitude $\Phi_{J;II^{'}}(p)$ on the real axis, which can be obtained by analytically
continuing   $\Phi_{J;II^{'}}(p)$, defined by the Eq.~(\ref{eq:on-the-contour})
to the real axis,
whereas the argument $q$, over which the integration is performed, still stays on
the the contour $C$. In doing so, it is important to check that one does not hit the
singularities of the kernel $Z$ during this analytic continuation. Below, we shall address
this issue in detail.

Let us define the quantity
\begin{align}
p_{0} & \equiv\sqrt{\left(\frac{E-M_\pi}{2}\right)^{2}-M_\pi^{2}}
\end{align}
It turns out that there are three different regions of $p$~\cite{Schmid,Cahill:1971ddy}:

\begin{enumerate}

\item

When
$p<p_{0}$, the singularities of $Z(p,q;E)$ lie just above and below the real axis,\footnote{Remember that the energy $E$ has infinitesimal positive imaginary part.} see
Fig.~\ref{fig:back-I}. Our contour does not intersect with any of these branch cuts and
the integration can be safely performed.

\item
  For $p_{0}<p<p_{t}$, The branch cuts move into the complex plane and start to intersect
  with the chosen contour, see Fig. \ref{fig:back-II}.
  In order to avoid the singularities, one has to deform the integration contour
  as well. The new contour, $C^{+}$,  consists of two parts: 

  \bigskip
  
{\em Part 1:} Starting from the origin, it follows the positive real axis
  until the beginning of the branch cut located at the first singularity of the kernel
  at $q=q_1$, where $q_1$ is given by
\begin{align}
  q_1 & =\frac{p}{2}-\frac{\varepsilon(p)}{2}\,\biggl(1-\frac{4M_\pi^2}{\varepsilon^2(p)-p^2}\biggr)^{1/2}\, ,\quad\quad\varepsilon(p)=E-w(p)\, .
        \end{align}
After reaching $q_1$, the contour
dives to the second Riemann sheet and
  goes back to the origin along the real axis. The contribution of this part to the
  integral is given by
\begin{align}
  &\quad\quad\Phi_{J;II^{'}}^{(1)}(p)
  \nonumber\\
  &=\sum_{I^{''}}\left(\int_{0+i\varepsilon}^{q_{1}+i\varepsilon}
    +\int_{q_{1}-i\varepsilon}^{0-i\varepsilon}\right)\frac{q^2dq}{2\pi^{2}2w(q)}
    Z_{J;II^{''}}(p,q;E)\tau_{I^{''}}(q;E)
  \Big(\delta_{I^{''}I^{'}}+\Phi_{J;I^{''}I^{'}}(q)\Big)\nonumber \\
  & =\sum_{I^{''}}\int_{0}^{q_{1}}\frac{q^2dq}{2\pi^{2}2w(q)}
    \Big[Z_{J;II''}(p,q;E)-Z_{J;II''}^{({\sf II})}(p,q;E)\Big]\tau_{I''}(q;E)
\Big(\delta_{I''I'}+\Phi_{J;I''I'}(q)\Big).
\end{align}
The kernel $Z_{J,II'}$, projected on any partial wave, is expressed 
through  a logarithmic function. Since the discontinuity 
is caused solely by this logarithm,
we conclude that the kernel on the second Riemann sheet, $Z^{({\sf II})}$
is obtained from $Z$ by replacing
the $\log z$ with $\log z+2\pi i$. Furthermore, since $q_1<p_0$, the function
$\Phi(q)$ is known along the real axis and the integral is completely defined.

\bigskip

{\em Part 2:} The second part of the path starts from the origin of the second Riemann sheet,
following path $C$ to reach the intersection point $q_{2}$ with
the branch cut. Returning through $q_{2}$ to the first Riemann sheet,
it continues along path $C$ to reach the integration endpoint $\Lambda$,
that is 
\begin{align}
 \Phi_{J;II^{'}}^{(2)}(p)
   & =\sum_{I''}\int_{0}^{q_{2}}\frac{q^2dq}{2\pi^{2}2w(q)}Z_{J;II''}^{({\sf II})}(p,q;E)
    \tau_{I''}(q;E)\Big(\delta_{I''I'}+\Phi_{J;I''I'}(q)\Big)\nonumber \\
  & +\sum_{I''}\int_{q_{2}}^{\Lambda}\frac{q^2dq}{2\pi^{2}2w(q)}
    Z_{J;II''}(p,q;E)\tau_{I''}(q;E)\Big(\delta_{I''I'}+\Phi_{J;I''I'}(q)\Big)\nonumber \\
  & =\sum_{I''}\int_{C}\frac{q^2dq}{2\pi^{2}2w(q)}\tilde{Z}_{J;II''}(p,q;E)
    \tau_{I''}(q;E)\Big(\delta_{I''I'}+\Phi_{J;I''I'}(q)\Big)\, .
\end{align}
The quantity $\tilde{Z}$ is defined in the following way:
\begin{align}
\tilde{Z}_{J;II'}(p,q;E) & =\begin{cases}
Z_{J;II'}^{({\sf II})}(p,q;E), & f(p,q;E)<0\\
Z_{J;II'}(p,q;E), & f(p,q;E)\geq0
\end{cases}.
\end{align}
Here, $f(p,q;E)$ is defined by 
\begin{align}
  f(p,q;E) & = \left((\text{Re}q)^{2}+(\text{Im}q)^{2}\right)^2
             +m^2(d^2(p)-b^2(p))\left(\frac{(\text{Re}q)^{2}}{d^2(p)}
             +\frac{(\text{Im}q)^{2}}{b^2(p)}\right),
\end{align}
with 
\begin{align}
b(p)=\frac{\varepsilon(p)}{p}\, ,\quad\quad d(p)=\frac{\varepsilon^2(p)-p^{2}}{2mp}.
\end{align}
Finally, in the region $p_0< p< p_t$, the full amplitude is given by 
\begin{align}
\Phi_{J;II'}(p) & =\Phi_{J;II'}^{(1)}(p)+\Phi_{J;II'}^{(2)}(p).
\end{align}

\begin{figure}[t]
\begin{centering}
  \subfigure[\label{fig:Contour-and-singularity}]{\begin{centering}
\includegraphics[scale=0.35]{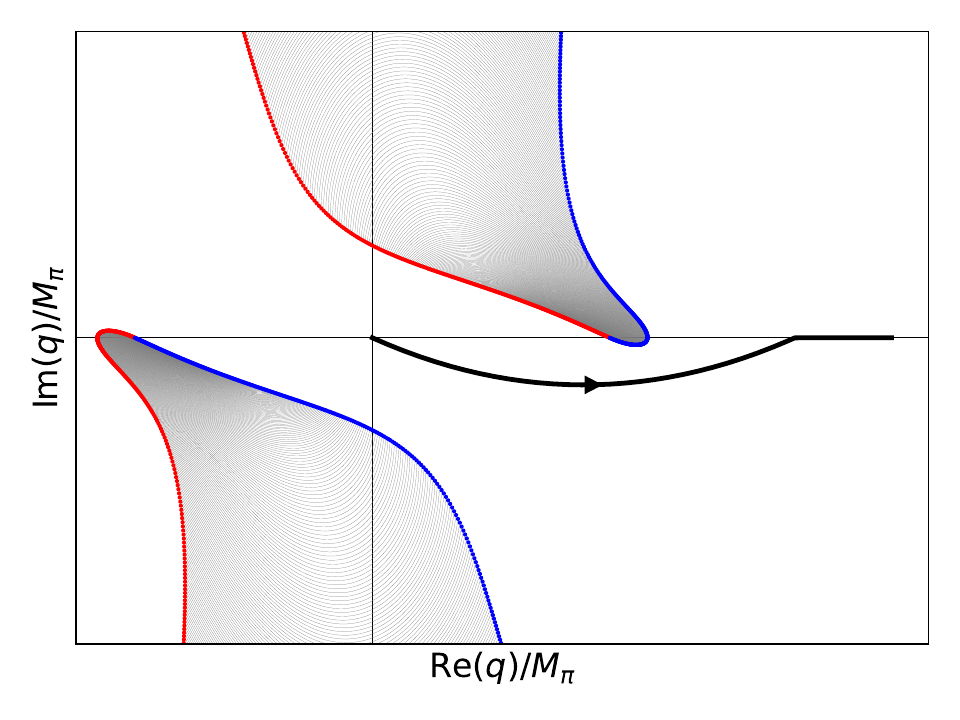}
\par\end{centering}
}$\qquad$\subfigure[\label{fig:back-I}]{\begin{centering}
\includegraphics[scale=0.35]{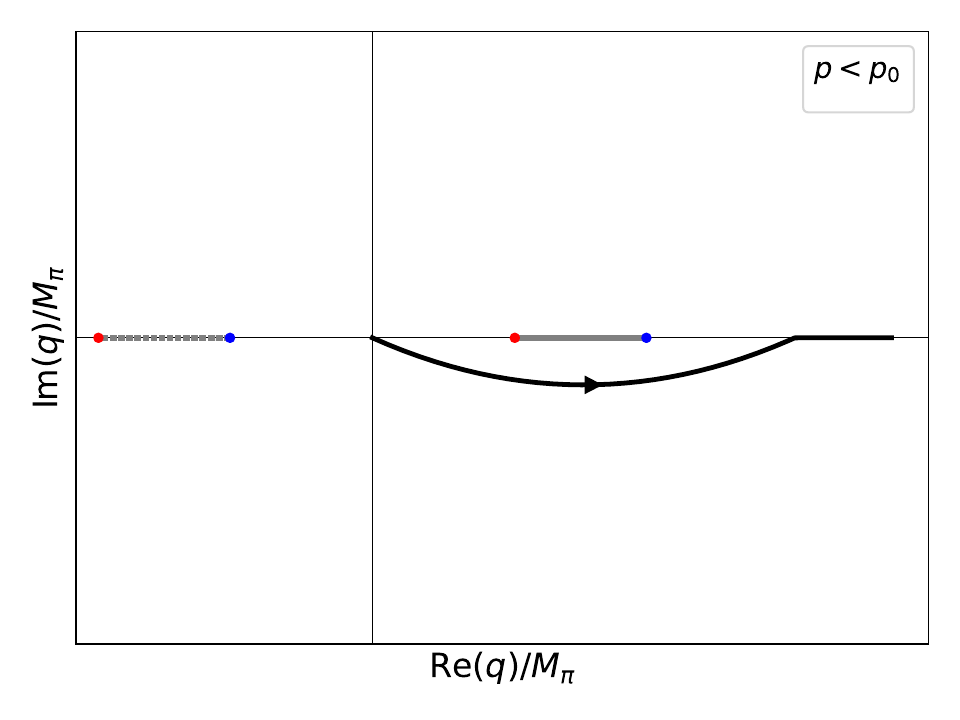}
\par\end{centering}
}
\par\end{centering}
\begin{centering}
\subfigure[\label{fig:back-II}]{\begin{centering}
\includegraphics[scale=0.35]{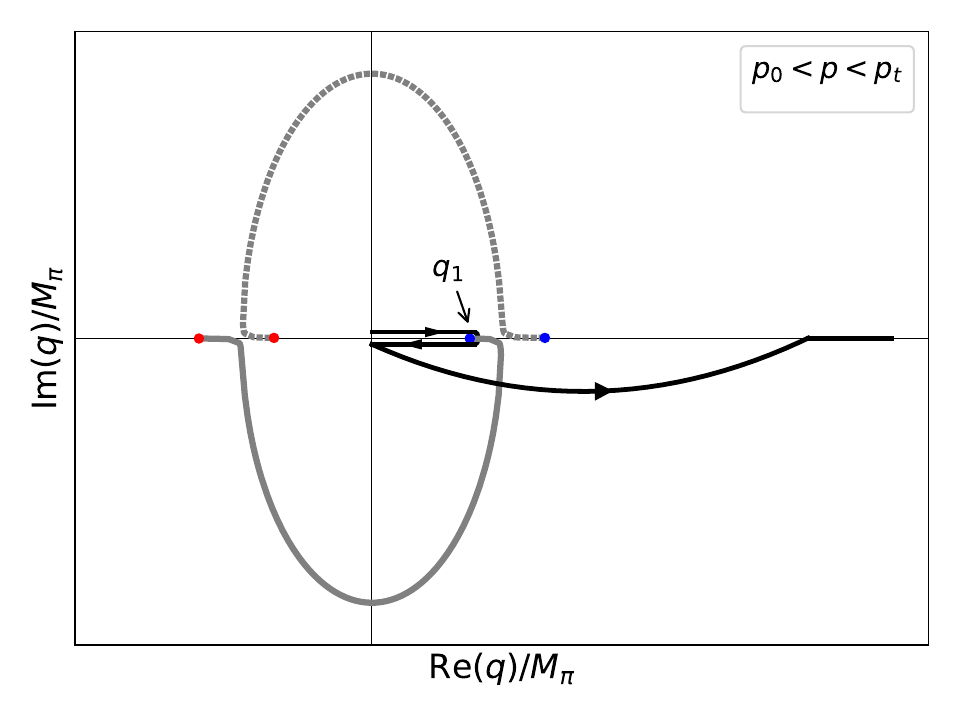}
\par\end{centering}
}$\qquad$\subfigure[\label{fig:back-III}]{\begin{centering}
\includegraphics[scale=0.35]{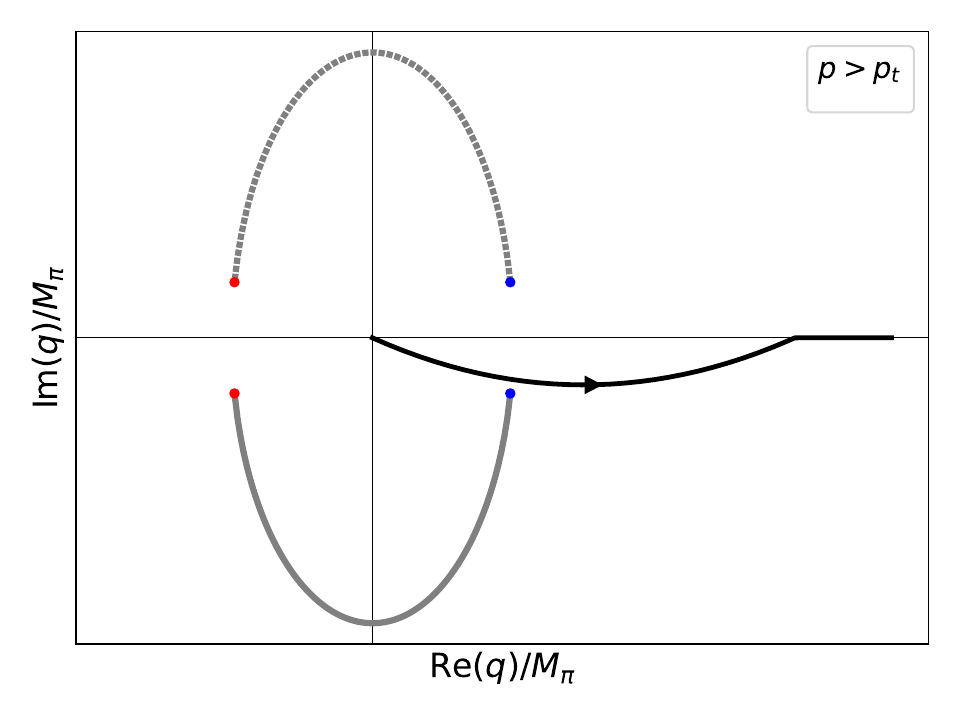}
\par\end{centering}
}
\par\end{centering}
\caption{In panel (a), the contour and the domain of the singularity of the kernel
  $Z$ is shown. It is clear that, when both momenta are located on the deformed contour,
  the kernel does not become singular. In the panels (b,c,d), the choice of the integration
  contour for the calculation of the amplitude $\Phi$ on the real axis is
  displayed for three different choices,  $p<p_{0}$, $p_{0}<p<p_{t}$,
  and $p>p_{t}$, respectively. In the second case, the path should be deformed from
  the original one, in order to avoid the singularities of the kernel (see the discussion in
  the text).}
\end{figure}

\item
  When $p>p_{t}$, the branch cuts of the function $Z$ move into the complex
  plane and do not intersect with our contour $C$ anymore,\footnote{In principle, when $p>p_{t}$ but very close
  to $p_{t}$, the branch cut may still intersect with the contour. Strictly speaking,
  for any given $p>p_t$, one may choose the parameter $\mu$ small enough that
  there is no intersection. On the contrary, for any fixed finite $\mu$, one may find
  an interval between $p_t$ and $p_t+\varepsilon'$, when the branch cut and the
  contour intersect. The quantity $\varepsilon'$ depends on $\mu$ and tends to
  zero at $\mu\to 0$. In practical terms, it means that one is not able to determine
  the solution for $p\in [p_t,p_t+\varepsilon']$ by using the method described here.
  The length of this interval can be however made arbitrary small with the choice
  of $\mu$. This does not create problems in actual calculations, while
  $\varepsilon'$ turns out to be extremely small. 
}
see Fig.~\ref{fig:back-III}.
Therefore, it is again safe to integrate over the initial contour.

\end{enumerate}

\begin{figure}[t]
\begin{centering}
\includegraphics[scale=0.4]{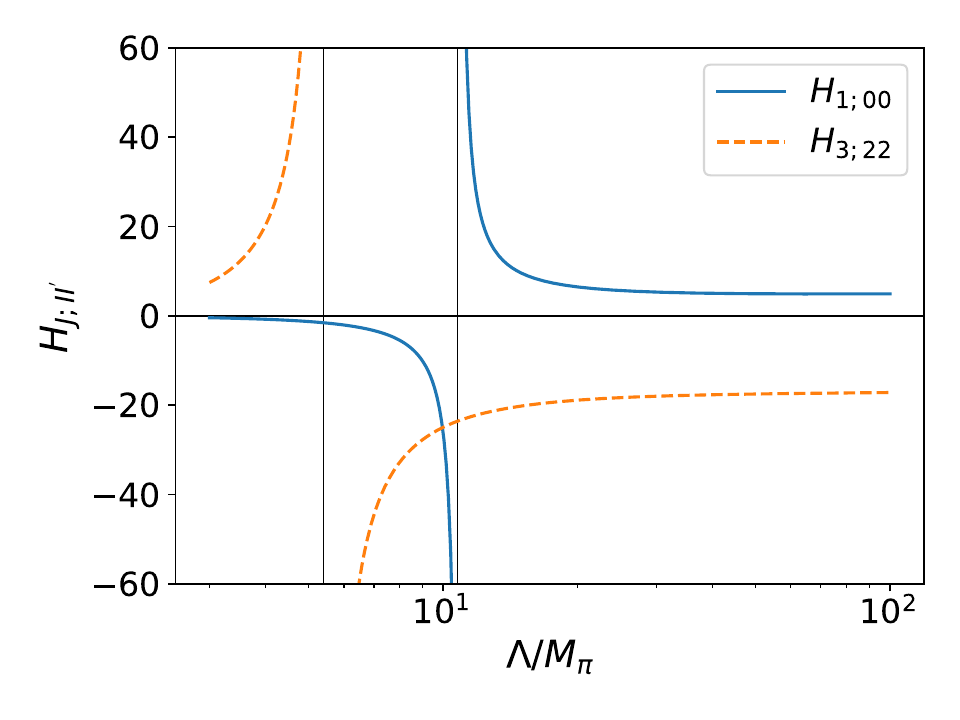}
\par\end{centering}
\caption{\label{fig:H-running}The running
  of the three-body couplings $H_{1;00}$ and $H_{3;22}$ with respect to the cutoff $\Lambda$.}
\end{figure}

\begin{figure}[t]
\begin{centering}
\subfigure[]{\begin{centering}
\includegraphics[scale=0.35]{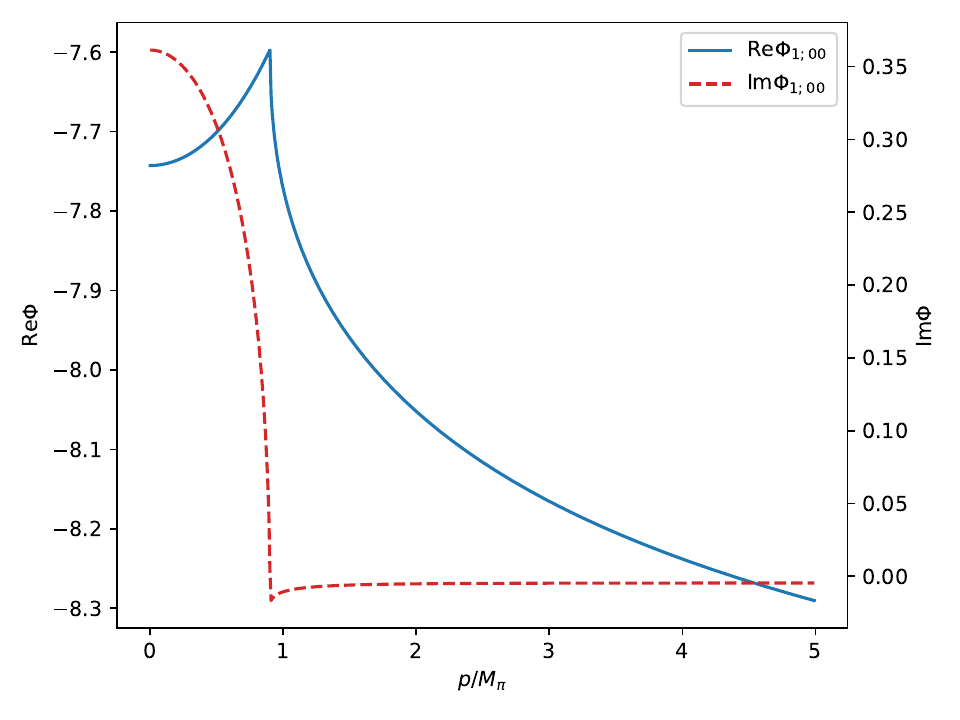}
\par\end{centering}
}\subfigure[]{\begin{centering}
\includegraphics[scale=0.35]{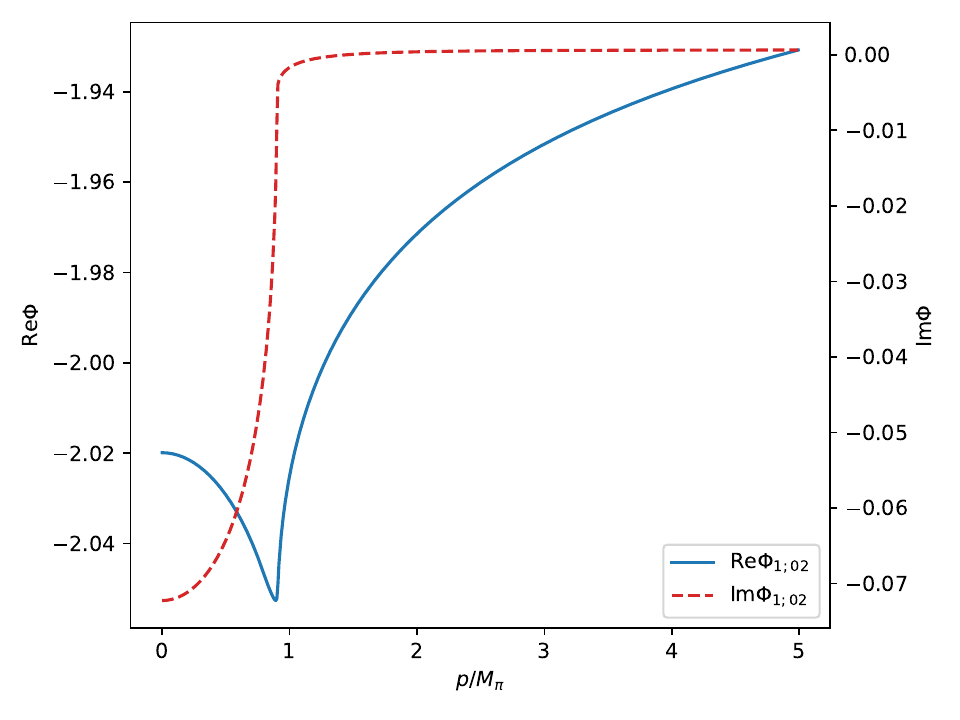}
\par\end{centering}
}
\par\end{centering}
\begin{centering}
\subfigure[]{\begin{centering}
\includegraphics[scale=0.35]{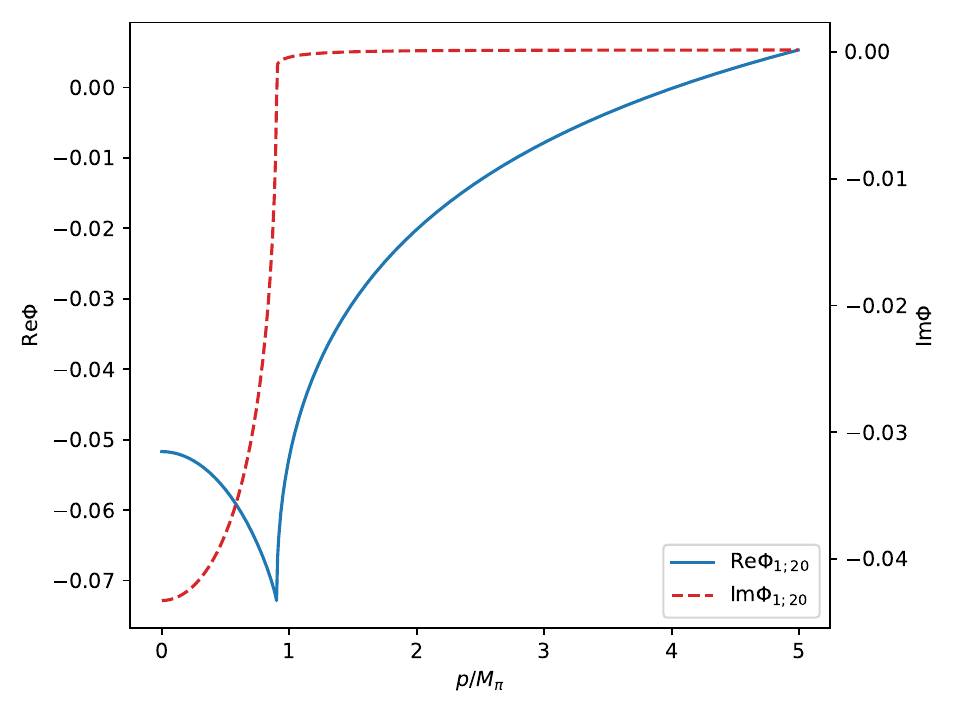}
\par\end{centering}
}\subfigure[]{\begin{centering}
\includegraphics[scale=0.35]{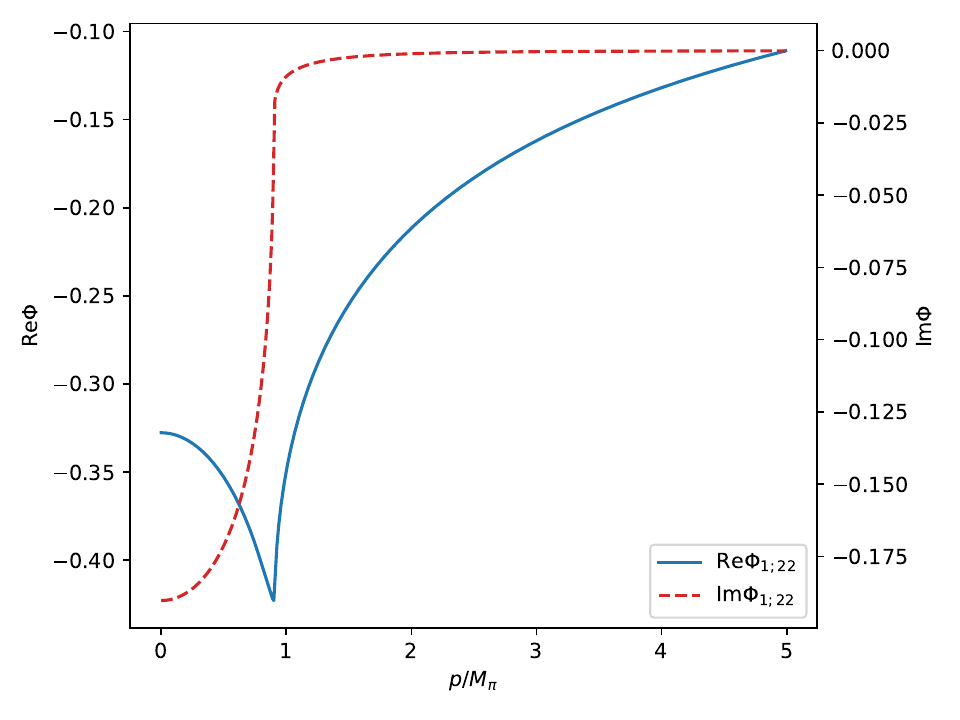}
\par\end{centering}
}
\par\end{centering}
\begin{centering}
\subfigure[]{\begin{centering}
\includegraphics[scale=0.35]{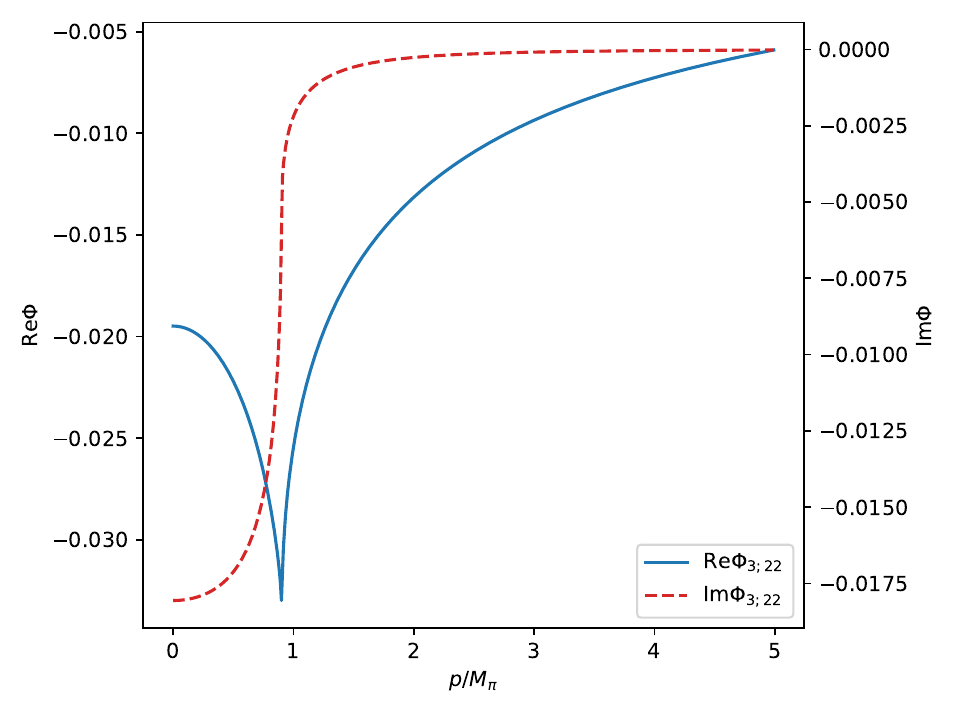}
\par\end{centering}
}
\par\end{centering}
\caption{\label{fig:Phi-amp}The solutions
 of  Eq.~(\ref{eq:infinite-volume-equation}), $\Phi_{1;00}$, $\Phi_{1;02}$,
$\Phi_{1;20}$, $\Phi_{1;22}$ and $\Phi_{3;22}$.
The blue solid line and the red dashed line represent the real and imaginary parts,
respectively. The cutoff $\Lambda=15M_\pi$ was chosen.}
\end{figure}

By using the techniques described above, we can calculate
the amplitudes $\Phi_{1;00}$, $\Phi_{1;02}$,
$\Phi_{1;20}$, $\Phi_{1;22}$ and $\Phi_{3;22}$ on the real axis.
These solutions are displayed in Fig.~\ref{fig:Phi-amp}.
We choose the cutoff  $\Lambda=15M_\pi$ and matched the
couplings $H_{1;00}$ and $H_{3;22}$ to the 
tree-level threshold amplitudes in ChPT as described above.
Varying the cutoff, but keeping the threshold
amplitude fixed, it can be shown that the three-body couplings exhibit singular behavior
that resembles the log-periodic
running in the unitary limit, see Fig.~\ref{fig:H-running}. Note also that the LL factor should be
cutoff-independent that will be explicitly demonstrated below.

\begin{figure}[t]
\begin{centering}
\subfigure[]{\begin{centering}
\includegraphics[scale=0.3]{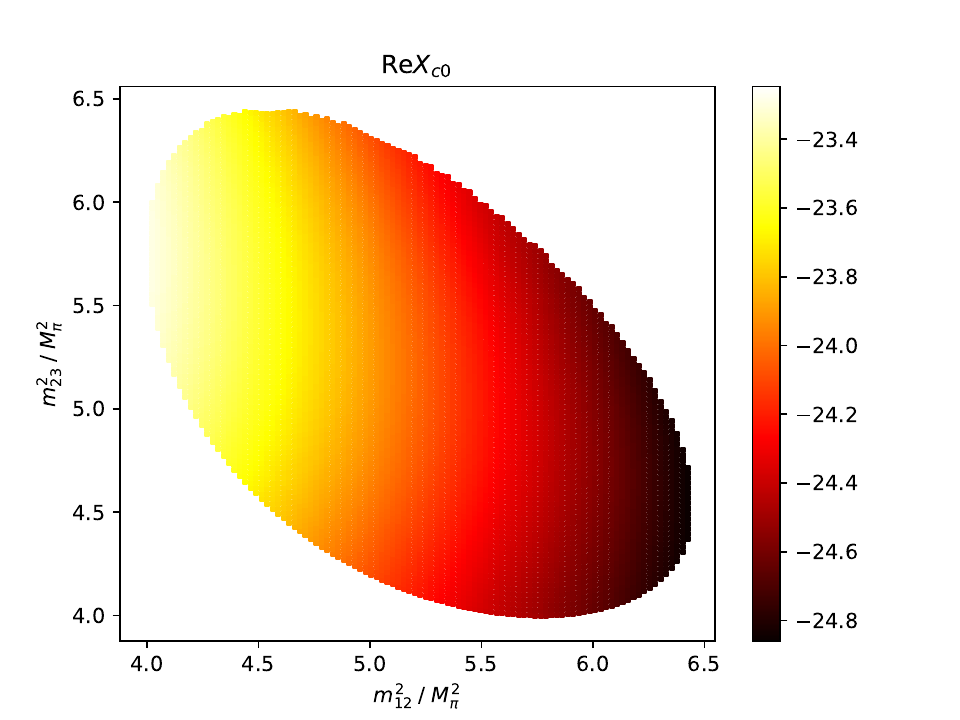}
\par\end{centering}
}\subfigure[]{\begin{centering}
\includegraphics[scale=0.3]{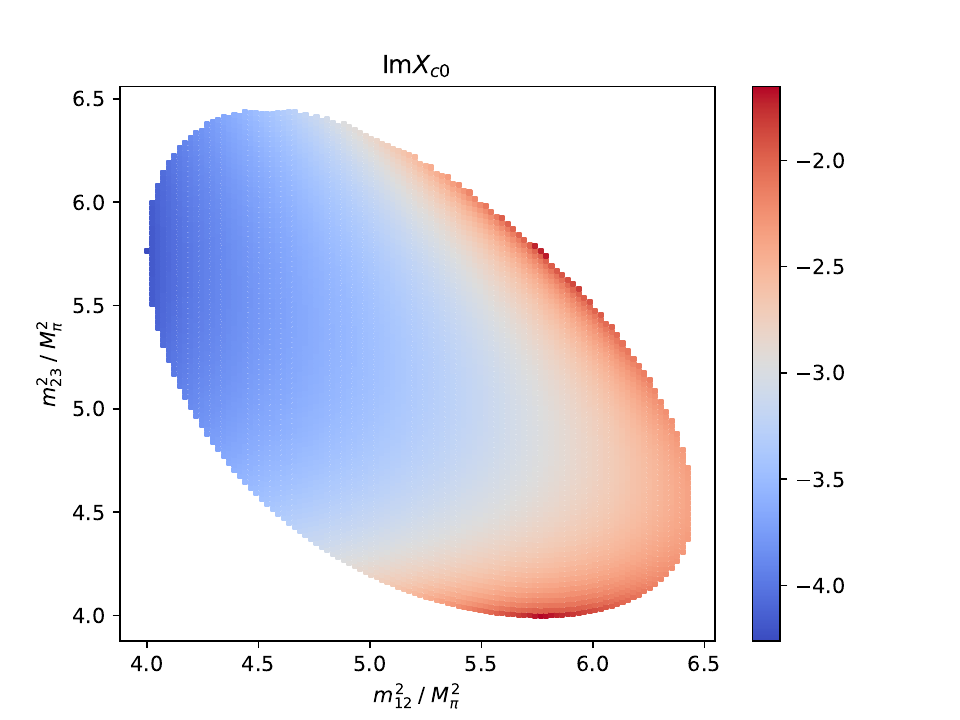}
\par\end{centering}
}
\par\end{centering}
\begin{centering}
\subfigure[]{\begin{centering}
\includegraphics[scale=0.3]{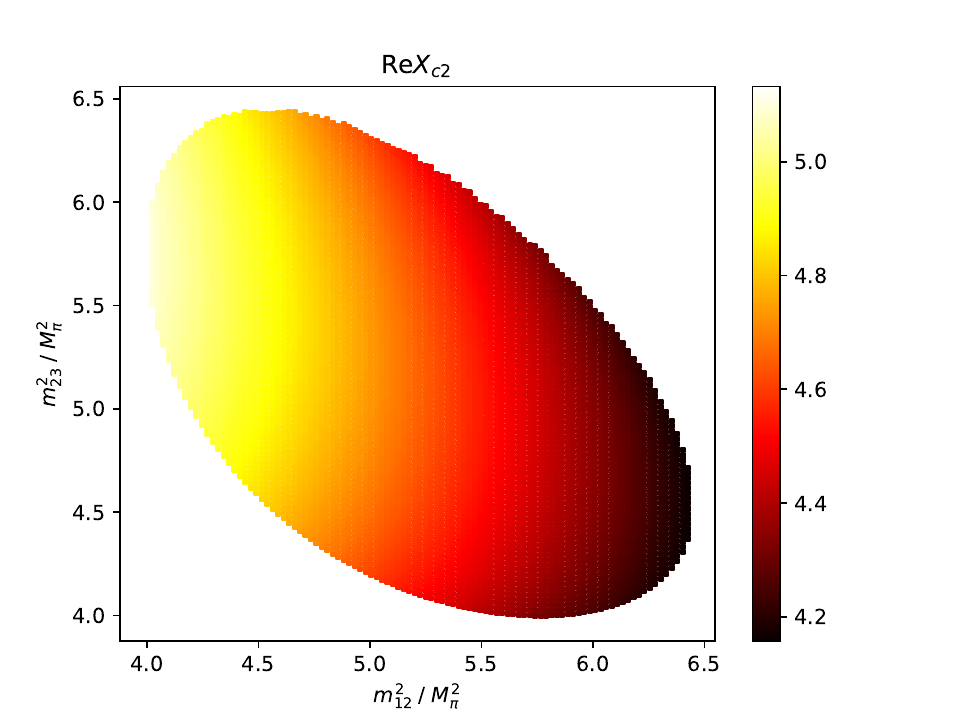}
\par\end{centering}
}\subfigure[]{\begin{centering}
\includegraphics[scale=0.3]{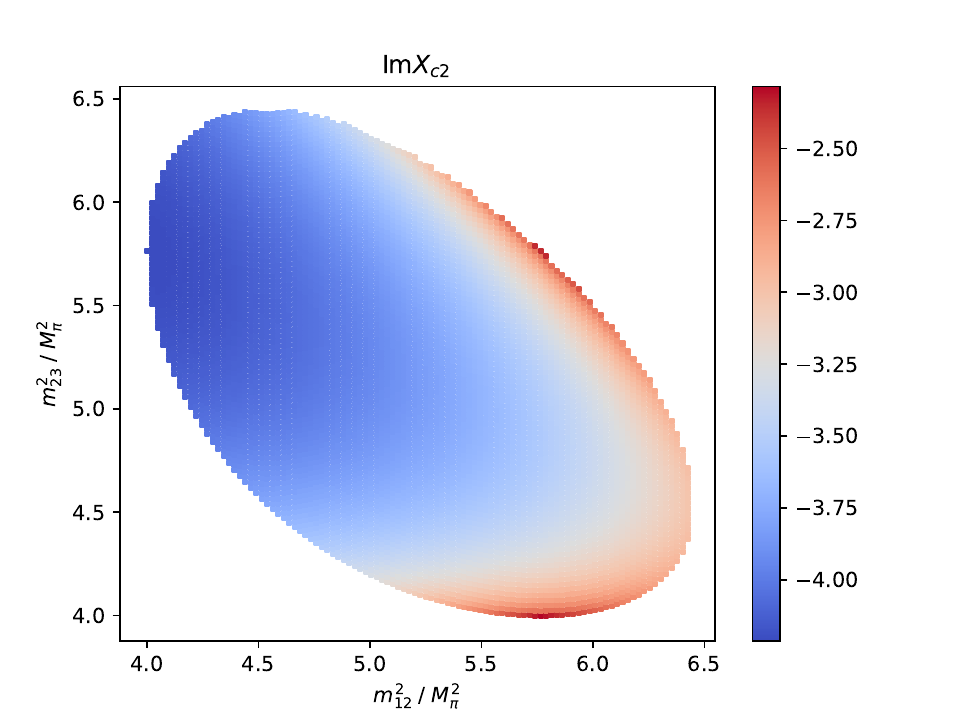}
\par\end{centering}
}
\par\end{centering}
\begin{centering}
\subfigure[]{\begin{centering}
\includegraphics[scale=0.3]{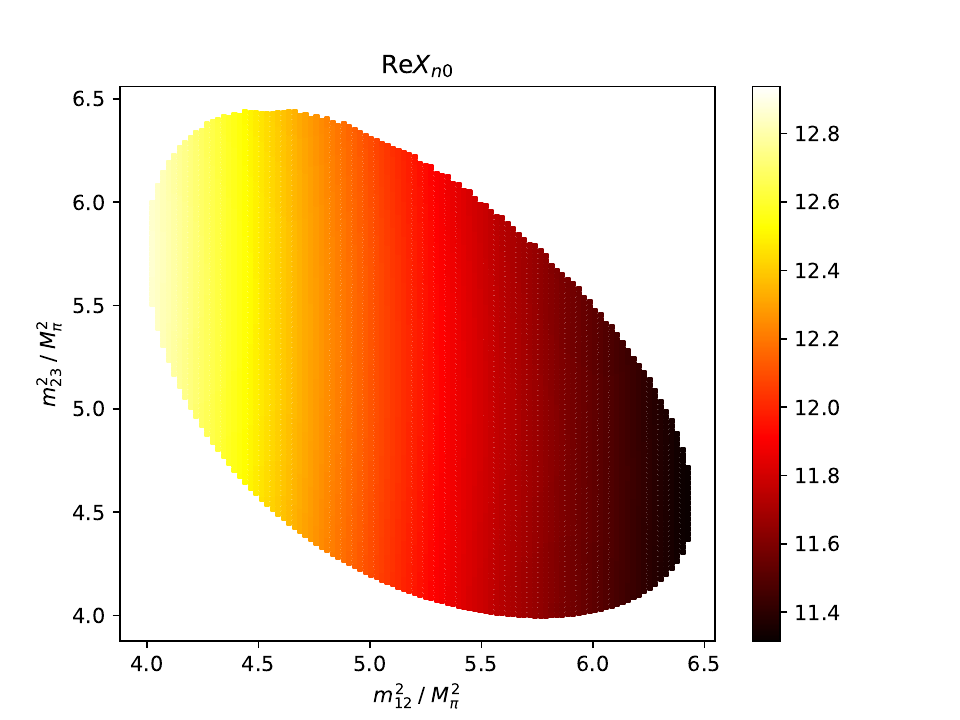}
\par\end{centering}
}\subfigure[]{\begin{centering}
\includegraphics[scale=0.3]{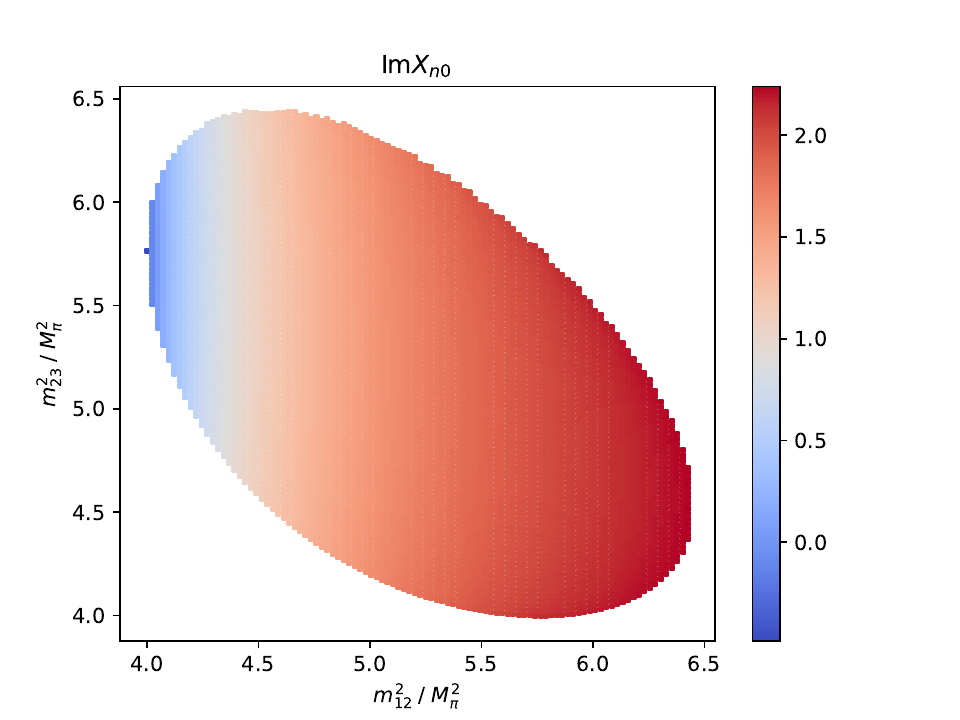}
\par\end{centering}
}
\par\end{centering}
\begin{centering}
\subfigure[]{\begin{centering}
\includegraphics[scale=0.3]{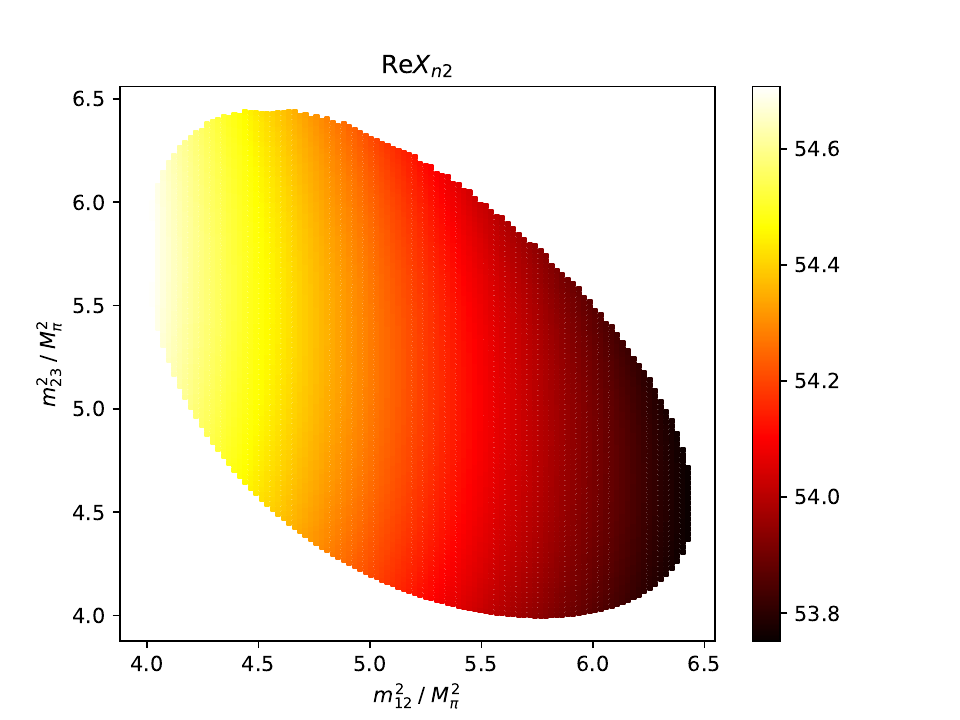}
\par\end{centering}
}\subfigure[]{\begin{centering}
\includegraphics[scale=0.3]{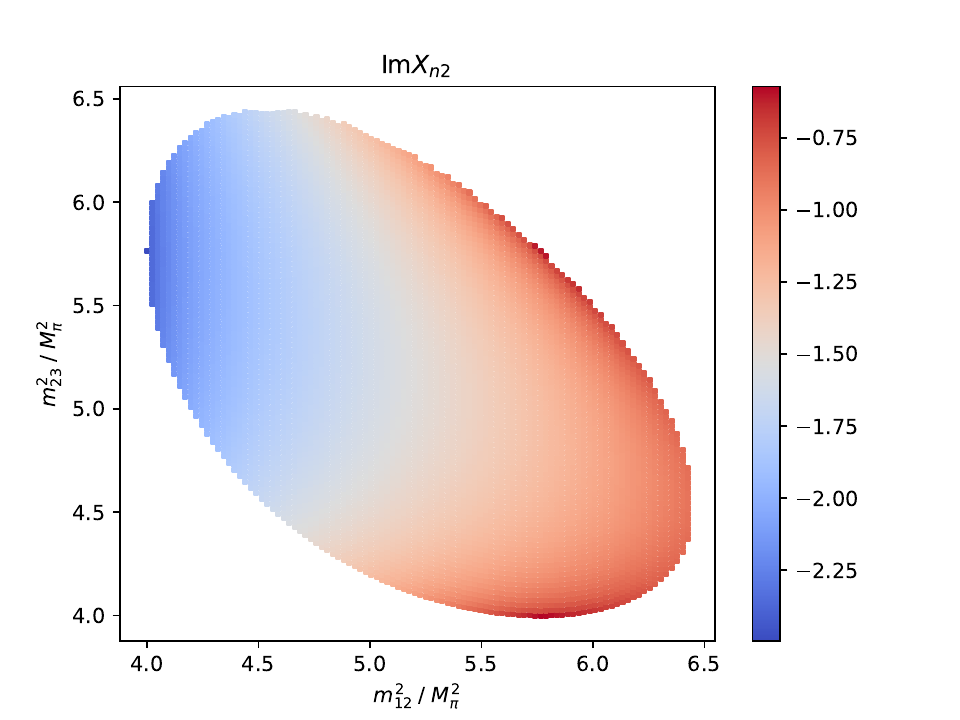}
\par\end{centering}
}
\par\end{centering}
\caption{\label{fig:XY-Dalitz}
  The real and imaginary parts of the quantities $X_{c0}$, $X_{c2}$, $X_{n0}$ and $X_{n2}$
  in the $m_{12}^2$, $m_{23}^2$-plane. The cutoff $\Lambda=15M_\pi$ is chosen.
}
\end{figure}

Next, in order to estimate the dependence on the kinematical variables,
we shall calculate the quantities  $X_{c0}$, $X_{c2}$, $X_{n0}$ and $X_{n2}$,
defined by Eq.~(\ref{eq:X}),
which enter the expression of the LL factor.
In the CM frame, the total energy is fixed to be the mass
of the kaon, i.e., $E=M_K\simeq 3.54M_\pi$.
These quantities depend on the momenta of outgoing pions that can be parameterized by
two invariant variables
$m_{12}^{2}=(p_{1}+p_{2})^{2}$ and
$m_{23}^{2}=(p_{2}+p_{3})^{2}$.
In Fig.~\ref{fig:XY-Dalitz}, we display the Dalitz plots for the quantities
$X_{c0}$, $X_{c2}$, $X_{n0}$, $X_{n2}$ that emerge in a result of such calculations.

\subsection{Finite-volume wave function}

\begin{figure}[t]
\begin{centering}
\subfigure[]{\begin{centering}
\includegraphics[scale=0.44]{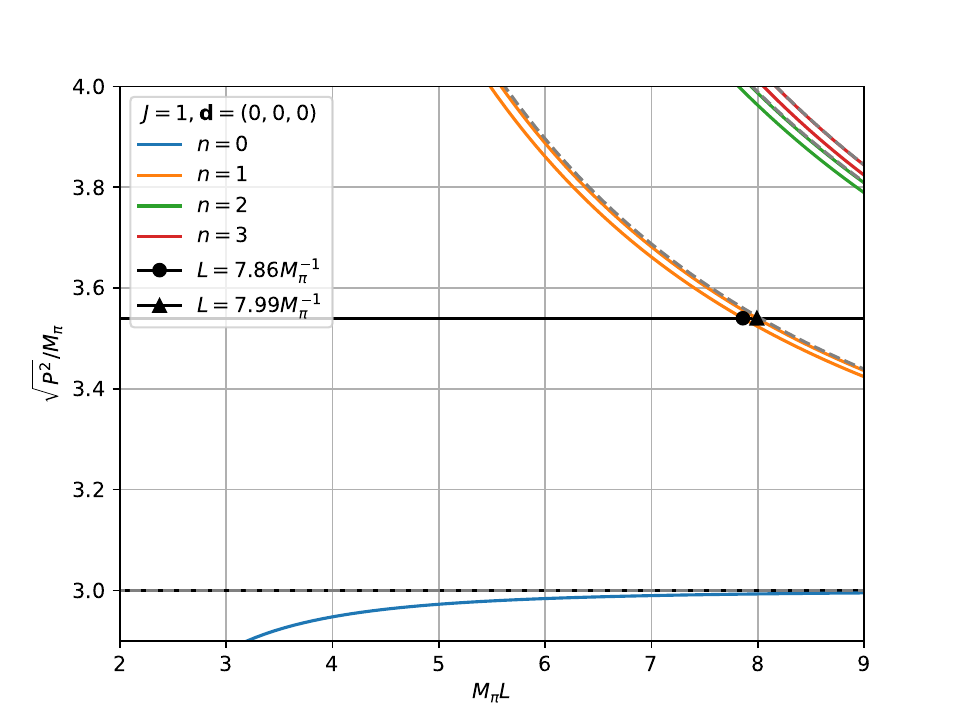}
\par\end{centering}
}\subfigure[]{\begin{centering}
\includegraphics[scale=0.44]{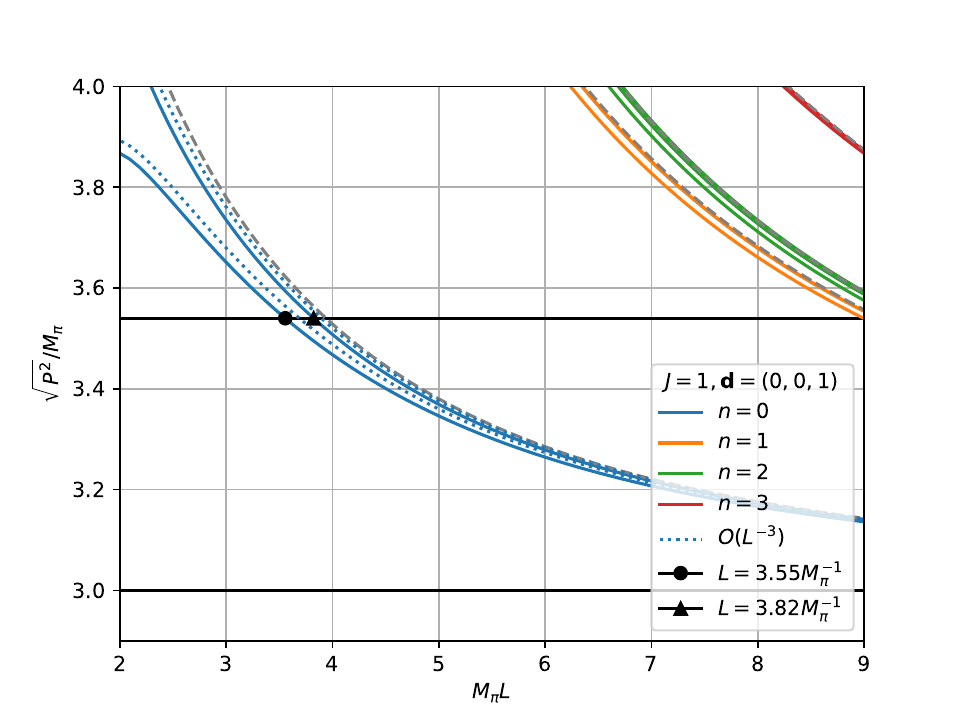}
\par\end{centering}
}
\par\end{centering}\vspace*{-.3cm}
\begin{centering}
\subfigure[]{\begin{centering}
\includegraphics[scale=0.44]{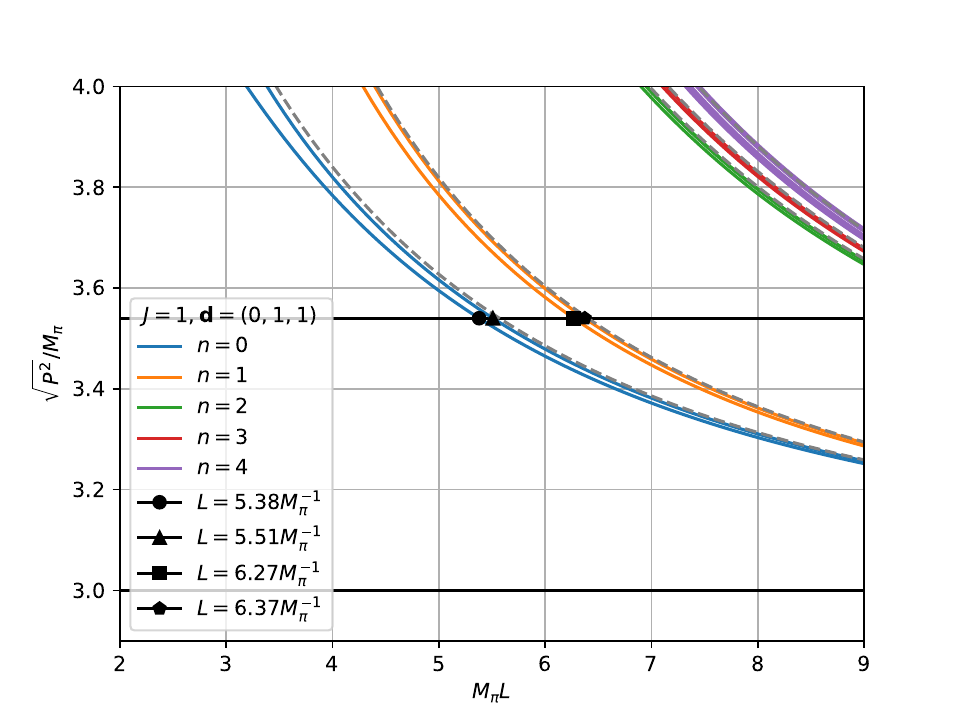}
\par\end{centering}
}\subfigure[]{\begin{centering}
\includegraphics[scale=0.44]{fig_spectrum_011_iso1}
\par\end{centering}
}
\par\end{centering}
\caption{\label{fig:Lattice-spectrum1}Finite-volume spectra of the $3\pi$
  system with the total isospin $J=3$. a) is obtained in the rest frame, irrep
  $\Gamma=A_1^-$. b), c), d) 
  show the spectra in the moving frame ${\bf d}=(0,0,1)$, ${\bf d}=(0,1,1)$ and
  ${\bf d}=(1,1,1)$, respectively, in the irrep $\Gamma=A_2$
  (the naming scheme of the irreps from Ref.~\cite{Gockeler:2012yj}
  is used here). To compute the LL factor,
  we determine values of the lattice size $L$ for which
  the invariant mass $\sqrt{P^{2}}=M_K$ (denoted by the solid black
  line around $\sqrt{P^{2}}/M_\pi\simeq 3.54$). In the subfigure b), we show
  perturbative energy shifts at $O(L^{-3})$ (see Eq.~\eqref{eq:energy_shift_001}).
  These are denoted by
blue dotted lines and give a clear understanding of the fine structure of the spectrum, namely, the splitting of the unperturbed level into two levels, when the interactions are switched on.
}
\end{figure}

\begin{figure}[t]
\begin{centering}
\subfigure[]{\begin{centering}
\includegraphics[scale=0.44]{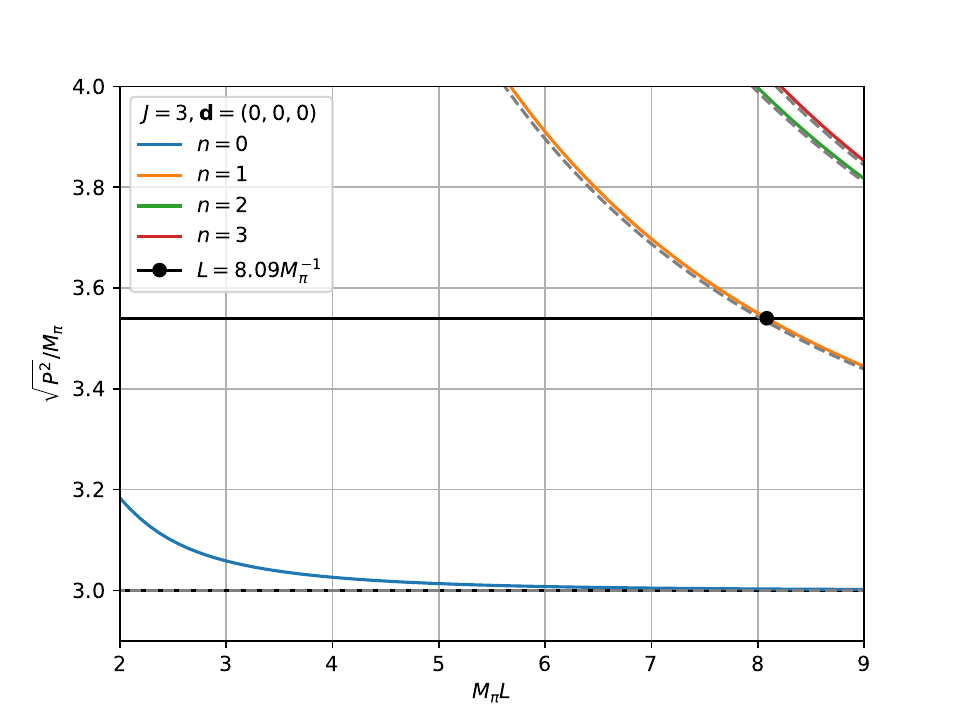}
\par\end{centering}
}\subfigure[]{\begin{centering}
\includegraphics[scale=0.44]{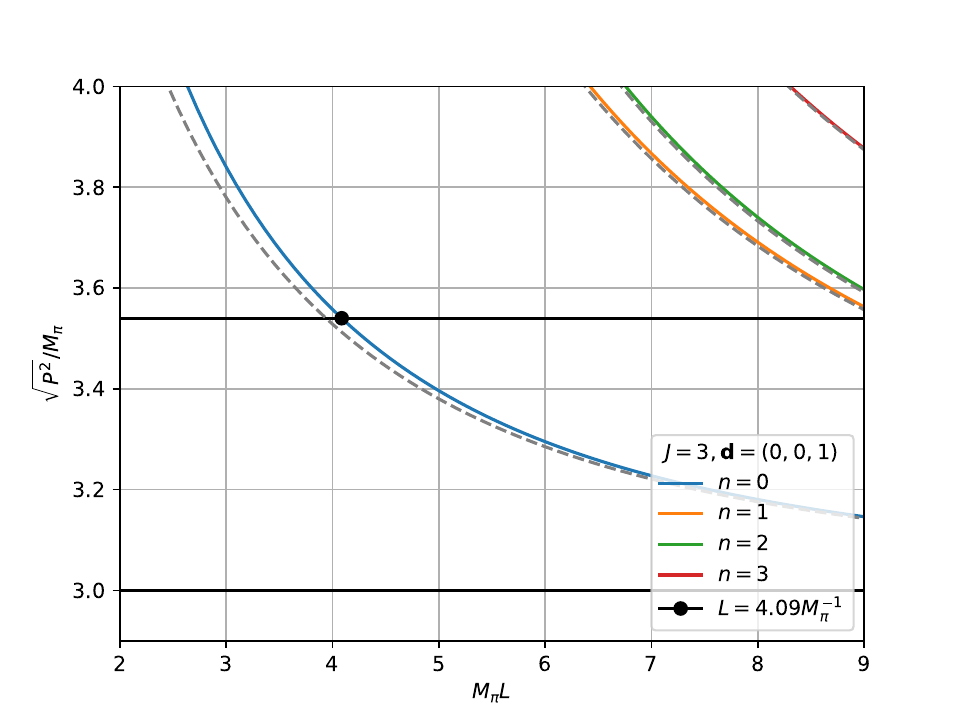}
\par\end{centering}
}
\par\end{centering}\vspace*{-.3cm}
\begin{centering}
\subfigure[]{\begin{centering}
\includegraphics[scale=0.44]{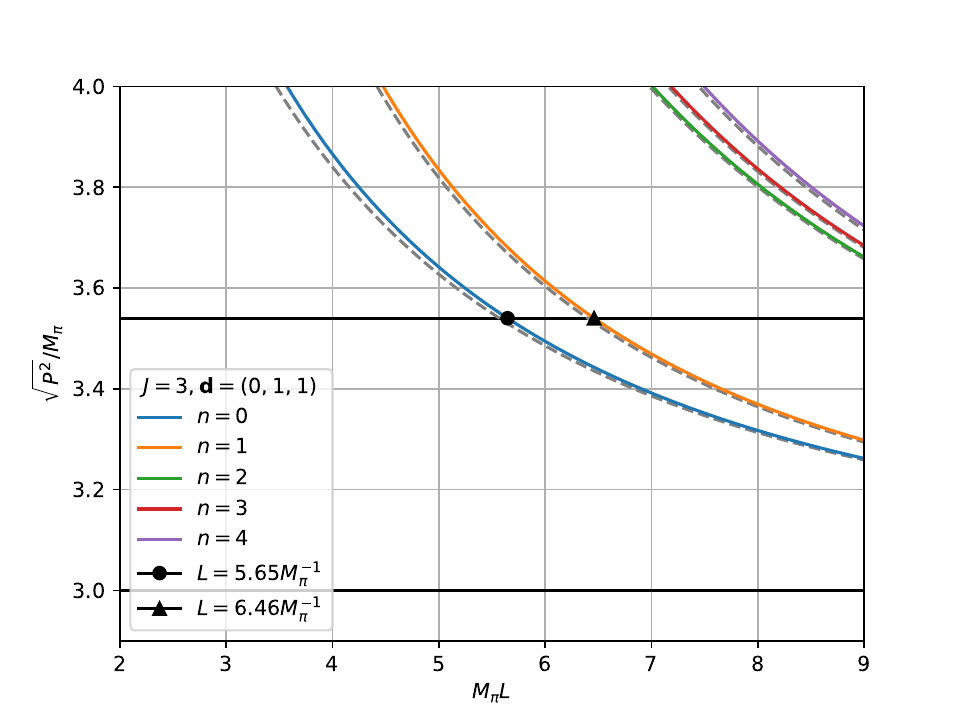}
\par\end{centering}
}\subfigure[]{\begin{centering}
\includegraphics[scale=0.44]{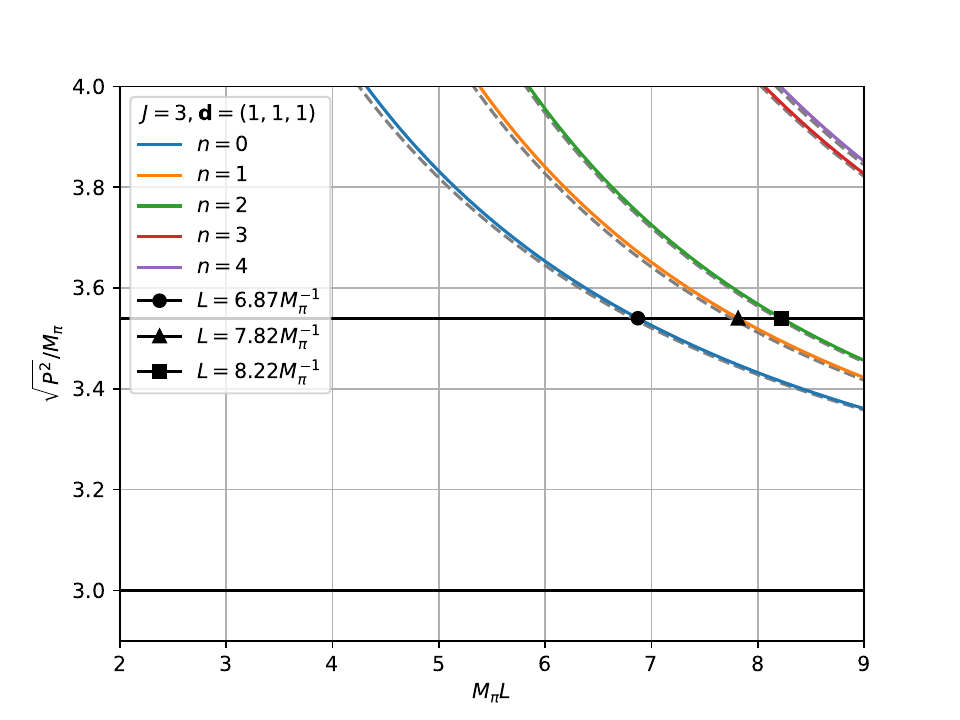}
\par\end{centering}
}
\par\end{centering}
\caption{\label{fig:Lattice-spectrum3} The same as in Fig.~\ref{fig:Lattice-spectrum1},
  but for the total isospin $J=3$.
  }
\end{figure}

In order to calculate $\Phi$ in a finite volume, we first need to
adjust the lattice size $L$ so that one has the energy level with the energy
exactly equal to the kaon mass. 
Solving the three-body quantization condition in the rest frame and in the
different moving frames, we obtain the energy spectrum,
see Fig.~\ref{fig:Lattice-spectrum1} and Fig.~\ref{fig:Lattice-spectrum3}. For demonstration,
we choose the moving frame
with the total momentum $(0,0,1)$ (in units of $2\pi/L$). 
In the channels with the total isospin
$J=1$ and $J=3$, respectively, the adjusted lattice size is given by  
\begin{align}
J=1: & \quad L=3.55M_\pi^{-1};\\
J=3: & \quad L=4.09M_\pi^{-1}.
\end{align}

\noindent
Once the lattice size is determined, we proceed to solve the Eq.~(\ref{eq:homogenous})
in order to obtain the (properly normalized) wave function in a finite volume. In its turn,
this wave function will be substituted into Eq.~(\ref{eq:PhinJI}) to find the amplitudes $\Phi^{(n)}_{J;I}$ that enter the expression of the LL factor, see Eq.~(\ref{eq:AJI}).

\subsection{The LL factor}

Putting pieces together, in this section we present the results of the calculation
of the LL factor, check its cutoff-dependence and the sensitivity to the input
two-body scattering lengths and the
three-particle amplitudes. To this end, we shall carry out calculations for three different
values of the cutoff. Furthermore, the LL factor
depends on the momenta of the final pions. We carry our calculations
at an arbitrary chosen point near the center of the Dalitz plot
$m_{12}^2=m_{23}^2=5M_\pi^2$.

\begin{figure}[t]
\begin{centering}
\subfigure[\label{fig:cut-off-indept}]{\begin{centering}
\includegraphics[scale=0.35]{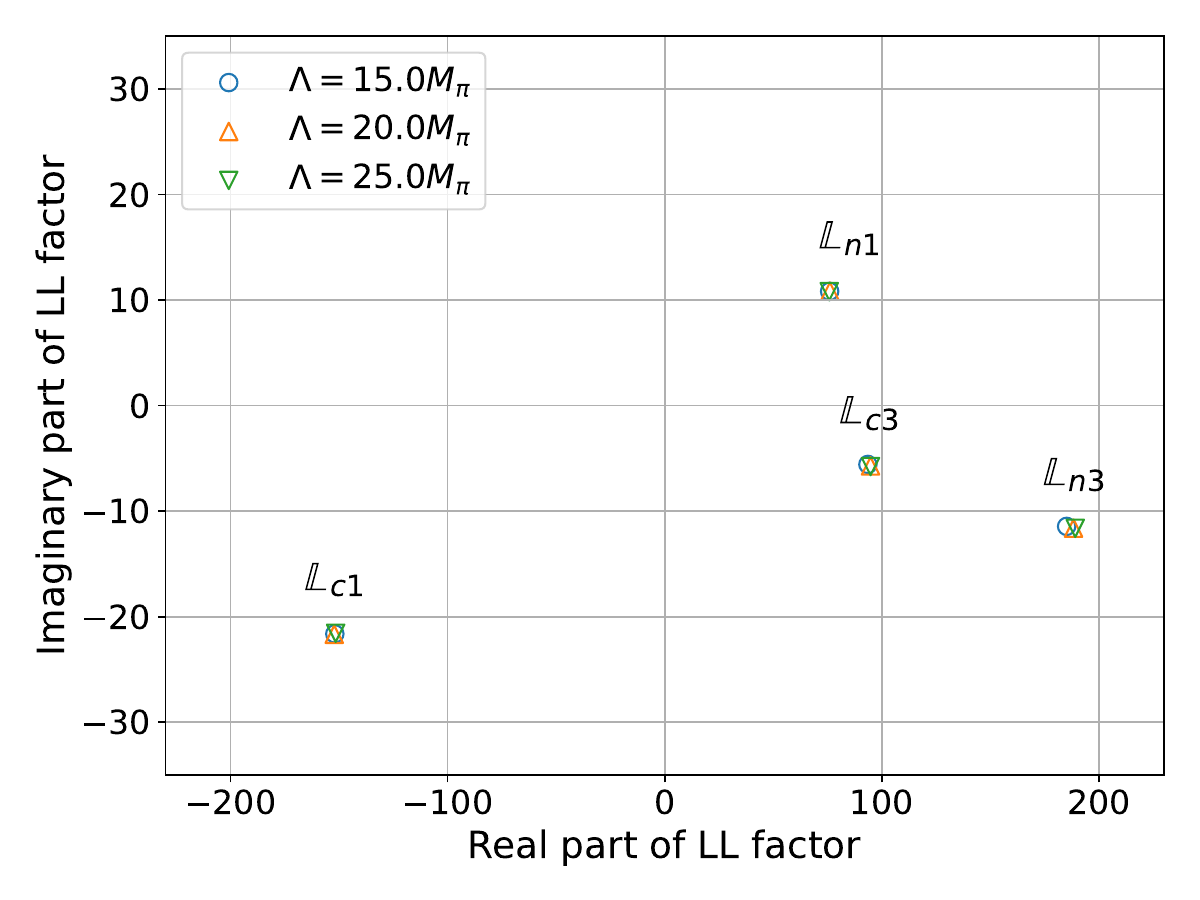}
\par\end{centering}
}\subfigure[\label{fig:XY-matrix}]{\begin{centering}
\includegraphics[scale=0.35]{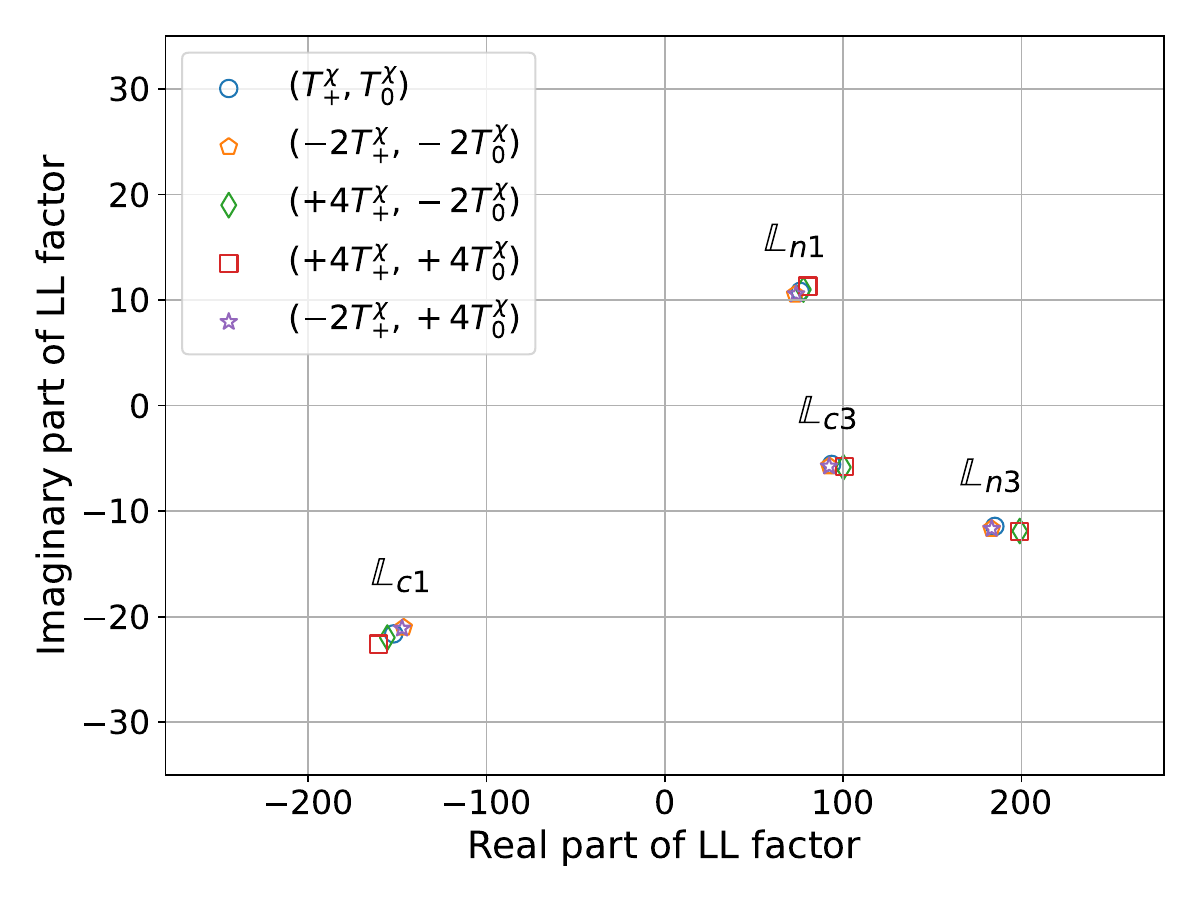}
\par\end{centering}
}
\par\end{centering}
\begin{centering}
\subfigure[\label{fig:A-matrix}]{\begin{centering}
\includegraphics[scale=0.35]{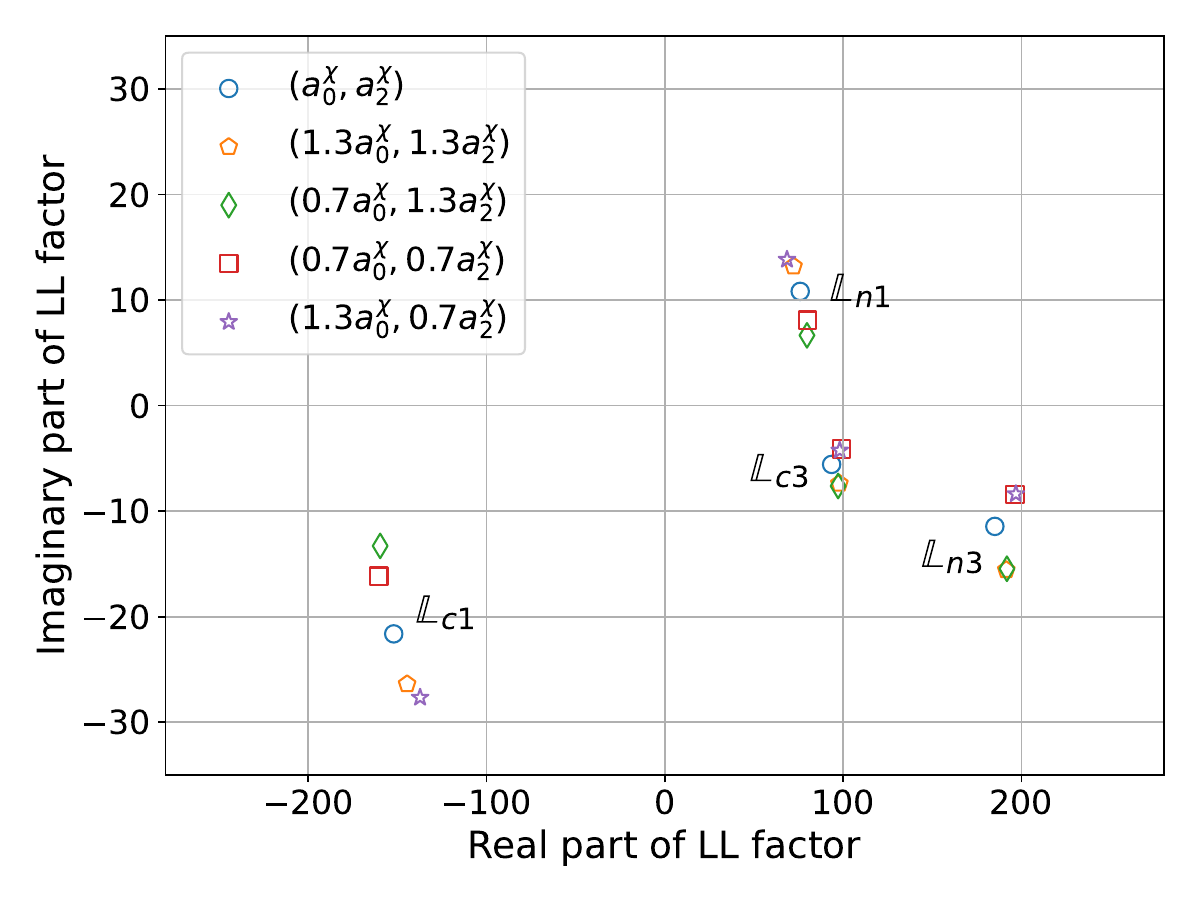}
\par\end{centering}
}
\par\end{centering}
\caption{\label{fig:result}a) The cutoff
independence of the LL factor. Here, the physical two-body scattering lengths
are used and the three-body threshold amplitudes are  fixed at $(T_+^\chi,T_0^\chi)$.
The quantities $\mathbb{L}_{c1}$, $\mathbb{L}_{c3}$, $\mathbb{L}_{n1}$
and $\mathbb{L}_{n3}$ are obtained for three different cutoff values:
$\Lambda=15M_\pi, 20M_\pi,25M_\pi$, represented by blue circles, brown
triangles, and green inverted triangles, respectively. In b) and c), the dependence
of the LL factor on the three-pion threshold amplitude and two-pion
scattering lengths are illustrated.}

\end{figure}

The Fig.~\ref{fig:result} summarizes our findings. First,  Fig.~\ref{fig:cut-off-indept}
shows the calculated LL factor for different values of the cutoff
$\Lambda=15M_\pi,20M_\pi,25M_\pi$. The difference is hardly seen by a bare eye, confirming our expectations. Next, in Fig.~\ref{fig:XY-matrix}, we show the results of calculations
for varying three-particle threshold amplitudes by 300\%, namely,
for pairs $(T_+^\chi,T_0^\chi)$, and $(T_+^\chi\pm 3\times T_+^\chi,T_0^\chi\pm 3\times T_0^\chi)$. The cutoff is fixed at $\Lambda=15M_\pi$. Again, the differences are small.
Finally, in Fig.~\ref{fig:A-matrix}, we show the result of variation of scattering lengths
by 30\% only, for the pairs $(a_0,a_2)$ and
$(a_0\pm 0.3\times a_0,a_2\pm 0.3\times a_2)$,
  with the three-body input $(T_+^\chi,T_0^\chi)$
and the cutoff $\Lambda=15M_\pi$ fixed. Now, the changes are sizable (despite the fact that the changes in the scattering lengths are factor 10 smaller than the changes in the three-body amplitudes), proving that the LL factor is much more
sensitive to the two-body input than the three-body threshold amplitudes. This result
constitutes the major finding of the present paper.

It is extremely important to understand the reason of such a behavior. Naively, according
to the NREFT counting introduced in Sect.~\ref{sec:TheLagrangian}, the exchange term
in the kernel $Z$ and the particle-dimer coupling count as $\mathcal{O}(\delta^{-2})$
and $\mathcal{O}(1)$, respectively. This counting, however, is known not to be valid non-perturbatively~\cite{Bedaque:1998kg,Bedaque:1998km}. Namely, the particle-dimer coupling should be promoted to the leading order to cope with the singular
dependence of the solutions on the cutoff. Hence, the power-counting arguments cannot directly explain a very little
sensitivity of the calculated LL factor on the three-body input. In order to study this
problem in more detail, we have carried out calculations for many different values of
the cutoff $\Lambda$ and noticed an interesting pattern: as soon as $\Lambda$ came close
to the critical values where the particle-dimer coupling becomes critical and flips the sign, the
dependence on the three-body input grows and becomes comparable with the
dependence on the values of the two-body scattering lengths, whereas away from the
critical cutoffs, the dependence on the three-body input was negligible. On the basis
of this observation one may argue that the formal promotion of the particle-dimer
coupling to the leading order is essential in the vicinity of critical cutoffs, whereas for other
values of the cutoff the arguments based on the naive counting still apply, for what concerns
the numerical estimate of the relative size of different contributions. This observation also shows the importance of a proper choice of cutoff in the calculations (away from singularities), albeit the results are formally cutoff-independent.

\subsection{The Weak Hamiltonian}

Up to now, in the derivation of the LL factor, we did not concentrate on the weak input
in the $K\to 3\pi$ decays. In other words, the couplings $G_1$ and $G_2$ are taken
to be completely arbitrary. In Nature, however, these couplings are subject to further
restrictions. Namely, at the lowest order in the Fermi-constant $G_F$, the weak
decays are described by the effective weak Hamiltonian that contains $\Delta I=1/2$ and
$\Delta I=3/2$ pieces, see e.g.,~\cite{Buchalla:1995vs}. Assuming conservation of isospin
in strong interaction, one immediately arrives at the conclusion that the total isospin
$J=3$ decay amplitude should vanish (in real world, it is strongly suppressed by one
power of $G_F$ or isospin-breaking parameters: the fine structure constant $\alpha$ or
$m_d-m_u$).

In order to see the consequences of this fact on the relative size of the couplings
$G_1$ and $G_2$, let us explicitly
write down the vectors $|J,J_3\rangle$ with $J=1,2,3$ and $J_3=1$ in the three-pion
space:
\begin{align}\label{eq:eigenstatesJ}
  |1,1\rangle^{(1)}&=\frac{1}{2}\,\left(|\pi^+\pi^0\pi^ 0\rangle-
                     |\pi^0\pi^+\pi^0 \rangle  -|\pi^+\pi^-\pi^+ \rangle+|\pi^-\pi^+\pi^+ \rangle  \right)\,,
\nonumber\\
  |1,1\rangle^{(2)}&= \frac{1}{\sqrt{3}}\, \left( |\pi^+\pi^-\pi^+ \rangle
                     -|\pi^0\pi^0\pi^+ \rangle+|\pi^-\pi^+\pi^+ \rangle\right)\, ,
\nonumber\\
  |1,1\rangle^{(3)}&=  \frac{2}{\sqrt{15}}\,\left(
6|\pi^+\pi^+\pi^- \rangle+|\pi^+\pi^-\pi^+ \rangle+|\pi^-\pi^+\pi^+ \rangle\right.
                     \nonumber\\
                   &\left.-3|\pi^0\pi^0\pi^+ \rangle-3   |\pi^+\pi^0\pi^0 \rangle+2           |\pi^0\pi^+\pi^0 \rangle \right)\, ,    
\nonumber\\
  |2,1\rangle^{(1)}&=  \frac{1}{2}\,\left(|\pi^+\pi^0\pi^ 0\rangle
                     -|\pi^0\pi^+\pi^0 \rangle  +|\pi^+\pi^-\pi^+ \rangle-|\pi^-\pi^+\pi^+ \rangle  \right)\,,
\nonumber\\
|2,1\rangle^{(2)}&=  \frac{1}{2\sqrt{3}}\,\left(
2|\pi^+\pi^+\pi^- \rangle-|\pi^+\pi^-\pi^+ \rangle-|\pi^-\pi^+\pi^+ \rangle\right.
\nonumber\\
                   &\left.-2|\pi^0\pi^0\pi^+ \rangle+ |\pi^+\pi^0\pi^0 \rangle+ |\pi^0\pi^+\pi^0 \rangle \right)\, ,    
                   \nonumber\\
  |3,1\rangle&=\frac{1}{\sqrt{15}}\,\left(
|\pi^+\pi^+\pi^- \rangle+|\pi^+\pi^-\pi^+ \rangle+|\pi^-\pi^+\pi^+ \rangle\right.
\nonumber\\
                   &\left.+2|\pi^0\pi^0\pi^+ \rangle+ 2|\pi^+\pi^0\pi^0 \rangle+ 2|\pi^0\pi^+\pi^0 \rangle \right)\, .    
\end{align}
At threshold, when momenta of all particles exactly vanish, the positions of the pions in the
vectors can be exchanged.\footnote{The threshold cannot be reached in physical decay
process. However, one arrives at the same conclusion at the center of the
Dalitz plot $m_{12}^2=m_{23}^2=m_{31}^2$.} This leads to the fact that the vectors which are antisymmetric
with respect to the exchange of any pair of pions, namely, $|1,1\rangle^{(1)}$,
$|2,1\rangle^{(1)}$ and $|2,1\rangle^{(2)}$, vanish at threshold. From this, one can immediately conclude that the $J=2$ amplitude does not contribute at threshold. Furthermore, it is
straightforward to check that, at threshold,
$|1,1\rangle^{(2)}=\frac{\sqrt{5}}{2}\,|1,1\rangle^{(3)}$. Reverting now
Eq.~(\ref{eq:eigenstatesJ}), expressing
the physical states $|\pi^0\pi^0\pi^+\rangle$ and $|\pi^+\pi^+\pi^-\rangle$ through
the eigenstates of the total isospin and dropping the contributions from $J=2,3$, we
finally arrive at a simple relation at threshold:\footnote{This relation directly follows from the last line in Eq.~(\ref{eq:eigenstatesJ}) and the symmetry of the states at threshold
with respect to the permutation of the particles 1,2,3.}
\begin{align}
  |\pi^+\pi^+\pi^-\rangle=-2|\pi^0\pi^0\pi^+\rangle\, .
  \end{align}
  Consequently, $G_2=-2G_1$ at leading order.

  It should be pointed out that our result is in agreement
  with the explicit calculation of this decay amplitude at next-to-leading order in
  ChPT~\cite{Bijnens:2002vr}. As seen from Eq.~(37) of that paper, our relation exactly
  holds (modulo the overall sign) for the amplitudes evaluated at the center of the
  Dalitz plot. Furthermore, the latest fit to the experimental data by NA48 collaboration
  yields the value $A_+=1.925\pm 0.015$ for this ratio (see
  Ref.~\cite{Batley:2009ubw}, Eqs.~(3,4)), which is quite close to our value
  $A_+=2$ (the overall sign is undefined in this analysis).
  The small deviation is due to the contribution
  of the higher-order terms in the NREFT power counting. Note finally that the difference
  in sign is due to the use of a particular convention for the basis states in the irreducible
  representations of the isospin. We consistently use Condon-Shortley phase convention
  in our calculations.\footnote{Here it should be pointed out that in Eq.~(2.7) of
  Ref.~\cite{Zemah}, which is consistent with~\cite{Bijnens:2002vr}, a convention
  different from the Condon-Shortley phase convention is used for the definition of
  the isospin eigenstates.}

  In conclusion, we wish to point out that the additional restriction  $G_2=-2G_1$
  does not affect our calculations of the LL factor since the latter, by definition, does not
  depend on the weak interactions that lead to the decay. The restriction simply means that
  only two linear combinations of $\mathbb{L}_{c1},\mathbb{L}_{c3},\mathbb{L}_{n1},\mathbb{L}_{n3}$ will be needed in the final
  result.


  \section{Conclusions}\label{sec:concl}

  Below, the results of our findings are briefly summarized:

\begin{itemize}  
  
\item[i)]
  We have performed an explicit calculation of the LL factor in the $K\to 3\pi$ decays at
  the leading order. Albeit the general framework has been already
  set up~\cite{Hansen:2021ofl,Muller:2022oyw}, an explicit numerical implementation
  of this framework is still a non-trivial exercise and represents a very useful
  endeavor on the way of the actual use of this framework in the analysis of the lattice
  QCD data. The message that we want to convey with this article, is clear:
  the framework 
  for the calculation of the $K\to 3\pi$ decay amplitudes on the lattice is now ready.
  \textcolor{black}{Moreover, higher-order terms can be systematically included
  in the expressions, when the accuracy of lattice data renders this inclusion necessary.}

\item[ii)] From the problems to be addressed we would like to single out the issue of the
  renormalization of the solutions of the Faddeev equations and the matching to the
  threshold three-pion amplitudes. At the first glance, the number of independent couplings
  that are needed to render all particle-dimer scattering amplitudes cutoff-independent,
  exceeds the number of three-particle threshold amplitudes at our disposal. However,
  as shown, the particle-dimer
  amplitudes that may (potentially) still have the singular cutoff
  dependence, do not contribute to the physical decay processes and are therefore
  harmless.

\item[iii)] The main finding of the present paper is the fact that the calculated LL factor
  has a very weak dependence on the input three-body threshold amplitudes, even if the
  latter change by a factor of 3 or so. This fact has a crucial importance for the future
  application of this framework in the studies of kaon decays on the lattice. Namely,
  one may, at the first stage, avoid extracting the three-pion coupling from lattice data
  and use instead the rough estimate, obtained from ChPT. From our findings
  we conclude that even an error of 100\% in the threshold amplitudes does not lead to
  a significant effect in the calculated LL factor, provided the cutoff $\Lambda$ is chosen away from the critical values.

\end{itemize}

\begin{acknowledgments}
  The authors would like to thank J. Bijnens, S. Dawid,
  H.-W. Hammer, B. Kubis and F. Romero-Lopez
  	for interesting discussions. 
	The work of R.B., F.M. and A.R. was  funded in part by
	the Deutsche Forschungsgemeinschaft
	(DFG, German Research Foundation) – Project-ID 196253076 – TRR 110 and by the Ministry of Culture and Science of North Rhine-Westphalia through the NRW-FAIR project.
	A.R., in addition, thanks Volkswagenstiftung 
	(grant no. 93562) and the Chinese Academy of Sciences (CAS) President's
	International Fellowship Initiative (PIFI) (grant no. 2024VMB0001) for the
	partial financial support.
	The work of J.-Y.P. and J.-J.W. was supported by the National Natural Science Foundation of China (NSFC) under Grants No. 12135011, 12175239, 12221005.  The work of J.-J.W. was also supported by the National Key R\&D Program of China under Contract No. 2020YFA0406400, and by Chinese Academy of Sciences under Grant No. YSBR-101.
\end{acknowledgments}
        
\appendix

\section{Integrating out the dimer fields}\label{app:integrating_out_dimers}
In order to integrate out the dimer fields, note that the dimer Lagrangian can be conveniently written as
\begin{align}
	\tilde{\mathcal{L}}_d &= \sum_{I, I_3} \sigma_{I}\, T_{II_3}^\dagger T_{I I_3} + \sum_{I, I_3} \left( T_{II_3}^\dagger O_{I I_3} + \text{h.c.} \right) \nonumber\\
	&+  \sum_{I ,I_3, i_3} \sum_{I', I'_3, i'_3}  T_{II_3}^\dagger \pi_{i_3}^\dagger \, h_{I I_3 i_3; I' I'_3 i'_3} \, \pi_{i'_3} T_{I' I'_3} \nonumber\\
	&+ \sum_{I, I_3, i_3} \left( g_{I I_3 i_3} K_+^\dagger T_{II_3} \pi_{i_3} + \text{h.c.} \right) \,,
\end{align}
where the coefficients $h_{I I_3 i_3; I' I'_3 i'_3}$ and $g_{I I_3 i_3}$ can be read off from Eq.~\eqref{eq:PD_Lagrangian_3body} and Eq.~\eqref{eq:tildeLK} respectively -- they are
linear combinations of the couplings $h_J^{(I,I')}$ and $g^{(J,I)}$. Defining
\begin{align}
	\tilde{T}_{I I_3} = \tilde{T}_{I I_3} + \sum_{I' I'_3} A^{-1}_{I I_3; I' I'_3} B_{I' I'_3} \,,
\end{align}
where
\begin{align}
	A_{I I_3; I' I'_3} = \sigma_I \delta_{I I'} \delta_{I_3 I'_3} + \sum_{i_3, i'_3} \pi_{i_3}^\dagger \, h_{I I_3 i_3; I' I'_3 i'_3} \, \pi_{i'_3}\, ,
\end{align}
and 
\begin{align}
	B_{I I_3} = O_{I I_3} + \sum_{i_3} g_{I I_3 i_3} \pi^\dagger_{i_3} K_+ \,,
\end{align}
the dimer Lagrangian can be brought into a quadratic form:
\begin{align}\label{eq:PD_Lagrangian_quadratic}
	\tilde{\mathcal{L}}_d =\sum_{I, I_3} \sum_{I', I'_3} \tilde{T}_{I I_3}^\dagger \, A_{I I_3; I' I'_3}\, \tilde{T}_{I' I'_3} - \sum_{I, I_3} \sum_{I', I'_3} B_{I I_3}^\dagger\,A^{-1}_{I I_3; I' I'_3} \, B_{I' I'_3} \,.
\end{align}
The dimer can then be integrated out in the path integral formalism in a standard way.
The second term on the right-hand side thus should be matched to the Lagrangian in
the particle picture at tree level. Noting that 
\begin{align}
	A^{-1}_{I I_3; I' I'_3} = \sigma^{-1}_I \delta_{I I'} \delta_{I_3 I'_3} - \sigma^{-1}_I \pi_{i_3}^\dagger \, h_{I I_3 i_3; I' I'_3 i'_3} \, \pi_{i'_3} \sigma^{-1}_{I'} + \dots \,,
\end{align}
where the ellipsis contains at least four pion fields, the second term on the right hand side of Eq.~\eqref{eq:PD_Lagrangian_quadratic} can be written as
\begin{align}
  -\sum_{I, I_3} \sum_{I', I'_3} B_{I I_3}^\dagger\,A^{-1}_{I I_3; I' I'_3} \, B_{I' I'_3}
  &= \sum_{I, I_3}  (-\sigma^{-1}_I O^\dagger_{I I_3}  O_{I I_3}) \nonumber\\
	&+  \sum_{I ,I_3, i_3} \sum_{I', I'_3, i'_3}  (-\sigma^{-1}_I O^\dagger_{I I_3}) \pi_{i_3}^\dagger \, h_{I I_3 i_3; I' I'_3 i'_3} \, \pi_{i'_3} (-\sigma^{-1}_{I'} O_{I' I'_3}) \nonumber\\
	& + \sum_{I, I_3, i_3} \left( g_{I I_3 i_3} K_+^\dagger (-\sigma^{-1}_I O_{I I_3}) \pi_{i_3} + \text{h.c.} \right) + \dots\,,
\end{align}
where the ellipsis contains at least eight pion- or two kaon fields. These terms do not contribute to the three-body sector. We can conclude that, integrating out the dimer fields, one merely has to replace $T_{I I_3} \to -\sigma^{-1}_I O_{I I_3}$.

Calculating the Green function
\begin{align}
	G_{i'_1 i'_2 i'_3}^{I I_3 i_3}(x_1, x_2, x_3) = \langle 0 | T \left[ O_{3\pi}^{i'_1 i'_2 i'_3}(x_1, x_2 x_3)  T^\dagger_{I I_3}(0) \pi_{i_3}^\dagger(0)\right] |0 \rangle
\end{align}
in the particle-dimer picture at tree level, one performs the functional integral
\begin{align}
  &\hspace*{1.cm}\int \mathcal{D}T \mathcal{D}T^\dagger \, T^\dagger_{I I_3}(0) \, \exp\left\{- i \int d^4x\,\tilde{ \mathcal{L}}_d \right\}
\nonumber\\
  &= -\sum_{I' I'_3} A^{-1}_{I I_3; I' I'_3} B_{I' I'_3}
    \exp\left\{ i \int d^4x\,\sum_{I', I'_3} \sum_{I'', I''_3} B_{I' I'_3}^\dagger\,A^{-1}_{I' I'_3; I'' I''_3} \, B_{I'' I''_3} \right\} \,.
\end{align}
The remaining functional integration is over the pion and kaon fields. At tree level, only $A^{-1}_{I I_3; I' I'_3} = \sigma^{-1}_I \delta_{I I'} \delta_{I_3 I'_3}$ contributes and the term containing the kaon can be dropped from $B_{I' I'_3}$ in the term
$A^{-1}_{II_3; I' I'_3} \, B_{I' I'_3}$ in front of the exponential. Thus, one may conclude that,
at tree level, the above Green function in the particle picture corresponds to  
\begin{align}
		G_{i'_1 i'_2 i'_3}^{I I_3 i_3}(x_1, x_2, x_3) = \langle 0 | T \left[ O_{3\pi}^{i'_1 i'_2 i'_3}(x_1, x_2 x_3)  (-\sigma_I^{-1} O^\dagger_{I I_3}(0))  \pi_{i_3}^\dagger(0)\right] |0 \rangle \,,
\end{align}
such that again the replacement $T_{I I_3} \to -\sigma_{I}^{-1} O_{I I_3}$ is justified.

Similarly, the vertex function
\begin{align}
V_{i'_1 i'_2 i'_3; j'_1 j'_2 j'_3}^{I I_3 i_3; J J_3 j_3}(x_1,x_2, x_3; y_1, y_2, y_3) &= \langle 0 | T \Big[ O^{i'_1 i'_2 i'_3}_{3\pi}(x_1, x_2, x_3) \left(O^{j'_1 j'_2 j'_3}_{3\pi}(y_1, y_2, y_3)\right)^\dagger \nonumber \\ 
&\times T^\dagger_{II_3}(0)\pi^\dagger_{i_3}(0) \pi_{j_3}(0) T_{J J_3}(0) \Big] | 0 \rangle \,,
\end{align}
can be evaluated in the particle picture by integrating out the dimer fields. At tree level,
this gives
\begin{align}
	V_{i'_1 i'_2 i'_3; j'_1 j'_2 j'_3}^{I I_3 i_3; J J_3 j_3}(x_1,x_2, x_3; y_1, y_2, y_3) &= \langle 0 | T \Big[ O^{i'_1 i'_2 i'_3}_{3\pi}(x_1, x_2, x_3) \left(O^{j'_1 j'_2 j'_3}_{3\pi}(y_1, y_2, y_3)\right)^\dagger \nonumber \\ 
	&\times  (-\sigma_I^{-1} O^\dagger_{I I_3}(0))\pi^\dagger_{i_3}(0) \pi_{j_3}(0)  (-\sigma_J^{-1} O^\dagger_{J J_3}(0)) \Big] | 0 \rangle \,.
\end{align}
Note that, in both cases, we implicitly discard disconnected pieces in the Green functions,
calculated in the particle picture.

\section{Three-pion amplitude in Chiral Perturbation Theory}
\label{sec:xpt}
\begin{figure}[t]
\begin{centering}
\subfigure[\label{fig:contact}Three-pion contact diagram]{\begin{centering}
\includegraphics[width=0.3\textwidth]{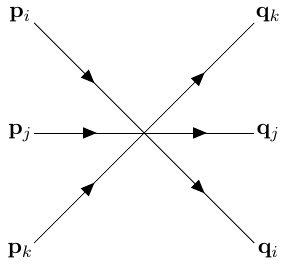}
\par\end{centering}
}$\qquad$\subfigure[\label{fig:ope}One-pion exchange diagram]{\begin{centering}
\includegraphics[width=0.3\textwidth]{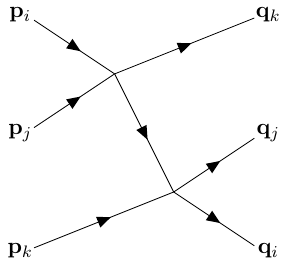}
\par\end{centering}
}
\par\end{centering}
\caption{\label{fig:xpt}Two kinds of diagrams that contribute to the three-pion scattering amplitude at leading order in ChPT. Diagram (b) is singular at threshold, whereas diagram (a) is finite. }
\end{figure}

For the matching of the particle-dimer coupling, we need to calculate the three-pion threshold amplitude in ChPT. Here, we give only a brief sketch
of this calculation, carried out at tree level, which follows the pattern outlined
in Ref.~\cite{Baeza-Ballesteros:2023ljl}. In general, there are only two types of contributions, shown in Fig.~\ref{fig:xpt}: The contact term, which emerges from the six-pion
Lagrangian, Fig.~\ref{fig:contact}, and the exchange term that features a single pion propagator (all possible
permutations of external lines in $in$- and $out$-states), Fig.~\ref{fig:ope}. We are interested in the
processes $3\pi^+\to 3\pi^+$ and $3\pi^0\to 3\pi^0$. The contact contributions
to these processes at threshold are given by
\begin{align}
  T^\chi_{\sf cont}(3\pi^+\to 3\pi^+) &
  =-\frac{18 M_{\pi}^{2}}{F^{4}_{\pi}}\, ,
\nonumber\\
T^\chi_{\sf cont}(3\pi^0\to 3\pi^0) &
=\frac{27 M_{\pi}^{2}}{F^{4}_{\pi}}\, .
\end{align}
Symbolically, the exchange contribution can be written down as follows
\begin{align}
  T^\chi_{\sf ex}(3\pi \to 3\pi) = \sum_{\sf{permutations}}
    \mathcal{M}(2\pi \to 2\pi)\frac{1}{M_\pi^2-k^2}
        \mathcal{M}(2\pi \to 2\pi)\,,\label{eq:exchange}
\end{align}
where $k$ denotes the four-momentum of the exchanged pion. Furthermore, the two-body
amplitudes at leading order are linear functions of the pertinent Mandelstam variables
$s,t,u$ and $M_\pi^2$:
\begin{align}\label{eq:symbolic}
\mathcal{M}(2\pi \to 2\pi)=\frac{\alpha s+\beta t+\gamma u +\delta M_\pi^2}{F_\pi^2 }\, .
\end{align}
Here, $\alpha,\beta,\gamma,\delta$ stand for some numerical coefficients. Furthermore,
  one can always choose these variables so that, say, $s$ and $t$ depend only on the
  external momenta (which are, by definition, on shell). Then, one could use the relation
  $s+t+u=3M_\pi^2+k^2$ and rewrite the expression for the two-body amplitude as
\begin{align}
  \mathcal{M}(2\pi \to 2\pi)=\frac{\alpha s+\beta t+\gamma (4M_\pi^2-s-t)
      +\gamma(k^2-M_\pi^2)+\delta M_\pi^2}{F_\pi^2 }\, .
  \end{align}
  The part of the amplitude that is proportional to $k^2-M_\pi^2$ will cancel with the propagator and will contribute to the regular part of the threshold amplitude. Separating the pole
  terms from the non-pole ones in all diagrams by using Eq.~(\ref{eq:symbolic}),
  in a result one gets for the regular part:
\begin{align}
  T^\chi_{\sf ex,reg}(3\pi^+\to3\pi^+) &
  =\frac{36 M_{\pi}^{2}}{F^{4}_{\pi}}\, ,
\nonumber\\
T^\chi_{\sf ex,reg}(3\pi^0\to 3\pi^0) &
=-\frac{225 M_{\pi}^{2}}{8F^{4}_{\pi}}\, .
\end{align}
Adding up these two contributions, we finally get
\begin{align}
  T^\chi_{\sf reg}(3\pi^+\to 3\pi^+) &\doteq T_+^\chi
  =\frac{18 M_{\pi}^{2}}{F^{4}_{\pi}}\, ,
\nonumber\\
  T^\chi_{\sf reg}(3\pi^0\to 3\pi^0) &\doteq T_0^\chi
=-\frac{9M_{\pi}^{2}}{8F^{4}_{\pi}}\, .
\end{align}

\section{Leading-order energy shift of the $n=1$ state}

In this appendix, we shall evaluate the energy shift of the states with the lowest energy
in different moving frames as well as in the rest frame. In particular,
our aim will be to demonstrate that the level splitting, which we
observe in a result of the solution of the quantization condition, can be interpreted
with the use of the perturbation theory as well.

We start from the
states with total isospin $J = 1$ and
choose $J_3 = 1$. One can define three orthogonal states with these quantum
numbers~\cite{Muller:2020vtt}:
\begin{align}
	|f_1\rangle &= \frac{1}{\sqrt{15}}\,(2|\pi^+\pi^+\pi^-\rangle + 2|\pi^+\pi^-\pi^+\rangle + 2|\pi^-\pi^+\pi^+\rangle \nonumber\\
	&\phantom{0000000}-|\pi^0\pi^0\pi^+\rangle - |\pi^0\pi^+\pi^0\rangle - |\pi^+\pi^0\pi^0\rangle) \,, \nonumber\\
	|f_2\rangle &= \frac{1}{2\sqrt{3}}\,(2|\pi^+\pi^+\pi^-\rangle - |\pi^+\pi^-\pi^+\rangle - |\pi^-\pi^+\pi^+\rangle \nonumber\\
	&\phantom{0000000}+2|\pi^0\pi^0\pi^+\rangle - |\pi^0\pi^+\pi^0\rangle - |\pi^+\pi^0\pi^0\rangle)\,, \nonumber\\
	|f_3\rangle &= \frac{1}{2}\,(|\pi^+\pi^0\pi^0\rangle - |\pi^0\pi^+\pi^0\rangle - |\pi^+\pi^-\pi^+\rangle  + |\pi^-\pi^+\pi^+\rangle) \,.
\end{align}
We consider moving frames with the total three-momentum
$\mathbf{P} = 2\pi\mathbf{d} / L$, where $\mathbf{d}=(0,0,1),(1,1,0),(1,1,1) $.
For these values of $\mathbf{d}$, the states with the lowest energy are given by
\begin{align}
	|\mathbf{p}_1, \mathbf{p}_2, \mathbf{p}_3 \rangle \in \left\{ |\mathbf{P},\mathbf{0},\mathbf{0}\rangle\,,~ |\mathbf{0},\mathbf{P},\mathbf{0}\rangle\,,~ |\mathbf{0},\mathbf{0},\mathbf{P}\rangle \right\} \,.
\end{align}
From these we construct the total wave functions of the three-pion system denoted by
\begin{align}
	|f_i; \mathbf{p}_1, \mathbf{p}_2, \mathbf{p}_3 \rangle \,,
\end{align}
where the momenta $\mathbf{p}_1$, $\mathbf{p}_2$ and $\mathbf{p}_3$ are assigned to the first, second and third pion in the isospin wave function $|f_i\rangle$ respectively. Naively, in the non-interacting case there would be nine degenerate states with energy
\begin{align}
	E_0 = 2 m + \sqrt{m^2 + \left(\frac{2\pi\mathbf{d}}{L}\right)^2} \,.
\end{align}
On the other hand, not all states are independent. Using Bose-symmetry we find that:
\begin{align}
	& |f_1; \mathbf{0}, \mathbf{P}, \mathbf{0} \rangle = |f_1; \mathbf{P}, \mathbf{0}, \mathbf{0} \rangle\,, \quad  \phantom{-}|f_1; \mathbf{0}, \mathbf{0}, \mathbf{P} \rangle = |f_1; \mathbf{P}, \mathbf{0}, \mathbf{0} \rangle\,, \nonumber \\
	&|f_2; \mathbf{0}, \mathbf{P}, \mathbf{0} \rangle = |f_2; \mathbf{P}, \mathbf{0}, \mathbf{0} \rangle \,, \quad\phantom{-} |f_2; \mathbf{0}, \mathbf{0}, \mathbf{P} \rangle = -2\,|f_2; \mathbf{P}, \mathbf{0}, \mathbf{0} \rangle \,, \nonumber \\
	&|f_3; \mathbf{0}, \mathbf{P}, \mathbf{0} \rangle = -|f_3; \mathbf{P}, \mathbf{0}, \mathbf{0} \rangle \,, \quad |f_3; \mathbf{0}, \mathbf{0}, \mathbf{P} \rangle = 0\,, \nonumber \\
	&|f_3; \mathbf{P}, \mathbf{0}, \mathbf{0} \rangle = -\sqrt{3}\, |f_2; \mathbf{P}, \mathbf{0}, \mathbf{0} \rangle\,.
\end{align}
Defining the states
\begin{align}
	&|X; \mathbf{P}, \mathbf{0}, \mathbf{0} \rangle = -\frac{1}{2}\, |f_2; \mathbf{P}, \mathbf{0}, \mathbf{0} \rangle + \frac{\sqrt{3}}{2}\, |f_3; \mathbf{P}, \mathbf{0}, \mathbf{0} \rangle \,, \nonumber\\
	&|Y; \mathbf{P}, \mathbf{0}, \mathbf{0} \rangle = \frac{\sqrt{3}}{2}\, |f_2; \mathbf{P}, \mathbf{0}, \mathbf{0} \rangle + \frac{1}{2}\, |f_3; \mathbf{P}, \mathbf{0}, \mathbf{0} \rangle = 0\,,
\end{align}
it can be seen that the lowest energy level in the frame $\mathbf{d} \neq \mathbf{0}$ with $J=1$ is twofold degenerate, corresponding to the states $|f_1;  \mathbf{P}, \mathbf{0}, \mathbf{0} \rangle$ and $|X; \mathbf{P}, \mathbf{0}, \mathbf{0} \rangle$.

The energy shift is calculated in the degenerate perturbation theory, where the potentials are obtained from the non-relativistic effective field theory. Contrary to \cite{Muller:2020vtt}, here we will use the covariant formulation. In this setup, the normalized one-particle states are given by:
\begin{align}\label{eq:normalized_covariant_states}
	|\pi_i(\mathbf{p}) \rangle =L^{-3/2}\,(2w(\mathbf{p}))^{-1/2}\, a_i^\dagger(\mathbf{p}) \, |0\rangle \,.
\end{align}
and the creation and annihilation operators obey the commutation relation
\begin{align}
	[a_i(\mathbf{p}) , a_j^\dagger(\mathbf{q})] = \delta_{ij}\,2w(\mathbf{p})L^3\,\delta_{\mathbf{p},\mathbf{q}} \,.
\end{align}
At the leading order, the energy shift $\Delta E^{J=1}_{\mathbf{d} \neq \mathbf{0}}$ is given by the eigenvalues of the potential
\begin{align}
	V = \frac{1}{3!} \begin{pmatrix}
	\langle f_1; \mathbf{P}, \mathbf{0}, \mathbf{0} | H_I | f_1; \mathbf{P}, \mathbf{0}, \mathbf{0} \rangle & \langle f_1; \mathbf{P}, \mathbf{0}, \mathbf{0} | H_I | X; \mathbf{P}, \mathbf{0}, \mathbf{0} \rangle \\
	\langle X; \mathbf{P}, \mathbf{0}, \mathbf{0} | H_I | f_1; \mathbf{P}, \mathbf{0}, \mathbf{0} \rangle & \langle X; \mathbf{P}, \mathbf{0}, \mathbf{0} | H_I | X; \mathbf{P}, \mathbf{0}, \mathbf{0} \rangle
	\end{pmatrix} \,,
\end{align}
where $H_I$ denotes the interaction Hamiltonian
\begin{align}
	H_I = -\int d^3\mathbf{x}\, \mathcal{L}_2 \,.
\end{align}
Here $\mathcal{L}_2$ is the two-body Lagrangian in the particle picture, as in Eq.~\eqref{eq:2body_Lagrangian_particle_picture}. The couplings $C_i$ are matched to the $I = 0,2$ scattering lengths $a_I$, according to
Eq.~\eqref{eq:matching_fI_aI} and
Eq.~\eqref{eq:matching_Ci_fI}.

Up to $O(L^{-3})$, the potential is given by:
\begin{align}
	V = \frac{4\pi}{27 m L^3}\begin{pmatrix}
		5(5 a_0 + 4 a_2) & 2\sqrt{5}(a_0-a_2) \\
		2\sqrt{5}(a_0-a_2) & 2(4a_0 + 5a_2) 
	\end{pmatrix}\,,
\end{align}
leading to an energy shift
\begin{align}\label{eq:energy_shift_001}
	\Delta E^{J=1}_{\mathbf{d} \neq \mathbf{0}} = \frac{2\pi}{9m L^3} \left( 11 a_0 + 10 a_2 \pm \sqrt{41 a_0^2 + 20 a_0 a_2 + 20 a_2^2} \right)\,,
\end{align}
and lifting the degeneracy of the non-interacting energy level.

Note that the same result is obtained in the non-covariant approach. The only difference in the covariant and non-covariant potentials, besides the use of different couplings $C_i$, is the normalization of states: instead of the factor
\begin{align}
	\delta_{\mathbf{p}_i, \mathbf{q}_l} \, \delta_{\mathbf{p}_j + \mathbf{p}_k, \mathbf{q}_m + \mathbf{q}_n}
\end{align}
which appears in Eqs.~(31) and (32) of Ref.~\cite{Muller:2020vtt} for the potentials,
sandwiched between the three-pion states that carry the momenta $\mathbf{p}_1, \mathbf{p}_2, \mathbf{p}_3$ and $\mathbf{q}_1, \mathbf{q}_2, \mathbf{q}_3$ respectively,
in the covariant approach the following factor emerges
\begin{align}
	\left( 2w(\mathbf{p}_j) 2w(\mathbf{p}_k) 2w(\mathbf{q}_m) 2w(\mathbf{q}_n) \right)^{-1/2}& \delta_{\mathbf{p}_i, \mathbf{q}_l} \, \delta_{\mathbf{p}_j + \mathbf{p}_k, \mathbf{q}_m + \mathbf{q}_n} \nonumber\\ 
	&= \frac{1}{4 m^2} \delta_{\mathbf{p}_i, \mathbf{q}_l} \, \delta_{\mathbf{p}_j + \mathbf{p}_k, \mathbf{q}_m + \mathbf{q}_n} +\cdots \,.
\end{align}
Here the multiplicative factors $2w(\mathbf{k}_i)$ account for the different
normalization of the one-particle states in the relativistic and non-relativistic framework,
see Eq.~\eqref{eq:normalized_covariant_states}. For the same reason, these factors
also arise in the matching condition in the non-covariant framework, while in the
covariant matching condition they are absent. Therefore the results after expressing
everything in terms of the scattering lengths are identical.

Next, note that, for $J=3$, there is only a single state with the lowest energy.
One can, for example, choose $J_3 = 3$ and consider the state
$|\pi^+ \pi^+ \pi^+; \mathbf{P}, \mathbf{0}, \mathbf{0}\rangle$.
The energy shift at the leading order can be calculated in the
non-degenerate perturbation theory and is given by
\begin{align}
	\Delta E^{J=3}_{\mathbf{d} \neq \mathbf{0}} = \frac{1}{3!} \langle \pi_+ \pi_+ \pi_+; \mathbf{P}, \mathbf{0}, \mathbf{0}| H_I|\pi_+ \pi_+ \pi_+; \mathbf{P}, \mathbf{0}, \mathbf{0}\rangle = \frac{20\pi}{3 m L^3}\, a_2 \,.
\end{align}
Finally, note that for the ground state $\mathbf{d} = \mathbf{0}$, the energy shifts
corresponding to the total isospin $J=1$ and $J=3$  can be obtained in the
non-degenerate perturbation theory again. For $J = 1$ and $J_3 = 1$,
only the state $| f_1; \mathbf{0}, \mathbf{0}, \mathbf{0}\rangle$ yields the
non-vanishing ground-state wave function due to Bose-symmetry.
The energy shift is then given by:
\begin{align}
		\Delta E^{J=1}_{\mathbf{d} = \mathbf{0}} = \frac{4 \pi}{3 m L^3}\left(5 a_0 +4 a_2\right) \,.
\end{align}
For $J=3$ and $J_3 = 3$, the ground-state wave function is given by $| \pi_+\pi_+\pi_+; \mathbf{0}, \mathbf{0}, \mathbf{0}\rangle$. At the leading order, the energy shift of this state is equal to 
\begin{align}
	\Delta E^{J=3}_{\mathbf{d} = \mathbf{0}} = \frac{12 \pi}{m L^3}\, a_2\, .
\end{align}

\bibliographystyle{unsrt}
\bibliography{ref}

\begin{thebibliography}{10}

\bibitem{Hansen:2014eka}
Maxwell~T. Hansen and Stephen~R. Sharpe.
\newblock {Relativistic, model-independent, three-particle quantization
  condition}.
\newblock {\em Phys. Rev.}, D90(11):116003, 2014.

\bibitem{Hansen:2015zga}
Maxwell~T. Hansen and Stephen~R. Sharpe.
\newblock {Expressing the three-particle finite-volume spectrum in terms of the
  three-to-three scattering amplitude}.
\newblock {\em Phys. Rev.}, D92(11):114509, 2015.

\bibitem{Hammer:2017uqm}
Hans-Werner Hammer, Jin-Yi Pang, and A.~Rusetsky.
\newblock {Three-particle quantization condition in a finite volume: 1. The
  role of the three-particle force}.
\newblock {\em JHEP}, 09:109, 2017.

\bibitem{Hammer:2017kms}
H.~W. Hammer, J.~Y. Pang, and A.~Rusetsky.
\newblock {Three particle quantization condition in a finite volume: 2. General
  formalism and the analysis of data}.
\newblock {\em JHEP}, 10:115, 2017.

\bibitem{Mai:2017bge}
M.~Mai and M.~{D\"{o}ring}.
\newblock {Three-body Unitarity in the Finite Volume}.
\newblock {\em Eur. Phys. J.}, A53(12):240, 2017.

\bibitem{Mai:2018djl}
Maxim Mai and Michael D{\"{o}}ring.
\newblock {Finite-Volume Spectrum of $\pi^+\pi^+$ and $\pi^+\pi^+\pi^+$
  Systems}.
\newblock {\em Phys. Rev. Lett.}, 122(6):062503, 2019.

\bibitem{Kreuzer:2008bi}
Simon Kreuzer and H.~W. Hammer.
\newblock {Efimov physics in a finite volume}.
\newblock {\em Phys. Lett. B}, 673:260--263, 2009.

\bibitem{Kreuzer:2009jp}
Simon Kreuzer and H.~W. Hammer.
\newblock {On the modification of the Efimov spectrum in a finite cubic box}.
\newblock {\em Eur. Phys. J. A}, 43:229--240, 2010.

\bibitem{Kreuzer:2010ti}
Simon Kreuzer and H.~W. Hammer.
\newblock {The triton in a finite volume}.
\newblock {\em Phys. Lett. B}, 694:424--429, 2011.

\bibitem{Kreuzer:2012sr}
Simon Kreuzer and Harald~W. Grie\ss{}hammer.
\newblock {Three particles in a finite volume: The breakdown of spherical
  symmetry}.
\newblock {\em Eur. Phys. J. A}, 48:93, 2012.

\bibitem{Briceno:2012rv}
Ra\'ul~A. Brice\~no and Zohreh Davoudi.
\newblock {Three-particle scattering amplitudes from a finite volume
  formalism}.
\newblock {\em Phys. Rev.}, D87(9):094507, 2013.

\bibitem{Polejaeva:2012ut}
K.~Polejaeva and A.~Rusetsky.
\newblock {Three particles in a finite volume}.
\newblock {\em Eur.\ Phys.\ J.\ A}, 48:67, 2012.

\bibitem{Jansen:2015lha}
M.~Jansen, H.~W. Hammer, and Yu~Jia.
\newblock {Finite volume corrections to the binding energy of the X(3872)}.
\newblock {\em Phys. Rev. D}, 92(11):114031, 2015.

\bibitem{Hansen:2015zta}
Maxwell~T. Hansen and Stephen~R. Sharpe.
\newblock {Perturbative results for two and three particle threshold energies
  in finite volume}.
\newblock {\em Phys. Rev.}, D93:014506, 2016.

\bibitem{Hansen:2016fzj}
Maxwell~T. Hansen and Stephen~R. Sharpe.
\newblock {Threshold expansion of the three-particle quantization condition}.
\newblock {\em Phys. Rev.}, D93(9):096006, 2016.
\newblock [Erratum: Phys. Rev. \textbf{D96}, 039901 (2017)].

\bibitem{Guo:2016fgl}
Peng Guo.
\newblock {One spatial dimensional finite volume three-body interaction for a
  short-range potential}.
\newblock {\em Phys. Rev.}, D95(5):054508, 2017.

\bibitem{Sharpe:2017jej}
Stephen~R. Sharpe.
\newblock {Testing the threshold expansion for three-particle energies at
  fourth order in $\phi^4$ theory}.
\newblock {\em Phys. Rev.}, D96(5):054515, 2017.

\bibitem{Guo:2017crd}
Peng Guo and Vladimir Gasparian.
\newblock {Numerical approach for finite volume three-body interaction}.
\newblock {\em Phys. Rev. D}, 97(1):014504, 2018.

\bibitem{Guo:2017ism}
Peng Guo and Vladimir Gasparian.
\newblock {A solvable three-body model in finite volume}.
\newblock {\em Phys. Lett.}, B774:441--445, 2017.

\bibitem{Meng:2017jgx}
Yu~Meng, Chuan Liu, Ulf-G Mei\ss{}ner, and A.~Rusetsky.
\newblock {Three-particle bound states in a finite volume: unequal masses and
  higher partial waves}.
\newblock {\em Phys. Rev. D}, 98(1):014508, 2018.

\bibitem{Briceno:2017tce}
R.~A. Brice\~no, Maxwell~T. Hansen, and Stephen~R. Sharpe.
\newblock {Relating the finite-volume spectrum and the two-and-three-particle
  $S$ matrix for relativistic systems of identical scalar particles}.
\newblock {\em Phys. Rev.}, D95(7):074510, 2017.

\bibitem{Guo:2018ibd}
Peng Guo, Michael D\"{o}ring, and Adam~P. Szczepaniak.
\newblock {Variational approach to $N$-body interactions in finite volume}.
\newblock {\em Phys. Rev.}, D98(9):094502, 2018.

\bibitem{Guo:2018xbv}
Peng Guo and Tyler Morris.
\newblock {Multiple-particle interaction in (1+1)-dimensional lattice model}.
\newblock {\em Phys. Rev. D}, 99(1):014501, 2019.

\bibitem{Klos:2018sen}
P.~Klos, S.~König, H.~W. Hammer, J.~E. Lynn, and A.~Schwenk.
\newblock {Signatures of few-body resonances in finite volume}.
\newblock {\em Phys. Rev.}, C98(3):034004, 2018.

\bibitem{Briceno:2018mlh}
Ra\'ul~A. Brice\~no, Maxwell~T. Hansen, and Stephen~R. Sharpe.
\newblock {Numerical study of the relativistic three-body quantization
  condition in the isotropic approximation}.
\newblock {\em Phys. Rev.}, D98(1):014506, 2018.

\bibitem{Briceno:2018aml}
Ra\'ul~A. Brice\~no, Maxwell~T. Hansen, and Stephen~R. Sharpe.
\newblock {Three-particle systems with resonant subprocesses in a finite
  volume}.
\newblock {\em Phys. Rev.}, D99(1):014516, 2019.

\bibitem{Mai:2019fba}
M.~Mai, M.~D\"{o}ring, C.~Culver, and A.~Alexandru.
\newblock {Three-body unitarity versus finite-volume $\pi^+\pi^+\pi^+$ spectrum
  from lattice QCD}.
\newblock {\em Phys.\ Rev.\ D}, 101:054510, 2020.

\bibitem{Guo:2019ogp}
Peng Guo and Michael D\"oring.
\newblock {Lattice model of heavy-light three-body system}.
\newblock {\em Phys. Rev. D}, 101(3):034501, 2020.

\bibitem{Guo:2020spn}
Peng Guo.
\newblock {Modeling few-body resonances in finite volume}.
\newblock {\em Phys. Rev. D}, 102(5):054514, 2020.

\bibitem{Blanton:2019igq}
Tyler~D. Blanton, Fernando Romero-L\'opez, and Stephen~R. Sharpe.
\newblock {Implementing the three-particle quantization condition including
  higher partial waves}.
\newblock {\em JHEP}, 03:106, 2019.

\bibitem{Pang:2019dfe}
Jin-Yi Pang, Jia-Jun Wu, H.~W. Hammer, Ulf-G. Mei{\ss}ner, and Akaki Rusetsky.
\newblock {Energy shift of the three-particle system in a finite volume}.
\newblock {\em Phys. Rev.}, D99(7):074513, 2019.

\bibitem{Jackura:2019bmu}
A.~W. Jackura, S.~M. Dawid, C.~Fern\'andez-Ram\'\i{}rez, V.~Mathieu,
  M.~Mikhasenko, A.~Pilloni, S.~R. Sharpe, and A.~P. Szczepaniak.
\newblock {Equivalence of three-particle scattering formalisms}.
\newblock {\em Phys. Rev. D}, 100(3):034508, 2019.

\bibitem{Briceno:2019muc}
Ra\'ul~A. Brice\~no, Maxwell~T. Hansen, Stephen~R. Sharpe, and Adam~P.
  Szczepaniak.
\newblock {Unitarity of the infinite-volume three-particle scattering amplitude
  arising from a finite-volume formalism}.
\newblock {\em Phys. Rev.}, D100(5):054508, 2019.

\bibitem{Romero-Lopez:2019qrt}
Fernando Romero-L\'opez, Stephen~R. Sharpe, Tyler~D. Blanton, Ra\'ul~A.
  Brice\~no, and Maxwell~T. Hansen.
\newblock {Numerical exploration of three relativistic particles in a finite
  volume including two-particle resonances and bound states}.
\newblock {\em JHEP}, 10:007, 2019.

\bibitem{Konig:2020lzo}
Sebastian K\"onig.
\newblock {Few-body bound states and resonances in finite volume}.
\newblock {\em Few Body Syst.}, 61(3):20, 2020.

\bibitem{Brett:2021wyd}
Ruair\'\i{} Brett, Chris Culver, Maxim Mai, Andrei Alexandru, Michael D\"oring,
  and Frank~X. Lee.
\newblock {Three-body interactions from the finite-volume QCD spectrum}.
\newblock {\em Phys. Rev. D}, 104(1):014501, 2021.

\bibitem{Hansen:2020zhy}
Maxwell~T. Hansen, Fernando Romero-L\'opez, and Stephen~R. Sharpe.
\newblock {Generalizing the relativistic quantization condition to include all
  three-pion isospin channels}.
\newblock {\em JHEP}, 07:047, 2020.

\bibitem{Blanton:2020gha}
Tyler~D. Blanton and Stephen~R. Sharpe.
\newblock {Alternative derivation of the relativistic three-particle
  quantization condition}.
\newblock {\em Phys. Rev. D}, 102(5):054520, 2020.

\bibitem{Blanton:2020jnm}
Tyler~D. Blanton and Stephen~R. Sharpe.
\newblock {Equivalence of relativistic three-particle quantization conditions}.
\newblock {\em Phys. Rev. D}, 102(5):054515, 2020.

\bibitem{Pang:2020pkl}
Jin-Yi Pang, Jia-Jun Wu, and Li-Sheng Geng.
\newblock {$DDK$ system in finite volume}.
\newblock {\em Phys. Rev. D}, 102(11):114515, 2020.

\bibitem{Hansen:2020otl}
Maxwell~T. Hansen, Ra\'ul~A. Brice\~no, Robert~G. Edwards, Christopher~E.
  Thomas, and David~J. Wilson.
\newblock {Energy-Dependent $\pi^+ \pi^+ \pi^+$ Scattering Amplitude from QCD}.
\newblock {\em Phys. Rev. Lett.}, 126:012001, 2021.

\bibitem{Romero-Lopez:2020rdq}
Fernando Romero-L\'opez, Akaki Rusetsky, Nikolas Schlage, and Carsten Urbach.
\newblock {Relativistic $N$-particle energy shift in finite volume}.
\newblock {\em JHEP}, 02:060, 2021.

\bibitem{Blanton:2020gmf}
Tyler~D. Blanton and Stephen~R. Sharpe.
\newblock {Relativistic three-particle quantization condition for nondegenerate
  scalars}.
\newblock {\em Phys. Rev. D}, 103(5):054503, 2021.

\bibitem{Muller:2020vtt}
Fabian M\"{u}ller, Akaki Rusetsky, and Tiansu Yu.
\newblock {Finite-volume energy shift of the three-pion ground state}.
\newblock {\em Phys. Rev. D}, 103(5):054506, 2021.

\bibitem{Blanton:2021mih}
Tyler~D. Blanton and Stephen~R. Sharpe.
\newblock {Three-particle finite-volume formalism for $\pi^+\pi^+K^+$ and
  related systems}.
\newblock {\em Phys. Rev. D}, 104(3):034509, 2021.

\bibitem{Muller:2021uur}
Fabian M\"uller, Jin-Yi Pang, Akaki Rusetsky, and Jia-Jun Wu.
\newblock {Relativistic-invariant formulation of the NREFT three-particle
  quantization condition}.
\newblock {\em JHEP}, 02:158, 2022.

\bibitem{Beane:2007es}
Silas~R. Beane, William Detmold, Thomas~C. Luu, Kostas Orginos, Martin~J.
  Savage, and Aaron Torok.
\newblock {Multi-Pion Systems in Lattice QCD and the Three-Pion Interaction}.
\newblock {\em Phys. Rev. Lett.}, 100:082004, 2008.

\bibitem{Detmold:2008fn}
William Detmold, Martin~J. Savage, Aaron Torok, Silas~R. Beane, Thomas~C. Luu,
  Kostas Orginos, and Assumpta Parreno.
\newblock {Multi-Pion States in Lattice QCD and the Charged-Pion Condensate}.
\newblock {\em Phys. Rev.}, D78:014507, 2008.

\bibitem{Detmold:2008yn}
William Detmold, Kostas Orginos, Martin~J. Savage, and Andre Walker-Loud.
\newblock {Kaon Condensation with Lattice QCD}.
\newblock {\em Phys. Rev. D}, 78:054514, 2008.

\bibitem{Blanton:2019vdk}
Tyler~D. Blanton, Fernando Romero-L\'opez, and Stephen~R. Sharpe.
\newblock {$I = 3$ three-pion scattering amplitude from lattice QCD}.
\newblock {\em Phys. Rev. Lett.}, 124(3):032001, 2020.

\bibitem{Horz:2019rrn}
Ben H\"{o}rz and Andrew Hanlon.
\newblock {Two- and three-pion finite-volume spectra at maximal isospin from
  lattice QCD}.
\newblock {\em Phys. Rev. Lett.}, 123(14):142002, 2019.

\bibitem{Culver:2019vvu}
Chris Culver, Maxim Mai, Ruair\'\i{} Brett, Andrei Alexandru, and Michael
  D\"{o}ring.
\newblock {Three pion spectrum in the $I=3$ channel from lattice QCD}.
\newblock {\em Phys. Rev. D}, 101(11):114507, 2020.

\bibitem{Fischer:2020jzp}
Matthias Fischer, Bartosz Kostrzewa, Liuming Liu, Fernando Romero-L\'opez,
  Martin Ueding, and Carsten Urbach.
\newblock {Scattering of two and three physical pions at maximal isospin from
  lattice QCD}.
\newblock {\em Eur. Phys. J. C}, 81(5):436, 2021.

\bibitem{Alexandru:2020xqf}
Andrei Alexandru, Ruair\'\i{} Brett, Chris Culver, Michael D\"{o}ring, Dehua
  Guo, Frank~X. Lee, and Maxim Mai.
\newblock {Finite-volume energy spectrum of the $K^-K^-K^-$ system}.
\newblock {\em Phys. Rev. D}, 102(11):114523, 2020.

\bibitem{Romero-Lopez:2018rcb}
Fernando Romero-L\'opez, Akaki Rusetsky, and Carsten Urbach.
\newblock {Two- and three-body interactions in $\varphi^4$ theory from lattice
  simulations}.
\newblock {\em Eur. Phys. J.}, C78(10):846, 2018.

\bibitem{Blanton:2021llb}
Tyler~D. Blanton, Andrew~D. Hanlon, Ben H\"orz, Colin Morningstar, Fernando
  Romero-L\'opez, and Stephen~R. Sharpe.
\newblock {Interactions of two and three mesons including higher partial waves
  from lattice QCD}.
\newblock {\em JHEP}, 10:023, 2021.

\bibitem{Mai:2021nul}
Maxim Mai, Andrei Alexandru, Ruair\'\i{} Brett, Chris Culver, Michael D\"oring,
  Frank~X. Lee, and Daniel Sadasivan.
\newblock {Three-Body Dynamics of the a1(1260) Resonance from Lattice QCD}.
\newblock {\em Phys. Rev. Lett.}, 127(22):222001, 2021.

\bibitem{Muller:2020wjo}
Fabian M\"uller and Akaki Rusetsky.
\newblock {On the three-particle analog of the Lellouch-L\"uscher formula}.
\newblock {\em JHEP}, 03:152, 2021.

\bibitem{Muller:2022oyw}
Fabian M\"uller, Jin-Yi Pang, Akaki Rusetsky, and Jia-Jun Wu.
\newblock {Three-particle Lellouch-L\"uscher formalism in moving frames}.
\newblock {\em JHEP}, 02:214, 2023.

\bibitem{Hansen:2021ofl}
Maxwell~T. Hansen, Fernando Romero-L\'opez, and Stephen~R. Sharpe.
\newblock {Decay amplitudes to three hadrons from finite-volume matrix
  elements}.
\newblock {\em JHEP}, 04:113, 2021.

\bibitem{Blanton:2021eyf}
Tyler~D. Blanton, Fernando Romero-L\'opez, and Stephen~R. Sharpe.
\newblock {Implementing the three-particle quantization condition for
  $\pi^+\pi^+ K^+$ and related systems}.
\newblock {\em JHEP}, 02:098, 2022.

\bibitem{Severt:2022jtg}
Daniel Severt, Maxim Mai, and Ulf-G. Mei\ss{}ner.
\newblock {Particle-dimer approach for the Roper resonance in a finite volume}.
\newblock {\em JHEP}, 04:100, 2023.

\bibitem{Baeza-Ballesteros:2023ljl}
Jorge Baeza-Ballesteros, Johan Bijnens, Tom\'a\v{s} Husek, Fernando
  Romero-L\'opez, Stephen~R. Sharpe, and Mattias Sj\"o.
\newblock {The isospin-3 three-particle K-matrix at NLO in ChPT}.
\newblock {\em JHEP}, 05:187, 2023.

\bibitem{Draper:2023xvu}
Zachary~T. Draper, Maxwell~T. Hansen, Fernando Romero-L\'opez, and Stephen~R.
  Sharpe.
\newblock {Three relativistic neutrons in a finite volume}.
\newblock {\em JHEP}, 07:226, 2023.

\bibitem{Bubna:2023oxo}
Rishabh Bubna, Fabian M\"uller, and Akaki Rusetsky.
\newblock {Finite-volume energy shift of the three-nucleon ground state}.
\newblock {\em Phys. Rev. D}, 108(1):014518, 2023.

\bibitem{Hansen:2019nir}
Maxwell~T. Hansen and Stephen~R. Sharpe.
\newblock {Lattice QCD and Three-particle Decays of Resonances}.
\newblock {\em Ann. Rev. Nucl. Part. Sci.}, 69:65--107, 2019.

\bibitem{Mai:2021lwb}
Maxim Mai, Michael D\"oring, and Akaki Rusetsky.
\newblock {Multi-particle systems on the lattice and chiral extrapolations: a
  brief review}.
\newblock {\em Eur. Phys. J. ST}, 230(6):1623--1643, 2021.

\bibitem{Lellouch:2000pv}
Laurent Lellouch and Martin Luscher.
\newblock {Weak transition matrix elements from finite volume correlation
  functions}.
\newblock {\em Commun. Math. Phys.}, 219:31--44, 2001.

\bibitem{Peterken:2023zwu}
Toby Peterken and Maxwell~T. Hansen.
\newblock {Higher partial wave contamination in finite-volume 1-to-2
  transitions}.
\newblock 4 2023.

\bibitem{Meyer:2011um}
Harvey~B. Meyer.
\newblock {Lattice QCD and the Timelike Pion Form Factor}.
\newblock {\em Phys. Rev. Lett.}, 107:072002, 2011.

\bibitem{Hansen:2012tf}
Maxwell~T. Hansen and Stephen~R. Sharpe.
\newblock {Multiple-channel generalization of Lellouch-Luscher formula}.
\newblock {\em Phys. Rev. D}, 86:016007, 2012.

\bibitem{Bernard:2012bi}
V.~Bernard, D.~Hoja, U.~G. Mei{\ss}ner, and A.~Rusetsky.
\newblock {Matrix elements of unstable states}.
\newblock {\em JHEP}, 09:023, 2012.

\bibitem{Cirigliano:2011ny}
Vincenzo Cirigliano, Gerhard Ecker, Helmut Neufeld, Antonio Pich, and Jorge
  Portoles.
\newblock {Kaon Decays in the Standard Model}.
\newblock {\em Rev. Mod. Phys.}, 84:399, 2012.

\bibitem{NA482:2007ucr}
J.~R. Batley et~al.
\newblock {Search for direct CP violating charge asymmetries in K+-
  ---\ensuremath{>} pi+- pi+ pi- and K+- ---\ensuremath{>} pi+- pi0 pi0
  decays}.
\newblock {\em Eur. Phys. J. C}, 52:875--891, 2007.

\bibitem{Colangelo:2006va}
Gilberto Colangelo, Juerg Gasser, Bastian Kubis, and Akaki Rusetsky.
\newblock {Cusps in $K\to3\pi$ decays}.
\newblock {\em Phys. Lett. B}, 638:187--194, 2006.

\bibitem{Gasser:2011ju}
Jurg Gasser, Bastian Kubis, and Akaki Rusetsky.
\newblock {Cusps in $K\to3\pi$ decays: a theoretical framework}.
\newblock {\em Nucl. Phys. B}, 850:96--147, 2011.

\bibitem{Ebert:2021epn}
M.~Ebert, H.~W. Hammer, and A.~Rusetsky.
\newblock {An alternative scheme for effective range corrections in pionless
  EFT}.
\newblock {\em Eur. Phys. J. A}, 57(12):332, 2021.

\bibitem{Pang:2022nim}
Jin-Yi Pang, Martin Ebert, Hans-Werner Hammer, Fabian M\"uller, Akaki Rusetsky,
  and Jia-Jun Wu.
\newblock {Spurious poles in a finite volume}.
\newblock {\em JHEP}, 07:019, 2022.

\bibitem{Ebert:2023aio}
M.~Ebert, H.~W. Hammer, and A.~Rusetsky.
\newblock {An Alternative Scheme for Pionless EFT: Neutron-Deuteron Scattering
  in the Doublet S-Wave}.
\newblock {\em Few Body Syst.}, 64(4):87, 2023.

\bibitem{Bedaque:1998kg}
Paulo~F. Bedaque, H.~W. Hammer, and U.~van Kolck.
\newblock {Renormalization of the three-body system with short range
  interactions}.
\newblock {\em Phys. Rev. Lett.}, 82:463--467, 1999.

\bibitem{Bedaque:1998km}
Paulo~F. Bedaque, H.~W. Hammer, and U.~van Kolck.
\newblock {The Three boson system with short range interactions}.
\newblock {\em Nucl. Phys. A}, 646:444--466, 1999.

\bibitem{Bedaque:1998mb}
Paulo~F. Bedaque, H.~W. Hammer, and U.~van Kolck.
\newblock {Effective theory for neutron deuteron scattering: Energy
  dependence}.
\newblock {\em Phys. Rev. C}, 58:R641--R644, 1998.

\bibitem{Doring:2018xxx}
M.~D\"oring, H.~W. Hammer, M.~Mai, J.~Y. Pang, \textsection{}~A. Rusetsky, and
  J.~Wu.
\newblock {Three-body spectrum in a finite volume: the role of cubic symmetry}.
\newblock {\em Phys. Rev. D}, 97(11):114508, 2018.

\bibitem{Glockle:1978zz}
W.~Glockle.
\newblock {S-matrix pole trajectory in a three-neutron model}.
\newblock {\em Phys. Rev. C}, 18:564--572, 1978.

\bibitem{Cahill:1971ddy}
R.~T. Cahill and I.~H. Sloan.
\newblock {Theory of neutron-deuteron break-up at 14.4 MeV}.
\newblock {\em Nucl. Phys. A}, 165:161--179, 1971.
\newblock [Erratum: Nucl.Phys.A 196, 632--632 (1972)].

\bibitem{Garofalo:2022pux}
Marco Garofalo, Maxim Mai, Fernando Romero-L\'opez, Akaki Rusetsky, and Carsten
  Urbach.
\newblock {Three-body resonances in the $\varphi^{4}$ theory}.
\newblock {\em JHEP}, 02:252, 2023.

\bibitem{Dawid:2023jrj}
Sebastian~M. Dawid, Md~Habib~E. Islam, and Ra\'ul~A. Brice\~no.
\newblock {Analytic continuation of the relativistic three-particle scattering
  amplitudes}.
\newblock {\em Phys. Rev. D}, 108(3):034016, 2023.

\bibitem{Schmid}
Erich~W. Schmid and Horst Ziegelmann.
\newblock {\em {Quantum Mechanical Three-body Problem}}.
\newblock Vieweg, 12 1974.

\bibitem{Gockeler:2012yj}
M.~G{\"o}ckeler, R.~Horsley, M.~Lage, U.~G. Mei{\ss}ner, P.~E.~L. Rakow,
  A.~Rusetsky, G.~Schierholz, and J.~M. Zanotti.
\newblock {Scattering phases for meson and baryon resonances on general
  moving-frame lattices}.
\newblock {\em Phys. Rev. D}, 86:094513, 2012.

\bibitem{Buchalla:1995vs}
Gerhard Buchalla, Andrzej~J. Buras, and Markus~E. Lautenbacher.
\newblock {Weak decays beyond leading logarithms}.
\newblock {\em Rev. Mod. Phys.}, 68:1125--1144, 1996.

\bibitem{Bijnens:2002vr}
Johan Bijnens, Pierre Dhonte, and Fredrik Borg.
\newblock {$K\to3\pi$ decays in chiral perturbation theory}.
\newblock {\em Nucl. Phys. B}, 648:317--344, 2003.

\bibitem{Batley:2009ubw}
J.~R. Batley et~al.
\newblock {Determination of the S-wave $\pi\pi$ scattering lengths from a study
  of $K^+\to \pi^+\pi^0 \pi^0$ decays}.
\newblock {\em Eur. Phys. J. C}, 64:589--608, 2009.

\bibitem{Zemah}
Charles Zemah.
\newblock {Three-pion decays of unstable particles}.
\newblock {\em Phys. Rev. B}, 133:1201, 1964.

\end{thebibliography}

\end{document}